\documentclass[qsecnum,amsmath,preprintnumbers,superscriptaddress,nofootinbib,aps,prd,10pt,a4paper]{revtex4-2}
\pdfoutput=1
\usepackage[utf8]{inputenc}
\usepackage{amsmath}
\usepackage{amsfonts}
\usepackage{amssymb}
\usepackage{amsthm}
\usepackage{hyperref}
\usepackage{xcolor}
\usepackage{bm}
\usepackage{graphicx}
\usepackage{feynmp-auto}
\usepackage{mathrsfs}
\usepackage{caption}
\usepackage{subcaption}
\usepackage{framed}

 \def\be{\begin{equation}}
\def\ee{\end{equation}}
 \def\ba{\begin{align}}
\def\ea{\end{align}}
\def\bea{\begin{eqnarray}}
\def\eea{\end{eqnarray}}

\def\a{\alpha}
\def\b{\beta}

\def\m{\mu}
\def\n{\nu}

\newcommand{\bseq}{\begin{subequations}}
\newcommand{\eseq}{\end{subequations}}

\begin{document}
\title{{\bf The shape of Scalar Gauss-Bonnet Gravity}}
\author{Mario Herrero-Valea}
\email[]{mherrero@sissa.it}
\address{SISSA, Via Bonomea 265, 34136 Trieste, Italy and INFN Sezione di Trieste}
\address{IFPU - Institute for Fundamental Physics of the Universe \\Via Beirut 2, 34014 Trieste, Italy}

\begin{abstract}
We study the consistency of Scalar Gauss-Bonnet Gravity, a generalization of General Relativity where black holes can develop non-trivial hair by the action of a coupling $F(\Phi){\cal G}$ between a function of a scalar field and the Gauss-Bonnet invariant of the space-time. When properly normalized, interactions induced by this term are weighted by a cut-off, and take the form of an Effective Field Theory expansion. By invoking the existence of a Lorentz invariant, causal, local, and unitary UV completion of the theory, we derive \emph{positivity bounds} for $n$-to-$n$ scattering amplitudes including exchange of dynamical gravitons. These constrain the value of \emph{all} even derivatives of the function $F(\Phi)$, and are highly restrictive. They require some of the scales of the theory to be of Planckian order, and rule out most of the models used in the literature for black hole scalarization. 
\end{abstract}

\maketitle

\newpage

\section{Introduction and Executive Summary}

Effective field theories (EFT) are wonderful tools to explore physics beyond the establishment. Their philosophy -- based on including new interactions respecting the symmetries of the low energy degrees of freedom -- has been extremely powerful on accounting for new physics within several fields, from particle physics to condensed matter \cite{Cohen:2019wxr}. In the recent years, the outstanding success of cosmological experiments and the beginning of the gravitational wave era, marked by the first detection of gravitational waves from an inspiral binary of black holes by LIGO \cite{Abbott:2016blz}, have opened a window into testing theories that modify or replace General Relativity (GR) in the strong field regime. EFTs of gravity \cite{Donoghue:1994dn} have since become standard in trying to predict future observations of gravitational wave observatories \cite{Barack:2018yly,TheLIGOScientific:2016src}.

One of the most popular, and minimal, family of theories beyond GR are scalar-tensor theories, obtained by extending the gravitational Lagrangian with the presence of a scalar field which interacts with the gravitational degrees of freedom but which is, completely or partially, decoupled from matter species. This idea was born long ago from seminal works by Jordan \cite{Jordan:1959eg}, Fierze \cite{Fierz:1956zz}, and Brans-Dicke \cite{Brans:1961sx}, but it has since evolved onto the discovery of a large family of scalar-tensor theories with second order equations of motion -- thus propagating no ghosts -- dubbed as degenerate higher order scalar-tensor theories (DHOST) \cite{Langlois:2015cwa,Langlois:2017mdk}. This contains not only the theories of Brans-Dicke and Jordan, but also all the terms in the Hordensky \cite{Horndeski:1974wa} and Beyond Hordensky \cite{Gleyzes:2014dya} families, and even more.

The parameter space of scalar-tensor theories, once promising as effective models for dark energy \cite{Clifton:2011jh}, has been strongly constrained in the recent years \cite{Ezquiaga:2017ekz,Baker:2017hug,Creminelli:2017sry,Monitor:2017mdv,TheLIGOScientific:2017qsa,Creminelli:2018xsv,Creminelli:2019kjy,Bezares:2020wkn}. However, most of these constraints are explicitly applicable only at cosmological scales and thus there is still hope that some scalar-tensor theories survive as EFTs to describe other gravitational phenomena. In particular, they have been proposed as interesting models to explore the physics of black holes beyond General Relativity (see \cite{Barausse:2019pri} and references therein). Among all the new physics introduced by DHOST terms, one of their most interesting features is the possibility of violating no-hair theorems \cite{Sotiriou:2015pka}. Black holes in some of these models can develop non-trivial profiles of the scalar field in their surroundings \cite{Damour:1991rd}, which can eventually lead to observational imprints in several observables \cite{Will:2014kxa,Yunes:2013dva}. Scalar hair can induce vacuum dipole gravitational wave emission \cite{Barausse:2016eii,Toubiana:2020vtf}, which could be observed in the low frequency inspiral of a binary system; deviations on the spectrum of quasinormal modes that get excited during the ringdown phase of such an inspiral \cite{Berti:2016lat}, and it might even impact the black hole shadow observed by the Event Horizon Telescope \cite{Akiyama:2019cqa,Psaltis:2020lvx,Volkel:2020xlc}.

In this work we focus on a single term in the DHOST Lagrangian, which has attracted a lot of interest recently -- a coupling between an arbitrary function of the scalar field $\Phi$ and the Euler density of the space-time, given by the Gauss-Bonnet invariant -- $F(\Phi) \cal G$ \cite{Kanti:1995vq,Sotiriou:2013qea,Sotiriou:2014pfa,Julie:2019sab,East:2021bqk}. The theory out-coming from appending this to the Einstein-Hilbert Lagrangian is known as \emph{Scalar Gauss-Bonnet Gravity} (SGBG). This term gives rise to interactions between the scalar field and the background space-time which, around non-trivial topologies, will influence the evolution of the system \cite{Antoniou:2017acq}. In particular, if one expands the function $F(\Phi)$ in a power series around a point $\Phi_0\equiv {\rm constant}$, the coupling with the Gauss-Bonnet term will give rise, among others, to an effective mass for the scalar field $m^2\propto - F''(\Phi_0) {\cal G}$, whose sign depends both on $F''(\Phi_0)$ and on the value of the Euler density. Whenever $m^2<0$, it signals the presence of a tachyonic instability in the dynamics of the scalar field. This implies that the background solution around which the scalar field was expanded is unstable, with a lifetime of order $\tau\sim m^{-1}$. For Schwarzchild black holes, which are solutions of the theory as long as $F'(\Phi_0)=0$, this lifetime can be within astrophysical scales. The final onset of the instability can be another black hole around which the scalar field has developed a non-trivial profile \cite{Silva:2017uqg,Doneva:2017bvd,Herdeiro:2020wei,Berti:2020kgk}, depending on the form of $F(\Phi)$. This process is dubbed as \emph{spontaneous scalarization} of the black hole. An equivalent phenomenon can also happen in neutron stars and other compact objects \cite{Barausse:2012da,Palenzuela:2013hsa,Sennett:2016rwa,Shibata:2013pra}, but we will not consider those here.
 
Since scalarization seems possible for astrophysical black holes, and the presence of scalar hair can lead to observational imprints in future experiments, its detection can serve as a smoking gun to constraint the parameter space of SGBG. Alas, the information that can be obtained from this is sparse. Mostly, scalarization can only be used to fix the sign of the second derivative of the function $F(\Phi)$ in a phenomenological dependent way, while its magnitude might be constrained from the observation of certain long--living populations of spinning black holes, but not much else can be said about the shape of the function $F(\Phi)$ beyond this. Although non-linear terms are important to correctly establish the endpoint of the scalarization process, since they quench the instability after the scalar field has grown enough \cite{Silva:2018qhn,Macedo:2019sem,Blazquez-Salcedo:2018jnn}, their analysis requires the use of numerical methods and can be model-dependent. Moreover, the presence of spontaneous scalarization -- the instability of hairless black holes -- and the possibility of stabilizing hairy black holes do not necessarily need to co-exist in a given theory. Furthermore, there is still the open question of whether one can place some constraints on $F(\Phi$) from pure theoretical arguments, without invoking experimental or observational data. 

When properly normalized, the terms arising from the expansion of the Gauss-Bonnet term around a constant scalar field configuration are suppressed by increasing powers of a dimensionful coupling. This structure matches that of an EFT, whose irrelevant interactions are 1) higher order in derivatives and 2) suppressed by an ultra-violet (UV) cut-off. Both conditions are satisfied by SGBG and therefore we must interpret it as an EFT valid only to describe physics at energies below such cut-off. Its Lagrangian thus corresponds to just the first terms in a low-energy expansion of an unknown high energy theory. This is in addition to the statement that any gravitational theory is naturally an EFT from first principles, due to the non-renormalizability of the Einstein-Hilbert action, with cut-off at the Planck Mass $M_P$.

Although this is not obvious from a bottom-up perspective, where an EFT Lagrangian is defined solely by the symmetries of the infra-red (IR) degrees of freedom, not all possible values of the coupling constants accompanying irrelevant EFT operators can be obtained from its UV completion, whenever it exists, in a top-down approach. Certain properties of the high-energy theory survive down to the IR and imply constraints on the interactions that can be added to the Lagrangian. In particular, assuming the UV completion to be Lorentz invariant, unitary, causal, and local, one can derive conditions on the forward limit of scattering amplitudes around flat space, called \emph{positivity bounds}. They constrain the sign of (a combination of) the couplings in the IR Lagrangian, implying that couplings violating this bound cannot be obtained from such a UV completion. Positivity bounds where first formulated for $2$-to-$2$ scattering amplitudes in massive theories \cite{Nicolis:2009qm,Adams:2006sv}, but have since been extended to more general grounds, including going beyond the forward limit \cite{deRham:2017avq,Bellazzini:2020cot}, beyond the 2-to-2 amplitude \cite{Elvang:2012st,Chandrasekaran:2018qmx}, beyond simple positivity \cite{Bellazzini:2017fep} and, in particular, they have been generalized to theories including gravity, where the results must be restricted to approximate positivity \cite{Hamada:2018dde,Tokuda:2020mlf,Herrero-Valea:2020wxz}. In the latter, the bounds must be satisfied up to the inclusion of gravitational UV effects, which are typically beyond the range of application of the EFT, but which can violate a strict positivity condition. Positivity bounds in the presence of gravity have been used to study the coupling of matter to gravitation in several models \cite{Alberte:2020jsk,Bellazzini:2019xts,Alberte:2020bdz,Aoki:2021ckh}, but their application to modified gravity has been mostly restricted to cases were dynamical gravitons are either neglected \cite{Dvali:2012zc, deRham:2017imi,  Melville:2019wyy,deRham:2021fpu, Davis:2021oce} or massive \cite{Cheung:2016yqr,  deRham:2017xox}.

One could a priori question if bounds obtained from perturbation theory around flat space are applicable in situations where the background space-time is curved, as it happens in the presence of a black-hole. However, it seems doubtful that both situations can actually be decoupled for a given action. Provided that we have a Lagrangian written as an EFT, it has to be able to describe any physics at energy scales below its cut-off, regardless of its curved or flat nature. If some parameters in the Lagrangian make processes within a certain range of energies problematic, the theory will not be healthy also above those energies. This is what would happen in the case at hand. If the SGBG action makes no sense around flat space, where the typical energy carried by a field is small, it cannot make sense either for highly curved space-times. Otherwise, this would imply a breakdown of the EFT approach, a possibility which is clearly not realistic or, at the very least, requires a very strong justification\footnote{As an additional argument to support this statement, we can always think on a dynamical way to build a black hole by accumulating perturbations in a small region of a flat space-time. In perturbation theory, this process can be properly defined in a continuous way \cite{Deser:1957zz}.}. 

In this work we derive positivity bounds for SGBG, thus constraining the parameter space of the theory. First, by looking at an appropiate $2$-to-$2$ scattering amplitude, we obtain a bound for the second derivative of the function $F(\Phi)$, which allows for spontaneous scalarization of black-holes, but tightly constraints the possibility of spin-induced scalarization. Later on, we generalize the derivation of positivity bounds to the case of $n$-to-$n$ amplitudes, which allows us to constrain the sign and size of \emph{all even derivatives} of $F(\Phi$). This imposes strong conditions on the feasibility of popular models of scalarizarion. When the models are not directly ruled out, satisfaction of positivity bounds imply constraints that severely condition the parameters in the theory. 

We start with an executive summary of our results, written in the same conventions used by the accumulated literature in black hole scalarization. If the reader is simply interested on knowing the implications of our results for particular choices of the function $F(\Phi)$, we recommend them to focus on this section and later move ahead to Sections \ref{sec:results} and \ref{sec:conclusions} for a deeper look on the results and conclusions. Otherwise, this paper is organized as follows. First, we introduce Scalar Gauss-Bonnet gravity in Section \ref{sec:SGBG}, describing its main features, our conventions, and discussing the possibility of scalarization of certain types of black-holes. Afterwards, we will discuss positivity of the $2$-to-$2$ scattering amplitude in Section \ref{sec:2to2}, focusing on gravitational Compton scattering and following the results of \cite{Herrero-Valea:2020wxz} in order to obtain a condition on $F''(\Phi)$. In Section \ref{sec:nton}, we discuss $n$-to-$n$ scalar amplitudes mediated by gravity, describing their kinematics. We generalize the derivation of \cite{Herrero-Valea:2020wxz} to these amplitudes under certain natural assumptions about the UV completion. That allows us to derive \emph{new positivity bounds} that we exploit in Section \ref{sec:results}, comparing our results with previous literature when possible. We draw our conclusions and discussion of future work directions on Section \ref{sec:conclusions}. Some details of the computation of interaction vertices and scattering amplitudes, as well as an example on the reduction of the kinematical variables, can be found in the appendices.

%%%%%%%%%%%%%%%%%%%%%%%%%%%%%%%%%%%%%%%%%%%%%%%%
%%%%%%%%%%%%%%%%%%%%%%%%%%%%%%%%%%%%%%%%%%%%%%%%
%%%%%%%%%%%%%%%%%%%%%%%%%%%%%%%%%%%%%%%%%%%%%%%%
%%%%%%%%%%%%%%%%%%%%%%%%%%%%%%%%%%%%%%%%%%%%%%%%
\subsection*{Executive Summary}
Let us start by summarizing our results here in a compact form, suitable for those which are interested only on knowing the consequences of positivity on SGBG, but which do not want to delve into the details of the derivation. 

Our findings can be stated as follows.We assume that SGBG, with action\footnote{Here we are using mostly minus signature. Notice that addition of a small mass $m$ and quartic coupling $\lambda \Phi^4$ to the action will not change our results here. The former will only correct the bounds below at sub-leading order ${\cal O}\left(m^2/M_P^4\right)$, while the latter is a renormalizable coupling. It never contributes to momentum dependent pieces in the scattering amplitudes at leading order, thus dropping from the bounds.}
\begin{align}\label{eq:action_other}
    S=M_P^2\int d^4x\sqrt{|g|}\left(-\frac{R}{2} + \frac{1}{2}\partial_\m \Phi \partial^\m \Phi + F(\Phi){\cal G}\right),
\end{align}
is the low-energy limit of a local, causal, unitary and Lorentz invariant UV theory, which lives above a certain cut-off $\Lambda\ll M_P$. 
Then, the following bounds must be satisfied\footnote{Of course,  the case of some derivatives $dF^{2n}/d\Phi^{2n}$ vanishing is allowed by our bounds.  In that case,  there are some models that might be able to evade the conclusions below.  However,  those would not lead to scalarization of black holes,  and their analysis is beyond the scope of this work.}
\begin{align}\label{eq:results}
\left.  \frac{d^2 F}{d\Phi^2}\right|_{\Phi=\Phi_0}>-{\cal O}\left(\frac{1}{M_*^2}\right),\quad  \left(\frac{M_P^2}{\Lambda^2}\right)^{n-1}\left. \frac{d^{2n} F}{d\Phi^{2n}}\right|_{\Phi=\Phi_0}<{\cal O}\left(\frac{1}{M_P^2}\right), \quad n\geq 2,
\end{align}
where $\Phi_0$ is any extreme of the function $F(\Phi)$ -- thus $F'(\Phi_0)=0$ -- and $M_*$ is the scale at which Quantum Gravity effects start to become important. In many models, such as String Theory, $M_*$ can be significantly lower than $M_P$. Notice that the option of vanishing derivatives is always allowed by the bounds.

While these conditions are simple, they are incredibly restrictive in most cases. For instance, provided that the function $F(\Phi)$ and its derivatives are analytic, and that the equation $F'(\Phi)=0$ has more than a single solution, we can safely put into question the validity of the model. The reason being that the combination of these two assumptions implies that $F''(\Phi_0)$ must be negative in at least one of the extremes of $F(\Phi)$. In that case, \eqref{eq:results} bounds the size of $F''(\Phi)$, which is a length scale squared, to be smaller than the corresponding length at which Quantum Gravity effects become appreciable. This implies that all the possible higher order gravitational operators in the EFT, that were ignored when writing the low energy action \eqref{eq:action_other}, cannot be neglected anymore, since its size becomes comparable to that of $F(\Phi){\cal G}$. Even if this were not the case, the typical size of the derivatives must be much smaller than the scales usually studied in scalarization works, which are of the order of the horizon radius of the black hole. This clearly puts the validity of these models in tension.

In the particular case of polynomial shapes of $F(\Phi)$, our results seem to imply that the only surviving possible option satisfying positivity bounds is the simple
\begin{align}
    F(\Phi)= F_0 + F_1 \Phi + F_2 \Phi^2,
\end{align}
where $F_2$ must satisfy the bound in \eqref{eq:results} for the second derivative. Any higher-order polynomial is either inconsistent or requires coefficients smaller than the Quantum Gravity scale, thus exiting the range of validity of the action \eqref{eq:action_other}.
%%%%%%%%%%%%%%%%%%%%%%%%%%%%%%%%%%%%%%%%%%%%%%%%
%%%%%%%%%%%%%%%%%%%%%%%%%%%%%%%%%%%%%%%%%%%%%%%%
%%%%%%%%%%%%%%%%%%%%%%%%%%%%%%%%%%%%%%%%%%%%%%%%
%%%%%%%%%%%%%%%%%%%%%%%%%%%%%%%%%%%%%%%%%%%%%%%%

\section{Scalar Gauss-Bonnet Gravity}\label{sec:SGBG}
In order to derive the bounds summarized previously, we  formulate Scalar Gauss-Bonnet gravity through the slightly different action
\begin{align}\label{eq:starting_action}
S=\int d^4x \sqrt{|g|}\ \left(-\frac{M_P^2}{2}R+\frac{1}{2}\nabla_\m \phi \nabla^\m \phi +f\left(\phi/\Lambda\right){\cal G}\right),
\end{align}
where $\phi$ is a scalar field, $M_P$ is the Planck Mass, $R$ refers to the scalar curvature of the manifold, and we are using mostly minus signature henceforward. We have also chosen units where $c=\hbar=1$. Here ${\cal G}$ is the Euler or Gauss-Bonnet (GB) density of the metric manifold, which is given in terms of the Riemann tensor $R_{\m\n\a\b}$ as
\begin{align}
    {\cal G}=R_{\m\n\a\b}R^{\m\n\a\b}+R^2-4 R_{\m\n}R^{\m\n}.
\end{align}

The function $f(x)$ is in principle arbitrary, but we have decided to explicit the scale $\Lambda\ll M_P$ so that $f(\phi/\Lambda)$ remains a dimensionless function of a dimensionless argument. Notice that this definition differs from the one in \eqref{eq:action_other}, and from those that can be found in the literature in black hole scalarization, which use a dimensionless scalar field. Both actions can be related by letting
\begin{align}\label{eq:change_var}
&\phi=M_P \Phi , \\
&f(\Phi M_P/\Lambda)= M_P^2 \ F(\Phi).
\end{align}
In the following we will stick however to the choice \eqref{eq:starting_action}, where the scalar field is canonically normalized, in order to be able to keep the suppression of different terms along our computation under control. Let us also point out that the case $\Lambda\gtrsim M_P$ is beyond the scope of this work, since that would require to know physics beyond the gravitational UV cut-off.

Varying the action \eqref{eq:starting_action} with respect to $\phi$, we get the equation of motion for the scalar field
\begin{align}\label{eq:eom_scalar}
    \square\phi=-\frac{f'(\phi/\Lambda) {\cal G}}{\Lambda},
\end{align}
where $f'(x)=\frac{df}{dx}$, while variation with respect to the metric leads to the modified Einstein equations
\begin{align}\label{eq:eom_g}
    {\cal E}_{\m\n}\equiv R_{\m\n}-\frac{1}{2}g_{\m\n}R-\frac{1}{M_P^2}\left(\nabla_\m\phi\nabla_\n \phi -\frac{1}{2}g_{\m\n}(\nabla \phi)^2\right)+\frac{2}{M_P^2}{\cal J}_{\m\n}=0,
\end{align}
where
\begin{align}
  \nonumber {\cal J}^{\a\b}&=\frac{f'(\phi/\Lambda)}{\Lambda}\left( -2 R \nabla^{\b}\nabla^{\a}\phi - 4 R^{\a\b} 
\nabla_{\m}\nabla^{\m}\phi + 2 g^{\a\b} R
\nabla_{\m}\nabla^{\m}\phi + 4 R^{\b}{}_{\m} 
\nabla^{\m}\nabla^{\a}\phi + 4 R^{\a}{}_{\m} 
\nabla^{\m}\nabla^{\b}\phi  \right.\\
\nonumber &\left.- 4 g^{\a\b} R_{\m\n} 
\nabla^{\n}\nabla^{\m}\phi+ 2 R^{\a}{}_{\m}{}^{\b}{}_{\n} 
\nabla^{\n}\nabla^{\m}\phi + 2 R^{\a}{}_{\n}{}^{\b}{}_{\m} 
\nabla^{\n}\nabla^{\m}\phi\right)+\frac{f''(\phi/\Lambda)}{\Lambda^2}\left(-2 R \nabla^{\a}\phi \nabla^{\b}\phi   \right.\\
\nonumber &+ 4 R^{\b}{}_{\m} \nabla^{\a}\phi \nabla^{\m}\phi + 4 R^{\a}{}_{\m} \nabla^{\b}\phi \nabla^{\m}\phi- 4 R^{\a\b} 
\nabla_{\m}\phi \nabla^{\m}\phi+ 2 g^{\a\b} R
\nabla_{\m}\phi \nabla^{\m}\phi - 4 g^{\a\b} R_{\m\n} 
\nabla^{\m}\phi \nabla^{\n}\phi \\
&\left.+ 4 R^{\a}{}_{\m}{}^{\b}{}_{\n} \nabla^{\m}\phi \nabla^{\n}\phi \right).
\end{align}

From this, we can easily note that solutions with constant scalar field $\phi =\phi_0$ and for which $f'(\phi_0/\Lambda)=0$, will also be solutions of the equations of motion of General Relativity, understood as the same action with the GB term absent. This is true, in particular, for flat space, which is a vacuum solution to the equations of motion as long as this condition is satisfied. Hereinafter we will label those values which extremise the function $f(\phi/\Lambda)$ as $\phi_0$. Notice that functions $f(x)$ with several extremes thus have several GR limits and several flat vacua, with no way to single out one of them as preferred.

For non-constant solutions and around backgrounds with non-trivial topology, the contribution of the GB term will influence the dynamics of the scalar field. In particular, let us assume that $f(x)$ is analytic around vacuum solutions. In that case we can always expand it in a power series around $\phi_0$, getting
\begin{align}
    f(\phi/\Lambda)=f_0 +\frac{f_2}{2} (\phi-\phi_0)^2+\frac{f_3}{6} (\phi-\phi_0)^3+\dots
\end{align}
where we are using the short-hand notation
\begin{align}
    f_j=\frac{1}{\Lambda^j}\left.\frac{d^j f(x)}{dx^j}\right|_{x=\frac{\phi_0}{\Lambda}}.
\end{align}

After a shift of the scalar field by $\sigma=\phi-\phi_0$, the action then becomes
\begin{align}\label{eq:action_sigma}
S=\int d^4x \sqrt{|g|}\ \left(-\frac{M_P^2}{2}R+\frac{1}{2}\nabla_\m \sigma \nabla^\m \sigma +\sum_{j=2}^\infty \frac{f_j {\cal G}}{j!}\sigma^j\right),
\end{align}
where we have got rid of the first two terms in the sum, since they correspond to a total derivative ($j=0$) and a term which vanishes around extremes of $f(x)$ ($j=1)$. The combinations $\lambda_j=-f_j {\cal G}$, whenever non-vanishing, have become couplings which control the behavior of the scalar field. In particular, $\sigma$ acquires an effective mass
\begin{align}
    m^2=\lambda_2=-f_2 {\cal G},
\end{align}
whose value and sign depend on those of ${\cal G}$. Notice that all interactions are suppressed by powers of $\Lambda$, thus behaving as an EFT expansion with cut-off $\Lambda$, as previously advertised. The action \eqref{eq:starting_action} thus corresponds only to the first few operators in the low-energy expansion of its unknown UV completion, with sub-leading terms suppressed by higher powers of the cut-off. An example of such kind of terms can be seen in the Appendix \ref{app:amplitudes}, as they are generated by the loop corrections shown there. The physics described by the SGBG action \eqref{eq:starting_action} is thus only dependable as long as all energy scales involved are significantly smaller than $\Lambda$.

An alternative argument supporting the interpretation of SGBG as an effective field theory comes from the separation of scales between $\Lambda$ and $M_P$. For all values of energy below the latter, gravitational interactions are sub-leading and suppressed by powers of $M_P$. However, they still leave an imprint on the dynamics of the scalar field. This can be formally accounted by integrating out the gravitational field, which will produce a tower of effective operators for the scalar field that survive even in the flat space limit, ordered by an increasing number of derivatives and with a cut-off suppression. Again, this has the form of an EFT expansion.

The value of $m^2$ leads to the signature feature of the theory -- the possibility of \emph{scalarization} of black holes. Since black holes are topologically non-trivial, their Gauss-Bonnet density is non-vanishing. In particular, for the Schwarzchild metric -- which we stress that it is a solution also to equations \eqref{eq:eom_scalar} and \eqref{eq:eom_g} as long as $f'(\phi_0/\Lambda)=0$ -- it reads
\begin{align}
    {\cal G}_{\rm Schwarzchild}=\frac{3}{4\pi^2}\frac{M^2}{M_P^4}\frac{1}{r^6},
\end{align}
where $M$ is the mass of the black-hole and $r$ is the radial coordinate. Notice that this is explicitly positive for all radii. Thus, for choices of the function $f(x)$ such that $f_2>0$, we find $m^2<0$ and the scalar field presents a tachyonic behavior around this space-time. As a consequence, the Schwarzchild solution does not remain stable and eventually decays onto a new space-time in a dynamical process in which the scalar field evolves into a non-trivial profile, developing hair around the newly formed black-hole, which is said to be scalarized.

Although scalarization only occurs on non-rotating black holes for the appropriate sign of $f_2$, the situation becomes more complicated for rotating ones, where the sign of the GB term depends not only on the spin $a$ and the mass $M$ of the black hole, but also on the distance from the horizon. In particular, for the Kerr metric we have
\begin{align}
    {\cal G}_{\rm Kerr}=\frac{3}{4\pi^2}\frac{M^2}{M_P^4}\left(\frac{r^6 - 15 r^4 \chi^2 +15 r^2 \chi^4 -\chi^6}{(r^2 +\chi^2)^6}\right),
\end{align}
where we are using Boyer-Lindquist coordinates and $\chi=a \cos\theta$. This opens up the possibility of scalarization existing for both signs of $f_2$ for an appropriate size of the angular momentum, a process that has been dubbed \emph{spin-induced scalarization}\cite{Herdeiro:2020wei,Dima:2020yac,Berti:2020kgk}. Indeed, it was shown in Ref. \cite{Dima:2020yac} that for $f_2<0$, scalarization is the main decay channel for Kerr black-holes -- dominating over superradiant instability. 

The possibility of scalarization, together with the fact that scalarised black holes produce a different imprint on possible observational quantities -- such as dipole gravitational wave emission -- implies that this process can be used to test the viability of the action \eqref{eq:starting_action} as a modified theory of Gravitation. Note however that all these possible experimental tests are bound to constraint only the parameter space of some of the couplings $\lambda_n$, but the full function $f(x)$ typically remains inaccessible. In particular, the mere possibility of scalarization only probes $\lambda_2$.

%%%%%%%%%%%%%%%%%%%%%%%%%%%%%%%%%%%%%%%%%%%%%%%%%
%%%%%%%%%%%%%%%%%%%%%%%%%%%%%%%%%%%%%%%%%%%%%%%%%
%%%%%%%%%%%%%%%%%%%%%%%%%%%%%%%%%%%%%%%%%%%%%%%%%
\section{Positivity of the $2\rightarrow 2$ amplitude}\label{sec:2to2}

Even though EFTs are very rich from the point of view of low energy phenomenology, where the Lagrangian is appended with operators under the only restriction of preserving symmetries, not every possible value of the extra couplings -- known as Wilson coefficients -- is actually allowed by UV physics. Compatibility of the EFT expansion with the existence of a Lorentz invariant, unitary, local and causal UV completion imposes constraints on the value of the Wilson coefficients.

In particular, as shown in \cite{Nicolis:2009qm,Adams:2006sv}, any EFT with such a completion must satisfy positivity bounds, which in their simplest form read
\begin{align}\label{eq:positivity}
    \left.\frac{d^2{\cal A}(s)}{ds^2}\right|_{s=0}>0,
\end{align}
where ${\cal A}(s)$ is the $ab\rightarrow ab$ scattering amplitude in flat space between some states in the EFT, evaluated in the forward limit $t\rightarrow 0$. Here $s$ and $t$ are the usual Mandelstam variables, describing the (squared) energy in the center of mass frame and the momentum transferred in the transverse channel, respectively.

We might be tempted to apply this to SGBG by looking at $\sigma \sigma \rightarrow \sigma \sigma$ amplitudes. However, nothing would come out of it due to two properties of the action \eqref{eq:starting_action}. First, the bound \eqref{eq:positivity} is not straightforwardly applicable to theories with massless particles in the intermediate channel, as it is the case with gravitons in SGBG. In this situation, the scattering amplitude in the $t$-channel will contain a pole $t^{-1}$ at tree-level and non-analiticities of the form $\log(t)$ at loop level, due to exchange and production of gravitons. These divergences impede to take the forward limit required to use \eqref{eq:positivity}. This problem can in principle be circumvented by subtracting the pole and the logarithm against the UV behaviour of the amplitude \cite{Herrero-Valea:2020wxz,Tokuda:2020mlf}, which renders a finite approximate positivity bound. However, we must note that the leading term for $\sigma \sigma \rightarrow \sigma \sigma$ in SGBG, which dominates the bound at low energies, is given by tree-level exchange of gravitons through the kinetic term of the scalar field, with no presence of the function $f(x)$. As a consequence, the positivity bound will only depend on the structure of the kinetic term, with any dependence on $f(x)$ laying on sub-leading terms whose sign cannot be constrained. Due to this, the positivity bound will be satisfied automatically, and it will not give us any new information whatsoever -- cf. Section IV of \cite{Herrero-Valea:2020wxz}.

A more interesting approach is to look instead at the channel $\sigma h \rightarrow \sigma h$, where $h_{\m\n}=g_{\m\n}-\eta_{\m\n}$ is the graviton perturbation, defined as the linear deviation of the gravitational field with respect to a flat background $\eta_{\m\n}$. The leading contribution to this amplitude involves both the kinetic term \emph{and} the GB term and therefore we will be able to obtain a condition on $f(x)$ from it. It reads
\begin{align}\label{eq:2to2}
    {\cal A}(s,t)=\frac{s}{M_P^2}\frac{s^2 + 3 st +3t^2}{t (s+t)}+\frac{f_2 s^2 + 2 f_2 t^2 + 2 s (f_2 t -1)}{M_P^2}.
\end{align}
The detailed computation can be found in Appendix \ref{app:amplitudes}.

As previously commented, this amplitude diverges as $t^{-1}$ in the forward limit due to graviton exchange, and thus requires regularization before computing any bound. Following \cite{Herrero-Valea:2020wxz,Tokuda:2020mlf}, this can be done simply by canceling the divergent term against a piece of the UV contribution of gravitational degrees of freedom in the Regge limit, which we assume here. As a consequence, the finite piece satisfies instead an approximate positivity bound
\begin{align}\label{eq:app_positivity}
    \left.\frac{d^2{\cal A}(s)}{ds^2}\right|_{s=0}>-{\cal O}\left(M_*^{-2}M_P^{-2}\right).
\end{align}
Here the scale $M_*$ controls the entrance of new \emph{gravitational} degrees of freedom, coming from unknown avatars of Quantum Gravity, into the spectrum of the EFT. Violations of positivity are induced by our lack of knowledge about the precise dynamics of these degrees of freedom. Depending on the details of the putative UV completion for gravity, $M_*$ can in principle be lower than the Planck Mass. For instance, if such completion were String Theory, $M_*$ would correspond to the string mass scale $M_*\sim \alpha_s^{-1/2}$, which can be as low as $10^{16}\ {\rm GeV}$. Through this work we will always consider that $M_*\geq \Lambda$. Otherwise, the EFT provided by SGBG would not capture all the relevant degrees of freedom from the beginning and it would be useless.

Taking the second derivative of the amplitude \eqref{eq:2to2} we thus find
\begin{align}
    \frac{f_2}{4M_P^2}>-{\cal O}\left(M_*^{-2}M_P^{-2}\right),
\end{align}
and, provided that the energy scales of interest remain within the range of application of the EFT, we conclude that
\begin{align}\label{eq:2to2bound}
    \left.\frac{d^2f(x)}{dx^2}\right|_{x=\frac{\phi_0}{\Lambda}}>-{\cal O}\left(\frac{\Lambda^2}{M_*^2}\right).
\end{align}

We see that possible violations of positivity in this bound are related to the separation of scales between $\Lambda$, the cut-off at which the dynamics of the scalar field demands the introduction of new operators, and $M_*$. As these scales become closer, there is a larger leaking of the unaccounted effects of hidden UV gravitational degrees of freedom onto the EFT dynamics, which allows for larger violations of positivity.

%%%%%%%%%%%%%%%%%%%%%%%%%%%%%%%%%%%%%%%%%%%%%%%%
%%%%%%%%%%%%%%%%%%%%%%%%%%%%%%%%%%%%%%%%%%%%%%%%
%%%%%%%%%%%%%%%%%%%%%%%%%%%%%%%%%%%%%%%%%%%%%%%%
%%%%%%%%%%%%%%%%%%%%%%%%%%%%%%%%%%%%%%%%%%%%%%%%

\section{Positivity of $n\rightarrow n$ Scalar Amplitudes in SGBG}\label{sec:nton}

Our strategy to further constrain the function $f(x)$ in the action \eqref{eq:starting_action} will be to focus now on the analytic properties of more general scattering amplitudes, extending positivity bounds to their case. Hereinafter we will focus on $n\rightarrow n$ scattering amplitudes ${\cal A}^n(p_1,p_2,...,p_{2n})$ in flat space, where the external legs will always correspond to scalar fields and $n>2$. In order to simplify the calculations below and get rid of problematic IR divergences in intermediate steps, we add to the action a mass for the scalar field perturbation
\begin{align}\label{eq:action_chi}
S=\int d^4x \sqrt{|g|}\ \left(-\frac{M_P^2}{2}R+\frac{1}{2}\nabla_\m \sigma \nabla^\m \sigma - \frac{{\mathfrak m}^2}{2} \sigma^2+\sum_{j=2}^\infty \frac{f_j {\cal G}}{j!}\sigma^j\right),
\end{align}
where ${\mathfrak m}\ll \Lambda$. Even though formally we are changing the dynamics of the theory by adding this mass, our results remain valid also in the limit $\mathfrak{m}\rightarrow 0$, which can be always taken smoothly without changing the number of degrees of freedom in the IR. We will do so after computing the relevant amplitudes, as discussed in Appendix \ref{app:amplitudes}.

%%%%%%%%%%%%%%%%%%%%%%%%%%%%%%%%%%%%%%%%%%
%%%%%%%%%%%%%%%%%%%%%%%%%%%%%%%%%%%%%%%%%%
%%%%%%%%%%%%%%%%%%%%%%%%%%%%%%%%%%%%%%%%%%
\subsection{UV completing SGBG}\label{sec:assumptions}

As previously discussed, the standard way to constrain EFTs through positivity bounds is to look at the analytic properties of the $2$-to-$2$ scattering amplitude in the forward limit. For this particular process, physical properties such as Lorentz invariance and causality completely determine the analytic structure of ${\cal A}^2(s,0)$, fixing the position of all poles and branch cuts and thus allowing to use Cauchy's integral theorem to define a dispersion relation. However, general $n$-to-$n$ amplitudes have a more complicated structure, due to the possible presence of inelastic contributions. For instance, in a $3$-to-$3$ scattering we could expect poles not only when the channel $123\rightarrow 456$ goes on-shell, but also when $12\rightarrow 3456$, and all other possible permutations, do so. This plagues the scattering amplitude with poles and cuts whose sign is not fixed and obstructs the derivation of positivity bounds. In order to proceed further, we thus need to make some extra assumptions about the UV completion of SGBG which, although they are well justified in our setting here, should be kept in mind when discussing the possibility of extending this theory up to arbitrary energy. Our final goal is to be able to derive a positivity bound equivalent to \eqref{eq:app_positivity}.

We start by noting that the low energy action \eqref{eq:starting_action} satisfies certain interesting properties that make SGBG special, i.e. there are no massive scalar poles and there are no tree-level contributions to scalar scattering amplitudes\footnote{Actually, in the absence of dynamical gravity, there are no contributions at all to scalar scattering amplitudes.}. This is of course a consequence of the structure of the interaction term $f(\phi/\Lambda){\cal G}$, where all powers of $\phi$ enter homogeneously and coupled only through gravitational operators. As an upshot of this, scattering amplitudes with scalar degrees of freedom in the external legs will contain no divergence or discontinuity other than those implied by graviton exchange and production. This implies, in particular, that worrying inelastic contributions, as well as connected terms in the amplitudes, will be absent, at least until UV gravitational degrees of freedom kick in. They could be constructed from disconnected terms though. However, as we argue in the next section, those will always be sub-leading, thus allowing us to discard them. This statement holds as long as the theory remains in weak coupling, where the loop expansion of the amplitudes can be trusted. A phase transition onto strong coupling could however introduce new massive poles by a rearranging of the degrees of freedom. We will thus hereinafter assume that the theory remains weakly coupled until the Regge scale $M_*$, which allows us to safely discard inelastic contributions. We will also assume that, for energies above $M_*$, the scattering amplitudes democratize, all of them following Regge trajectories analogous to the one of the $2$-to-$2$ amplitude \cite{Herrero-Valea:2020wxz}
\begin{align}
    {\cal A}(s,t)\sim \left(\frac{s}{M_*^2}\right)^{c_{R}+l(t)} \left(1+\dots\right),
\end{align}
where $c_R$ is constant and $l(t)<0$ and the dots stand for sub-leading terms in $s/M_*^2$. Later we will assume that the value of $c_R$ is such that the total cross-section of all processes satisfies a bound analogous to the Froissart-Martin bound $\left.\bar{\sigma}\right|_{s\rightarrow \infty}\sim \log^2(s)$. A consequence of this choice is that the positivity bounds that we will derive are only \emph{approximate}, with its applicability already diminished by the accounting of unknown gravitational UV effects analogous to those in the 2-to-2 amplitude (cf. formula \eqref{eq:app_positivity} and reference \cite{Herrero-Valea:2020wxz}).

This democratization of amplitudes is a natural assumption if we demand that all scalar-gravity interactions remain of the form
\begin{align}
    {\cal O}[f(\phi/\Lambda)]\times {\cal O}'[g_{\m\n}],
\end{align}
where ${\cal O}$ and ${\cal O}'$ are generic operators, at all energies. This property holds at weak coupling -- cf. Appendix \ref{app:amplitudes} -- and it is reasonable to demand it also above the cut-off.

%%%%%%%%%%%%%%%%%%%%%%%%%%%%%%%%%%%%%%%%%%%%%%%%%%%%%%%%
%%%%%%%%%%%%%%%%%%%%%%%%%%%%%%%%%%%%%%%%%%%%%%%%%%%%%%%%
%%%%%%%%%%%%%%%%%%%%%%%%%%%%%%%%%%%%%%%%%%%%%%%%%%%%%%%%
\subsection{$n\rightarrow n$ scattering amplitudes}

From now on we will consider scattering amplitudes for processes $\sigma\overbrace{\dots}^{n} \sigma\rightarrow \sigma\overbrace{\dots}^{n} \sigma$ with $n$ scalar fields $\sigma$ both in the initial and final states, and focus on the leading contributions from the interaction vertices contained in the action, including exchange of dynamical gravitons. Notice that from the Einstein-Hilbert term and the kinetic terms of the scalar fields we obtain vertices containing up to two legs with scalar fields, and as many graviton legs as we wish, including self-interactions of the latter. From the GB terms instead, we obtain vertices containing any number of $\sigma$ legs, labelled by the coupling $f_j$, with $j$ the number of scalars; and at least two graviton legs. The linear term in the graviton fluctuation vanishes around flat space, since $\delta(\sqrt{|g|}{\cal G})/\delta{g_{\m\n}}\propto {\rm Riemann}$. The explicit structure of the interaction vertices is shown in appendix \ref{app:perturbation}.

%%%%%%%%%%%%%%%%%%%%%%%%%
%%%%%%%%%%%%%%%%%%%%%%%%%
%%%%%%%%%%%%%%%%%%%%%%%%%
%%%%%%%%%%%%%%%%%%%%%%%%%
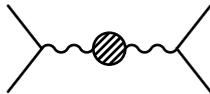
\begin{figure}
  \centering
\begin{fmffile}{ntree2} 
\begin{fmfgraph*}(80,80) 
\fmfsurroundn{v}{16}
\fmf{plain}{c1,v8}
\fmf{plain}{c1,v10}
\fmf{photon}{c1,c3}
\fmfblob{.15w}{c3}
\fmf{photon}{c3,c2}
\fmf{plain}{c2,v2}
\fmf{plain}{c2,v16}
\end{fmfgraph*}\vspace{.35cm}
\end{fmffile}
\caption{Tree level leading diagram contributing to $n\rightarrow n$ scattering processes.  Solid lines refer to $\sigma$, while curvy ones represent gravitons.  The solid blob represents a vertex coming from the GB term, with $2n-4$ external scalar legs attached to it. } \label{fig:tree}
\end{figure}
%%%%%%%%%%%%%%%%%%%%%%%%%
%%%%%%%%%%%%%%%%%%%%%%%%%
%%%%%%%%%%%%%%%%%%%%%%%%%
%%%%%%%%%%%%%%%%%%%%%%%%%
For $n\rightarrow n$ amplitudes,  the first contribution starts at tree-level,  from a diagram constructed by plugging a vertex from the GB term in between two vertices coming from the kinetic term, as shown in figure \ref{fig:tree}.  The resulting amplitude from this diagram depends,  from Lorentz invariance,  on the kinematical variables $s_{ij}=(p_i+p_j)^2$ only,  where the labels run over the different external legs.  Any other diagram that we can construct at this order will have an additional $M_P^2$ suppression coming from extra graviton propagators,  and we thus ignore it hereinafter.  Later on we will be interested in taking a forward limit in which the variables $s_{ij}$ will either vanish or be proportional to the center of mass energy $s$.  Simple dimensional analysis,  together with the assumption of locality of the amplitudes,  which restricts forward divergences to single poles,  unveils that the tree-level diagrams will contribute to the amplitude as
\begin{align}\label{eq:tree_n}
{\cal A}_{\rm tree}^{(n)}(s)=\lim_{t\rightarrow 0}M_P^{-4}\Lambda^{2n-4}\left(A_1 s^2 +A_2 \frac{s^3}{t}\right),
\end{align}
where the coefficients $A_i$ must be determined by direct computation, and $t$ is kept here as a regulator of the forward limit divergence with dimensions of energy squared.  As we will argue later -- cf.  the discussion after \eqref{eq:positivity_bounds} -- terms of this form give vanishing contributions to positivity bounds and therefore we will forget about them from now on,  refraining from even computing them\footnote{This is similar to what happens when one considers the contribution to $2\rightarrow 2$ positivity bounds of massless scalars and gauge fields.  The corresponding amplitude goes as $1/t$,  with no power of $s$ in the numerator,  and thus it cancels when computing the derivatives in \eqref{eq:app_positivity}.  We will show later here that $n \rightarrow n$  positivity bounds are sensitive to amplitudes of the form $\propto s^l$ only if $l\geq 6$.}.

%%%%%%%%%%%%%%%

We thus focus now in the next-to-leading order contribution,  coming from one-loop terms.  Regarding scale suppression, every graviton propagator introduces a power of $M_P^{-2}$, while every coupling $f_j$ contains $\Lambda^{-j}$. The leading contribution is given by a single insertion of $f_j$, coming from the GB term. We have several potential topologies that could contribute to the amplitude. First, we should worry about gravitational tadpoles, where we take a single vertex with $2n$ scalar legs and close the two graviton legs in a loop. However, since the graviton is massless, these are automatically vanishing in dimensional regularization and we thus disregard them. In a similar fashion, disconnected one-loop diagrams will vanish, because they can only be proportional to these tadpoles. The first non-vanishing disconnected contribution starts at two-loops andt thus will be sub-leading.

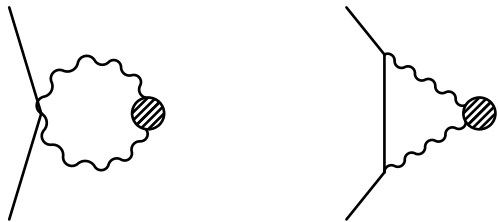
\begin{figure}
  \centering
\begin{fmffile}{fishF} \vspace{.5cm}
\begin{fmfgraph*}(80,80) 
\fmfleft{l1,l2} 
\fmfright{r1,r2} 
\fmf{plain}{c3,l1}
\fmf{plain}{c3,l2}
\fmf{phantom}{r1,c1}
\fmf{phantom}{r2,c1}
\fmfblob{.15w}{c1}
\fmf{photon,left,tension=.3}{c1,c3,c1}
\end{fmfgraph*}
\end{fmffile}\qquad \qquad \begin{fmffile}{triF} 
\begin{fmfgraph*}(80,80) 
\fmfleft{l1,l2} 
\fmfright{r1,r2} 
\fmf{phantom}{r1,c3}
\fmf{phantom}{r2,c3}
\fmf{plain}{c1,l1}
\fmf{plain}{c2,l2}
\fmf{plain, tension=.2}{c1,c2}
\fmf{photon,tension=.4}{c1,c3}
\fmf{photon,tension=.4}{c2,c3}
\fmfblob{.15w}{c3}
\end{fmfgraph*}\vspace{.35cm}
\end{fmffile}
\caption{Leading contributions to the one-loop scattering amplitudes.  Solid lines refer to $\sigma$, while curvy ones represent gravitons.  The outcoming amplitudes are suppressed by $M_P^{-4}\Lambda^{2-2n}.$ The solid blob represents a vertex coming from the GB term, with $2n-2$ external scalar legs attached to it.}\label{fig:diagrams}
\end{figure}

Next we wonder about radiating suns. These are successions of $N$ vertices and propagators until they close a whole loop. However, note that since there are no vertices with only scalar legs, and we are not considering external gravitons, we must introduce a graviton propagator every even number of vertices. We can now start counting the number of vertices. Let us start with $N=2$ vertices with $2n$ external scalar legs. For a single graviton propagator, it is impossible to built such a diagram. For two graviton propagators, it is suppressed by $M_P^{-4}$. Then we need to choose whether the external legs come from the kinetic term or not. The case with vertices only from the kinetic term is only possible if the amplitude is $2\rightarrow 2$. For $n>2$, the less suppressed diagram is obtained by taking one of the vertices from the kinetic term and the other from the GB term. They are suppressed by $M_P^{-4}\Lambda^{2-2n}$. For $N\geq 5$ we are forced to include three graviton propagators. This leads to a suppression of $M_P^{-6}$, so we disregard these diagrams as sub-leading. With $N=4$ vertices we are forced to take at least one GB vertex with a single graviton, which do not exist for $n>2$. Finally, for $N=3$ the leading term is obtained by taking two vertices from the kinetic term and one from the GB term, and it is again suppressed by $M_P^{-4}\Lambda^{2-2n}$. This leaves us with only two possible topologies at the leading order, shown in figure \ref{fig:diagrams}. Notice that these configurations are universal except for the number of external legs attached to the vertex coming from the Gauss-Bonnet term. This will allow us to compute all of them at once in a simple manner.

Regarding sub-leading terms, they can be estimated in a a similar fashion. The next less suppressed contribution is obtained by constructing a triangle with four external scalars in two of the vertices -- two on each vertex -- and thus $2n-4$ in the GB blob. It is suppressed by $\Lambda^{2n-4}M_P^{-6}$, thus carrying an extra suppression of $\Lambda^2/M_P^{2}$ with respect to the leading contribution. As long as the EFT scale is far enough from the gravitational UV cut-off, we can safely dismiss these contributions\footnote{There is, in principle, a less suppressed sub-leading correction obtained by taking a fish diagram with two GB vertices. However, the extra suppression in this case is $\sim P^2/\Lambda^2$, where $P$ depends on the external momentum. Due to this, its contribution to the bound \eqref{eq:pos_result} later will be proportional to $\delta^2/\Lambda^2$, which is always tiny and can be safely neglected. In particular, it will be always smaller than $\Lambda^2/M_P^2$, which thus dominates the sub-leading terms.}.

%%%%%%%%%%%%%%%%%%%%%%%%%%%%%%%%%%%%%%%%%%%%%%%%%%%
%%%%%%%%%%%%%%%%%%%%%%%%%%%%%%%%%%%%%%%%%%%%%%%%%%%
%%%%%%%%%%%%%%%%%%%%%%%%%%%%%%%%%%%%%%%%%%%%%%%%%%%
%%%%%%%%%%%%%%%%%%%%%%%%%%%%%%%%%%%%%%%%%%%%%%%%%%%
%%%%%%%%%%%%%%%%%%%%%%%%%%%%%%%%%%%%%%%%%%%%%%%%%%%
%%%%%%%%%%%%%%%%%%%%%%%%%%%%%%%%%%%%%%%%%%%%%%%%%%%
%%%%%%%%%%%%%%%%%%%%%%%%%%%%%%%%%%%%%%%%%%%%%%%%%%%
%%%%%%%%%%%%%%%%%%%%%%%%%%%%%%%%%%%%%%%%%%%%%%%%%%%
%%%%%%%%%%%%%%%%%%%%%%%%%%%%%%%%%%%%%%%%%%%%%%%%%%%

In general, scattering amplitudes for processes $\sigma\overbrace{\dots}^{n} \sigma\rightarrow \sigma\overbrace{\dots}^{n} \sigma$ in a Lorentz invariant theory depend on a finite set of kinematical invariants, constructed from the different momenta $p_i$ of all the particles in the initial and final states
\begin{align}
    s_{ij}=(p_i+p_j)^2,\quad i\neq j
\end{align}
where we are taking all $p_i$ outgoing. The number of independent variables can be counted -- when $2n\geq d$, with $d$ the spacetime dimension -- in the following way\footnote{For $2n\leq d$ this counting is not valid, since some of the conditions that we introduce become redundant. For a detailed counting cf. \cite{Chaichian:1987zt}.}. We have $d\times 2n$ kinematical components in $2n$ momenta in $d$ dimensions. Out of them, we can remove $2n$ by on-shell conditions $p_i^2=\mathfrak{m}^2$, and $d$ more from momentum conservation. Finally, we can always set $d(d-1)/2$ elements to vanish by acting with a Lorentz transformation -- an element of the $O(1,d-1)$ group. The final number of independent variables is thus ${\cal N}_d=2n(d-1)-d(d+1)/2$, which in $d=4$ dimensions gives ${\cal N}_4=6n-10$.

In a general case, the evaluation of the scattering amplitude would require to compute the Feynman diagrams in figure \ref{fig:diagrams} for a particular kinematic configuration, where we attach the external legs to the external momenta $p_1,p_2,\dots p_{2n}$, and afterwards apply crossing symmetry in order to obtain all the possible inequivalent channels in the process. For the $n\rightarrow n$ processes at hand, crossing symmetry corresponds to the symmetrization over all possible $s_{ij}$, a procedure that generates a number of terms which grows quickly with $n$. However, things are simpler for the case of SGBG. Due to the structure of the GB term and the aforementioned democratization of amplitudes, there is no distinction between the individual momenta of the different $\sigma$ legs that enter through vertices with $f_j$. Once the external legs are fixed to a particular momentum configuration, most permutations given by crossing symmetry will be completely equivalent. This implies that every particular channel will depend on a reduced set of $s_{ij}$ only.

In order to exploit this property, let us then choose a particular kinematical configuration by assigning the free legs -- those \emph{not} entering through the GB vertex -- to the momenta $p_1$ and $p_2$, while through the GB vertex we will have a momentum $p_f=\sum_{i=3}^{2n} p_i$. We will call this configuration \emph{12-channel}. In this channel, the amplitude can only depend on the following three independent variables \footnote{Exchange between the external free legs, corresponding to $1\Longleftrightarrow 2$ for the $12$-channel, is accounted by a global symmetry factor in the amplitude.}
\begin{align}
    s_{12}=(p_1+p_2)^2=p_f^2,\quad  \alpha=(p_f+p_1)^2, \quad \beta=(p_f+p_2)^2.
\end{align}
Notice however, that due to momentum conservation, both $\alpha=\beta=\mathfrak{m}^2$ and we are left with $s_{12}$ as the only independent variable. Thus, the scattering amplitude in the 12-channel depends only on this single kinematical variable, which identifies the channel ${\cal A}_{12}(s_{12})$, and contains only the momenta of the independent legs. Notice also that the computation of the diagram, which can be found in detail on Appendix \ref{app:amplitudes}, does not depend on $n$ at all, neither on the individual legs entering on the GB vertex, only on the sum of their momenta $p_f=-(p_1+p_2)$. Thus, the result ${\cal A}_{12}(s_{12})$ is universal and valid regardless the value of $n$. The latter will only become relevant at the time of performing crossing symmetry, which, since the amplitude of the 12-channel depends on a single variable, is realized simply by
\begin{align}\label{eq:amplitude}
    {\cal A}^n(s_{12}, s_{13},\dots )=\sum_{s_{ij}} {\cal A}_{12}(s_{ij}).
\end{align}
From now on we take the limit $\mathfrak{m}\rightarrow 0$ in all kinematic invariants since, as it can be seen in Appendix \ref{app:amplitudes}, it is regular for the amplitudes at hand. This will greatly simplify the following discussion.

At this point, we wish to connect with the standard derivation of positivity bounds, which we will later extend to our case here by following Ref. \cite{Herrero-Valea:2020wxz}. A first step in order to do that is to compute the scattering amplitude ${\cal A}^n(s_{12}, s_{13},\dots )$ in the forward limit. For a standard $2\rightarrow 2$ scattering, this is defined simply by letting $t\equiv s_{13}=p_1+p_3=0$, which due to momentum conservation also implies $p_2+p_4=0$. For amplitudes with more than two external legs, however, there exist multiple forward limits that we can define, corresponding to all possible ways in which we can identify incoming and outgoing legs. In here we choose to take \cite{Elvang:2012st} 
\begin{align}\label{eq:forward_limit}
s_{i,i+n}=(p_i+p_{i+n})^2=0,\quad 1\leq i\leq n, \quad s_{ij}=\frac{s}{n}, \quad i,j\leq n,
\end{align}
with all incoming particles sharing the same fraction of energy-momentum.

By taking this limit, $n$ of the $s_{ij}$ invariants will automatically vanish. The rest of them can be grouped in two sets. The first one are those with both labels satisfying either $i,j\leq n$ or $i,j >n$. In both cases, due to the forward limit condition, we have $s_{ij}=s/n$ and it is easy to check that there are $2\times n(n-1)/2=n(n-1)$ elements of these form. The remaining ones have one index in the range $i,j\leq n$ and the other in $i,j > n$ and take instead the value $s_{ij}=-s/n$. There are also $n(n-1)$ of these\footnote{So that the total number of elements adds up to $n(2n-1)$, which is the number of independent elements in an $2n\times 2n$ anti-symmetric matrix.}. In Appendix \ref{app:example} we workout a complete example of this classification in the case of a 3-to-3 scattering.

Upon these substitutions, the center of mass energy of the general $n$-to-$n$ process can be written as
\begin{align}\label{eq:ECM}
   E^2_{\rm CM}=\left(\sum_{i=1}^n p_i \right)^2=n\times \frac{s}{n}=s.
\end{align}
Thus, the variable $s$ takes the role of the center of mass energy squared, in an identical way to the usual Mandelstam variable $s_{12}$. We will thus use it to perform the analytic extension of the scattering amplitudes to the complex plane in what followss.

Even though this structure looks complicated at a first glimpse, the final result is really simple when particularized to the amplitude \eqref{eq:amplitude}. Summing over all independent variables $s_{ij}$ we simply have, in the forward limit
\begin{align}\label{eq:amplitude_n}
    {\cal A}^{n}(s)=n(n-1)\left({\cal A}_{12}(s)+{\cal A}_{12}(-s)\right),
\end{align}
where we have already taken into account that the amplitude ${\cal A}_{12}$ computed in Appendix \ref{app:amplitudes} has no divergences in the forward limit \eqref{eq:forward_limit}, since it vanishes polynomially when any of the $s_{ij}$ goes to zero. This final form for the amplitude is completely reminiscent of the structure obtained in the forward limit of a 2-to-2 scattering.

%%%%%%%%%%%%%%%%%%%%%%%%%%%%%%%%%%%%%%%%%%
%%%%%%%%%%%%%%%%%%%%%%%%%%%%%%%%%%%%%%%%%%
%%%%%%%%%%%%%%%%%%%%%%%%%%%%%%%%%%%%%%%%%%
\subsection{Amplitudes's Positivity}

We wish now to derive positivity bounds for the amplitudes that we have just constructed. As in the case of the $2\rightarrow 2$ scattering, this amplitude could potentially contain divergences in the forward limit. Those would impede to take it naively and instead require regularization against the UV behaviour of the amplitude, as discussed in \cite{Herrero-Valea:2020wxz,Tokuda:2020mlf}. However, due to the very particular properties of the GB term and the vertices that it induces, we find instead that this limit is completely regular for the amplitudes that we compute -- cf. Appendix \ref{app:amplitudes} --, which allows us then to avoid the process of regularization of the forward limit. This is a consequence of the particular structure of the Lagrangian, that we demand to be true up to the Regge scale, which is tantamount to say that we demand SGBG to be in weak coupling up to $M_*$. By taking this assumption, we can ensure that the regularity of the forward limit is preserved by further loop contributions.

We follow now the derivation of \cite{Herrero-Valea:2020wxz} in order to construct positivity bounds for our amplitudes. For the moment let us forget about the specific form of the SGBG action and imagine that we had at our disposal the full amplitude for $n\rightarrow n$ processes, up to arbitrary energy. We will assume, as already discussed, that such UV completion satisfies reasonable properties for a fundamental theory. In particular, we will assume that that it is Lorentz invariant, causal, local, and unitary, and that it maintains the democratization of amplitudes previously discussed.

Under these assumptions, we can define a family of dispersion relations
\begin{align}
    {\cal A}^n(s)=\frac{(s-\mu)^j}{2\pi i} \oint_{\gamma_s}dz \frac{{\cal A}^n(z)}{(z-s)(z-\mu)^j},
\end{align}
where $j>1$ and $\mu$ is arbitrary as long as it lays outside the integration contour $\gamma_s$, which is a small circle surrounding the point $z=s$, and in the same -- upper or down -- half of the complex plane. Upon integration, we recover the original amplitude by application of Cauchy's integral theorem.

\begin{figure}
  \includegraphics[scale=.46]{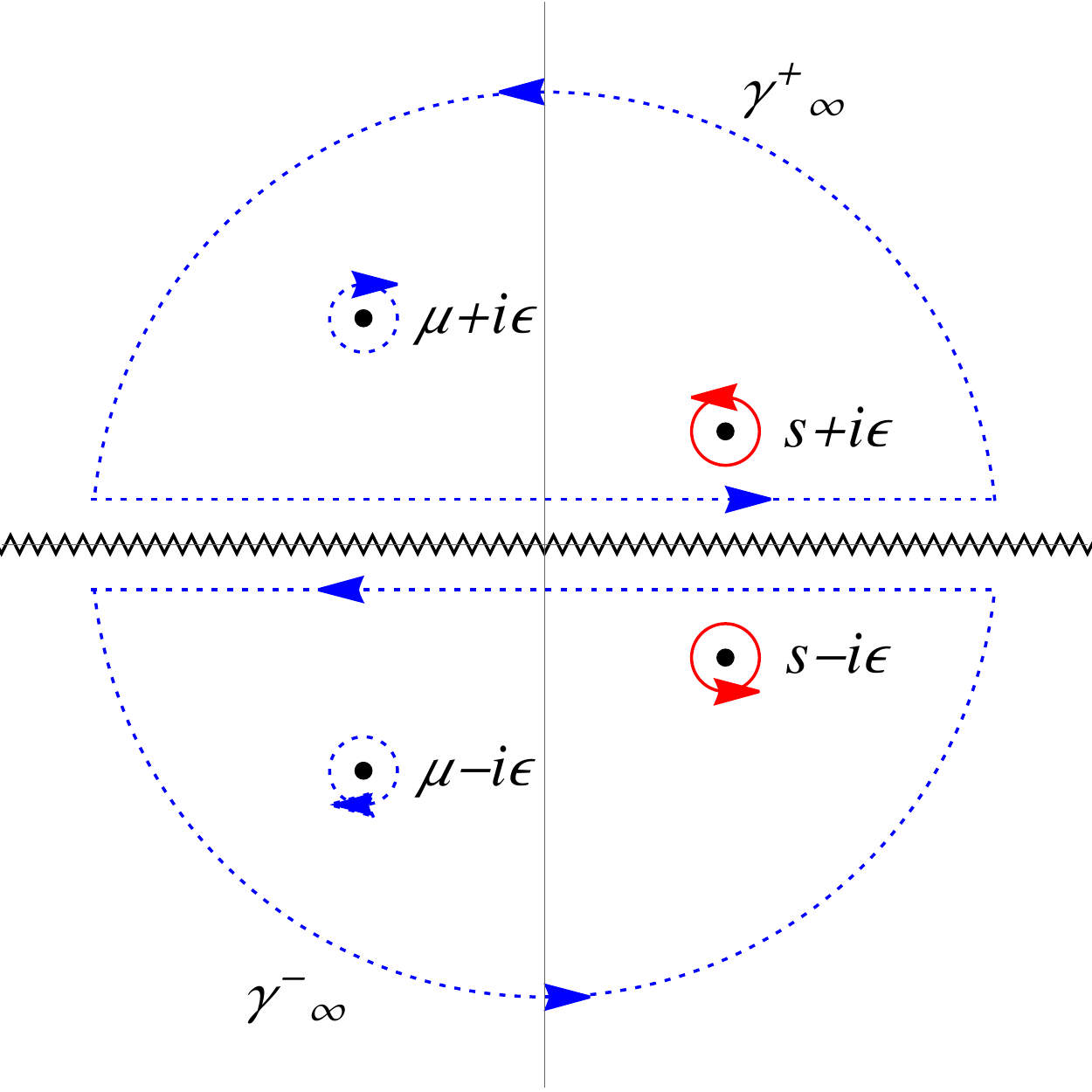}
\caption{Integration contours in the complex $s$-plane. The zigzag line corresponds to the branch cut. For points $s\pm i\epsilon$, the integration contour $\gamma_s$ in the corresponding half of the complex plane is shown in red. The equivalent contours used in \eqref{eq:two_expre} are dotted in blue, including the subtraction of the point $\mu\pm i\epsilon$. The radius of the semi-circumferences $\gamma_\infty^\pm$ is $|s|\rightarrow \infty.$}
\label{fig:contours}
\end{figure}

The analytic structure of the amplitude ${\cal A}^n(z)$ is completely fixed by causality. Due to the presence of massless gravitons, it exhibits a branch cut along the whole real line, but it is otherwise analytic in the rest of the complex plane. Thus, for any given value of $s$ in a half of the complex plane and close to the real line $s\rightarrow s\pm i\epsilon$, with $\epsilon \ll 1$, we can deform the integration contour as shown in figure \ref{fig:contours} to arrive to two different but equivalent expressions 
\begin{align}\label{eq:two_expre}
    &{\cal A}^{n}(s+i\epsilon)=\frac{(s-\mu)^j}{2\pi i} \oint_{\gamma^+_\infty}dz \frac{{\cal A}^n(z)}{(z-s)(z-\mu)^j}+\frac{(s-\mu)^j}{2\pi i} \int_{-\infty}^\infty dz \frac{{\cal A}^n(z+i\epsilon)}{(z-s)(z-\mu)^j},\\
    &{\cal A}^{n}(s-i\epsilon)=\frac{(s-\mu)^j}{2\pi i} \oint_{\gamma^-_\infty}dz \frac{{\cal A}^n(z)}{(z-s)(z-\mu)^j}+\frac{(s-\mu)^j}{2\pi i} \int_{\infty}^{-\infty} dz \frac{{\cal A}^n(z-i\epsilon)}{(z-s)(z-\mu)^j},
\end{align}
where we notice that we have removed the pole in $z=\mu$ in both cases by subtracting a small contour around it, and we have kept $\epsilon$ non-vanishing only on non-holomorphic terms.

The integrals over $\gamma^{\pm}$ can be neglected by invoking the Froissart-Martin bound \cite{Froissart:1961ux}, which states that for any massive theory, the 2-to-2 cross-section of the process $\bar{\sigma}$ will behave as 
\begin{align}\label{eq:froissart-martin}
    \left.\bar{\sigma}(s)\right|_{s\rightarrow \infty}\sim \log^2(s),
\end{align}
for large $s$. Although this property has not been proven in general for $n$-to-$n$ processes, we will assume it here by invoking one of the conditions demanded to our UV completion -- that amplitudes democratize, depend only on $f(\phi/\Lambda)$, and all of them Reggeize at high energies. Assuming \eqref{eq:froissart-martin} implies polynomial boundedness for the scattering amplitude. Taking into account the relation between the amplitude and the cross section $|{\cal A}^n(s)|^2\sim s^{n/2}\bar{\sigma}$ we see that at high energies $|{\cal A}^{(n)}|\sim s^{n/4}\log s$, from which we conclude
\begin{align}
    \lim_{s\rightarrow \infty}\left|\frac{{\cal A}^n(s)}{s^{\left[1+\frac{n}{4}\right]+1}}\right|=0,
\end{align}
where $[x]$ denotes the integer part of $x$. This is enough to show that the integrals over $\gamma^{\pm}$ vanish for $j>\left[1+\frac{n}{4}\right]+1$. We will thus assume this bound hereinafter. 

We now subtract both expressions in \eqref{eq:two_expre} to find
\begin{align}
    {\cal A}^n(s+i\epsilon)-{\cal A}^n(s-i\epsilon)=\frac{(s-\mu)^j}{2\pi i}\int_{-\infty}^\infty \frac{{\cal A}^n(z+i\epsilon) +{\cal A}^n(z-i\epsilon)}{(z-s)(z-\mu)^j},
\end{align}
from which we obtain
\begin{align}
    {\rm Im}{\cal A}^n(s\pm i \epsilon)=\mp \frac{(s-\mu)^j}{2\pi}\int_{-\infty}^\infty dz \frac{{\rm Re}{\cal A}^n(z\pm i\epsilon)}{(z-s)(z-\mu)^j},
\end{align}
after using the Schwartz reflection principle\footnote{Formally, the Schwartz reflection principle is only true if the function is analytic at least in an open set of the real line. This is not strictly the case here, but the issue can be circumvented by restoring a small non-vanishing $t$, which opens a gap in the branch cut, and showing that the limit $t\rightarrow 0$ is continuous.} ${\cal A}^n(s+i\epsilon)=\left({\cal A}^n(s-i\epsilon)\right)^*$. Replacing this back onto \eqref{eq:two_expre} we thus find
\begin{align}
    {\rm Re}{\cal A}^n(s+i \epsilon)=\frac{(s-\mu)^j}{2\pi}\int_{-\infty}^\infty dz\frac{{\rm Im}{\cal A}^n(z+i\epsilon)}{(z-s)(z-\mu)^j}.
\end{align}

The integral in the rhs runs over all possible values of $z$. However, for reasons that will be clear in a moment, we would like to restrict ourselves here to values of $s$ within the physical region for the $n\rightarrow n$ scattering, which corresponds in the massless limit to $s>0$. We can do this by splitting the integration regime in two regions $\{-\infty, 0\}$ and $\{ 0,\infty\}$ and performing a change of variables $z\rightarrow -z$ in the first one, arriving to
\begin{align}\label{eq:B_disp}
    {\rm Re}{\cal A}^n(s+i \epsilon)=\frac{(s-\mu)^j}{2\pi}\int_{0}^\infty dz\left( \frac{{\rm Im}{\cal A}^n(z+i\epsilon)}{(z-s)(z-\mu)^j} + \frac{(-1)^j  {\rm Im}{\cal A}^n(-z+i\epsilon)}{(z+s)(z+\mu)^j} \right),
\end{align}

Since the amplitude in the forward limit \eqref{eq:amplitude_n} that we consider here is invariant under $s\rightarrow -s$, we simply write ${\rm Im}{\cal A}^n(-z+i\epsilon)={\rm Im}{\cal A}^n(z+ i \epsilon)$ from now on.

We finally introduce the last ingredient in our derivation of positivity bounds. We define the following quantity
\begin{align}\label{eq:def_sigma}
    \Sigma_j^n=\frac{1}{2\pi i}\oint_{\gamma_\delta}ds \frac{s^3   {\rm Re}{\cal A}^n(s+i \epsilon)}{(s^2+\delta^2)^{j+1}},
\end{align}
where $\gamma_\delta$ is the sum of two small circumferences surrounding the points $s=\pm i\delta$, with $\delta>0$. Performing the integral over $s$ explicitly in the rhs of \eqref{eq:B_disp}, it becomes
\begin{align}\label{eq:sigma_int}
    \Sigma_j^n=\frac{1}{2\pi i}\oint_{\gamma_\delta}ds \frac{s^3   {\rm Re}{\cal A}^n(s+i \epsilon)}{(s^2+\delta^2)^{j+1}}=\frac{1+(-1)^j}{2\pi}\int_{0}^\infty  dz  \frac{z^3  {\rm Im}{\cal A}^n(z+i\epsilon)}{(z^2+\delta^2)^{j+1}},
\end{align}
where all dependence on $\mu$ has cancelled.

At this point, we must note an important property about the rhs of this expression. First, let us recall that the generalized optical theorem \cite{Peskin:1995ev} states that for any $n \rightarrow n$ scattering process
\begin{align}\label{eq:optical theorem}
    {\rm Im}{\cal A}^n(s)=\frac{1}{2} \sum_X \int d^X_{\rm LIPS}  (2\pi)^4 |{\cal A}(n\rightarrow X)|^2  \delta^{(4)}\left(\Sigma_i p_i  \right) ,
\end{align}
where ${\cal A}(n\rightarrow X)$ is the scattering amplitude from the state of $n$ particles of the $\sigma$ field to an intermediate configuration labeled by $X$. The rhs of the previous expression thus sum over all possible intermediate channels mediating the $n\rightarrow n$ scattering process. Here $d^X_{\rm LIPS}$ is the Lorentz invariant phase space of the intermediate channel
\begin{align}
    d^X_{\rm LIPS}=\prod_{k\in X}\frac{d^3p_k}{2(2\pi)^3E_k},
\end{align}
with $E_k$ the energy of the $k$-th particle.

A trivial consequence of the optical theorem is that, above the physical threshold $s>0$, the imaginary part of the amplitude is explicitly positive. If we note that all the other individual elements involved in the integral in the rhs of \eqref{eq:sigma_int} are also positive within the integration regime for even $j$, we can conclude that
\begin{align}
    \Sigma_j^n>0, \quad j \ \ {\rm even}.
\end{align}

However, for this to be true, the integral in \eqref{eq:sigma_int} must be finite, converging in the upper limit $z\rightarrow \infty$. Otherwise, the exchange of the order of integration which leads to \eqref{eq:sigma_int} is not possible. We can confirm whether this is the case or not by invoking again the optical theorem. From \eqref{eq:optical theorem} and the definition of the cross-section for a $n\rightarrow n$ process, we can get a rough estimation of the imaginary part of the amplitude at large energies
\begin{align}
    {\rm Im}{\cal A}^n(s)\sim s^{3n-2}\bar \sigma \sim s^{3n-2}\log^2(s),
\end{align}
where we have used the Froissart-Martin bound \eqref{eq:froissart-martin} and the power of $s$ comes from the integration measure -- which contributes with $s^n$ --, the delta function -- $s^2$ -- and the relation $|{\cal A}^n(s)|^2\sim s^{\frac{n}{2}}\bar{\sigma}$. We can see that for increasing values of $n$, the convergence of the integral worsens. From here, we thus find that we can only safely ensure convergence for values of $j$ satisfying\footnote{Once can check that this condition also ensure that the bounds are independent of the value of the IR regulator $\delta$.}
\begin{align}
    j>\frac{3(n-1)}{2}.
\end{align}

Collecting everything together, we thus have
\begin{align}\label{eq:positivity_bounds}
    \Sigma^n_{2k}>0,\quad 4k>3(n-1),
\end{align}
where we have redefined $j=2k$. This is the final version of the positivity bounds that we will exploit in the following, valid for $n\rightarrow n$ scattering amplitudes in the forward limit, provided that such a limit is regular.  Notice that the condition \eqref{eq:positivity_bounds} implies that amplitudes of the form \eqref{eq:tree_n} and in general any amplitude polynomial in $s$ with a power smaller than $s^6$ -- for $n=3$,  for larger values of $n$ the power must be larger -- will give a vanishing contribution to the bounds,  after taking the limit $\delta\rightarrow 0$.  This justifies our omission of the tree contributions to the amplitude.  On the other hand,  non-polynomial terms,  like the ones coming from loop amplitudes,  always contribute to $\Sigma_{2k}^n$  regardless of the value of $k$.

Up to here we have assumed that we knew the form of the amplitude up to arbitrary energy, which has allowed us to exploit certain analytic properties inherited from the UV completion. However, in real life we do not know such UV completion or even if it exist. Nevertheless, note that the value of $\Sigma_j^n$ in \eqref{eq:def_sigma} is defined solely by the value of the amplitude in the poles $s=\pm i\delta$, which is calculable within the EFT given by the SGBG Lagrangian \eqref{eq:action_chi}, provided that $\delta\ll \Lambda^2$. However, its positivity \eqref{eq:positivity_bounds} is required from properties of the UV completion. Since the amplitudes depend on the derivatives of $f(x)$ through the interaction vertices, this leads us to an advantageous connection between IR and UV physics which we can use to bound the form of the function $f(x)$. We just need to compute the value of $\Sigma^n_j$ and demand satisfaction of the positivity bounds. Whenever these are violated, that can be due to two reasons. The first option is that the value of the cut-off $\Lambda$ is larger than $M_P$. This would invalidate our power counting for the leading terms of the amplitude and would require to know a (possibly partial) UV completion in order to asses the problem. The second option is that the Lagrangian cannot be computed from a UV theory satisfying our assumptions. In both cases, the theory does not make sense at the scales at which it is formulated, provided our hypothesis are true.

We explicit now our result to the case of SGBG. As shown in Appendix \ref{app:amplitudes}, the $12$-channel amplitude for any $n$-to-$n$ process is universal and takes the form, after renormalization at a scale $\mu_R$ and setting $\mathfrak m\rightarrow 0$
\begin{align}
    {\cal A}_{12}(s_{12})=-\frac{f_n}{72\pi^2 M_P^4} s_{12}^3 \left(23-12\log \left(\frac{s_{12}}{\mu_R^2}\right)\right).
\end{align}

Using this, we can compute the whole amplitude ${\cal A}^n(s)$ for the kinematic configuration of interest, as shown in \eqref{eq:amplitude_n}
\begin{align}
    {\cal A}^{n}(s)=n(n-1)\left[{\cal A}_{12}(s)+{\cal A}_{12}(-s)\right].
\end{align}

Before going further, let us recall that here we are carrying only the leading terms in the amplitude, with subleading corrections suppressed by $\Lambda^{2-n} / M_P^2$ in the form
\begin{align}\label{eq:limit}
     {\cal A}_{12}(s_{12})=-\frac{\left[f_n+{\cal O}\left(\frac{\Lambda^{2-n}}{M_P^{2}}\right)\right]}{72\pi^2 M_P^4} s_{12}^3 \left(23-12\log \left(\frac{s_{12}}{\mu_R}\right)\right).
\end{align}
As a consequence of this, the strict positivity bound \eqref{eq:positivity_bounds} can only be applied in a solid way up to the order of magnitude at which these corrections become dominant. We should also take into account the sub-leading terms introduced by the assumption of a Regge scale above $M_*$. However, these enter with a factor ${\cal O}\left(\Lambda^2/M_*^2 \times \Lambda^{2-n}/M_P^{2}\right)$ and thus they can be dismissed with respect to the sub-leading correction in \eqref{eq:limit} as long as $\Lambda < M_*$.

Going back to the amplitude, we thus plug into \eqref{eq:def_sigma} and perform the integral, which is tantamount to computing the residues in the poles at $s=\pm i\delta$. Doing this, we get a tower of bounds which behave as
\begin{align}\label{eq:pos_result}
    \Sigma^n_{2k}=- H(n,j) f_{2n-2} \frac{\delta^{7-4k}}{M_P^4}, \quad 4k>3(n-1).
\end{align}
Since $k>2$ for all $n$, the previous result is bounded by the value of $\delta \ll \Lambda$. The function $H(n,j)$ is a pure numerical factor of definite sign $H(n,k)>0$ for all allowed values of $n$ and $k$, and that we have not been able to express in a closed form. Note that the renormalization scale has dropped from the result, erasing any arbitrariness that it could introduce, as expected since it is a non-physical artifact of the renormalization process.

Therefore, for any value of $\delta\ll \Lambda^2$, and taking into account the size of the sub-leading contributions, we get from \eqref{eq:pos_result}
\begin{align}
   \left.\frac{d^{2n}}{dx^{2n}}f(x)\right|_{x=\frac{\phi_0}{\Lambda}}<-{\cal O}\left(\frac{\Lambda^2}{M_P^2}\right), \quad n\geq 2.
\end{align}

Again, we find violations of positivity (negativity) which are controlled by the separation of scales between the naive cut-off $\Lambda$ of the EFT and the scale at which gravitational degrees of freedom -- gravitons from the Einstein-Hilbert term in the action, in this case -- become dominant in the amplitude. The closer the scales, the larger the allowed violation. Note that in setting the inequality in the previous bound, we are ignorant about the sign of the sub-leading contribution, so we have chosen the worst case scenario in order to estimate the size of the allowed violation and of the bound. The same was done when deriving \eqref{eq:2to2bound}.

%%%%%%%%%%%%%%%%%%%%%%%%%%%%%%%%%%%%%%%%%%%%%%%%%%%%%%%%
%%%%%%%%%%%%%%%%%%%%%%%%%%%%%%%%%%%%%%%%%%%%%%%%%%%%%%%%
%%%%%%%%%%%%%%%%%%%%%%%%%%%%%%%%%%%%%%%%%%%%%%%%%%%%%%%%

\section{Results}\label{sec:results}

We wish now to compare our results with the accumulated literature on SGBG, and in particular on scalarization of black holes within SGBG. We start by summarizing our findings so far. By assuming that the action \eqref{eq:starting_action} can be derived from a Lorentz invariant, causal, unitary, and local UV completion, we have derived positivity bounds on the scattering amplitudes of the theory around flat space. They imply that, for every extreme point of the function $f(x)$, the following conditions must be satisfied\footnote{Notice that the bounds for $n=1$ and $n\geq 2$ have different sign. This is a consequence of the fact that the 2-to-2 scattering amplitude that dominates the bound for $n=1$ starts at tree-level, while the fish diagram that controls the bound for $n\geq 2$ is one-loop and thus picks up an extra minus sign from second order perturbation theory. This reconciles our results with those of \cite{Chandrasekaran:2018qmx}, in which all bounds share the same sign since all of them are obtained from tree-level amplitudes.}
\begin{align}\label{eq:final_bounds}
    \left.\frac{d^2f(x)}{dx^2}\right|_{x=\frac{\phi_0}{\Lambda}}>-{\cal O}\left(\frac{\Lambda^2}{M_*^2}\right), \qquad  \left.\frac{d^{2n}}{dx^{2n}}f(x)\right|_{x=\frac{\phi_0}{\Lambda}}<{\cal O}\left(\frac{\Lambda^2}{M_P^2}\right), \quad n\geq 2.
\end{align}

Even derivatives of the function $f(x)$ must be positive or negative \emph{up to corrections depending on the UV physics, which can violate positivity (negativity)}. For functions with multiple extremes, we must demand the satisfaction of the bounds for all of them, since they represent, in principle, equivalent vacua of theory.

Before going further and in order to connect with the current literature in SGBG, let us change variables to the dimensionless scalar field $\Phi$, by using \eqref{eq:change_var}. In terms of this, our positivity bounds \eqref{eq:final_bounds} read
\begin{align}\label{eq:bounds_fbar}
  \left.  \frac{d^2 F}{d\Phi^2}\right|_{\Phi=\Phi_0}>-{\cal O}\left(\frac{1}{M_*^2}\right),\qquad  \left(\frac{M_P^2}{\Lambda^2}\right)^{n-1}\left. \frac{d^{2n} F}{d\Phi^{2n}}\right|_{\Phi=\Phi_0}<{\cal O}\left(\frac{1}{M_P^2}\right), \quad n\geq 2,
\end{align}
where $\Phi_0=\phi_0 M_P/\Lambda$.

%%%%%%%%%%%%%%%%%%
Since positivity bounds derived from $n$-to-$n$ amplitudes depend on extra -- although reasonable -- assumptions about the high energy behavior of the UV completion, we will in the following separate the consequences of $2$-to-$2$ and $n$-to-$n$ bounds for the different models under study. In general, most of the conclusions that we draw in the following are obtained solely from the $2$-to-$2$ bound, which is of course great because they are thus independent of the UV completion. Only when the $n$-to-$n$ bounds are required to obtain extra information, we will consider them.
%%%%%%%%%%%%%%%%%%

A first trivial result that can be derived from \eqref{eq:bounds_fbar} is the fact that although scalarization is possible for any black hole, the so-called spin-induced scalarization is highly constrained. The negative values of $F''(\Phi)$ allowed by positivity are constrained to be extremely small $|F''(\Phi_0)|<{\cal O}\left(M_*^{-2}\right)$, making the process impossible, at least under known mechanisms. 

But we can do better. Let us now invoke some models for the function $F(\Phi)$ previously used in the literature and understand what our bounds here imply for them. We will study three different popular models
\begin{itemize}
\item \emph{The exponential model} 
\begin{align}\label{eq:exp_model}
F(\Phi)=\frac{\lambda^2}{2\omega} (1-e^{-\omega \Phi^2}),
\end{align}
where $\omega >0$ and $\lambda$ is a length scale. Introduced in \cite{Doneva:2017bvd}, this is a model in which large deviations from General Relativity are possible for a large range of values in the parameter space. It was also used in \cite{Silva:2018qhn} to study the effect of non-linear terms in quenching the tachyonic instability.

In this case, the function $F(\Phi)$ has a single finite extreme at $\Phi_0=0$. Expanding the function around it we get
\begin{align}
  F(\Phi)=\frac{\lambda^2}{2\omega} \left(\omega \Phi^2 -\frac{1}{2}  \omega^2 \Phi^4 +\frac{1}{3!} \omega^3 \Phi^6 \dots\right),
\end{align}
so that the positivity bound for the $2$-to-$2$ amplitude demands
\begin{align}
    \frac{\lambda^2}{2} >-{\cal O}\left(\frac{1}{M_*^2}\right),
\end{align}
which is automatically satisfied.

However, let us also consider the bounds coming from $n$-to-$n$ amplitudes, and thus subjected to the assumptions outlined in subsection \ref{sec:assumptions}. They read
\begin{align}
\lambda^2 \left(\frac{\omega M_P^2}{\Lambda^2}\right)^{n-1}<{\cal O}\left(\frac{1}{M_P^2}\right),\quad n\geq 2,
\end{align}
and strongly constrain the value of $\omega$ and $\lambda^2$. They can only be satisfied for all $n$ if
\begin{align}
    \omega<{\cal O}\left(\frac{\Lambda^2}{M_P^2}\right),\qquad \lambda^2<{\cal O}\left(\frac{1}{M_P^2}\right).
\end{align}

Although formally there is a range of parameters for which the bounds are satisfied, they require the scale $\lambda$ to be sub-Planckian, which puts the GB term in the Lagrangian in the scale of Quantum Gravity.

\item \emph{The cubic model} 
\begin{align}
F(\Phi)=a_1 \Phi - a_3 \Phi^3,
\end{align}
Models of this type where studied numerically in \cite{Antoniou:2017hxj}. Notice that the existence of an extreme for $F(\Phi)$ demands $a_1/a_3>0$.

In this case we have two potential extremes, at $\Phi_{\pm}=\pm\sqrt{\frac{a_1}{3 a_3}}$. Expanding the function around them we get
\begin{align}
   F(\Phi)=(a_1 \Phi_\pm - a_3 \Phi_\pm^3) +(a_1 -3 a_3 \Phi_\pm^2) (\Phi-\Phi_\pm)- 3 a_3 \Phi_\pm  (\Phi-\Phi_\pm)^2-a_3 (\Phi-\Phi_\pm)^3 .
\end{align}

We find that satisfaction of the bounds \eqref{eq:final_bounds} requires, from $F''(\Phi_{\pm})$
\begin{align}
    \pm \sqrt{a_1 a_3}>-{\cal O}\left(\frac{1}{M_*^2}\right).
\end{align}

This is only possible if
\begin{align}
    \sqrt{a_1a_3}<{\cal O}\left(\frac{1}{M_*^2}\right).
\end{align}

Again, we find that although satisfaction of the bounds is possible, it requires one of the length scales in the Lagrangian to be below the scale at which effects of Quantum Gravity become dominant and cannot be ignored.

\item \emph{The quartic model} 
\begin{align}\label{eq:minimal_model}
\bar f(\Phi)=a_2 \Phi^2 + a_4 \Phi^4,
\end{align}

As a last example we study the simplest model which exhibits scalarization as well as a $\Phi\rightarrow -\Phi$ symmetry.

Again, we have several extremes, depending on the value of $a_2$ and $a_4$. We denote them $\Phi_0=0$ and $\Phi_{\pm}=\pm\sqrt{-\frac{a_2}{2 a_4}}$, with these existing only for $a_2/a_4<0$. Expanding around $\Phi_0$ leaves the function untouched and we find that the bound from the $2$-to-$2$ amplitude is automatically satisfied  if
\begin{align}\label{eq:boundsx0}
    a_2>-{\cal O}\left(\frac{1}{M_*^2}\right).
\end{align}

This can be satisfied easily by letting $a_2>0$, with $a_4$ unconstrained.  Its sign is fixed instead by either allowing or not the presence of the extra extremes $\Phi_{\pm}$, whose existence demands $a_2 a_4<0$.  Expanding around them we get
\begin{align}
   F(\Phi)=-\frac{a_2^2}{4a_4}-2a_2 (\Phi-\Phi_\pm)^2\pm 2\sqrt{-2a_2 a_4}(\Phi-\Phi_\pm)^3 +a_4 (\Phi-\Phi_\pm)^4.
\end{align}

In contrast to the cubic case, the original symmetry of the action under $\Phi\rightarrow -\Phi$ implies the same conclusion for both points $\Phi_\pm$. We get
\begin{align}
    a_2<{\cal O}\left(\frac{1}{M_*^2}\right),
\end{align}
where we have omitted order one coefficients. Together with the condition obtained from expanding around $\Phi_0=0$ we thus get
\begin{align}
    |a_2|<{\cal O}\left(\frac{1}{M_*^2}\right).
\end{align}

Again, we find that the length scales in the problem allow for satisfaction of positivity bounds only if they are smaller than the UV scale at which quantum gravitational effects enter into the spectrum of the theory.

A way out of this is then to choose $a_2 a_4>0$ so that the extremes $\Phi_{\pm}$ are not real. However, in this case we can now invoke the $n$-to-$n$ positivity bounds, which constrain precisely the value of $a_4$ to be
\begin{align}
a_4<{\cal O}\left(\frac{\Lambda^2}{M_P^2}\frac{1}{M_P^2}\right)\ll  {\cal O}\left(\frac{1}{M_P^2}\right),
\end{align}
and again rules out the model unless the scale $a_4$ is comparable to that of Quantum Gravity.

\end{itemize}
%%%%%%%%%%%%%%%%%%%%%%%%%%%%%%%%%%%%%

The examples that we have just explored show how powerful even the simplest positivity bounds can be. By just demanding satisfaction at all extremes of the function $F(\Phi)$ we can severely constraint or even rule out popular models found in the literature. 

The main conundrum for SGBG models, specially when $n$-to-$n$ amplitudes are considered, comes from the size of the allowed violation of positivity (negativity) in \eqref{eq:bounds_fbar}. For values of $\Lambda$ compatible with our initial assumption $\Lambda\ll M_P$, they typically require some of the length scales given by $L_{2n}=\sqrt{\frac{d^{2n}F}{d\Phi^{2n}}}$ to be significantly smaller than either the Quantum Gravity scale $M_*^{-1}$ or the Planck length. This puts a huge question mark on the validity of SGBG as a valid low energy description of gravitational phenomena. Due to the presence of this scale, there will be eventually physical effects which require the full understanding of the theory at energies larger than those described by General Relativity. Although we cannot barre out the possibility of writing a non-linear model that conspires to satisfy positivity bounds and, at the same time, avoids this issue; it seems to be a difficult task. 

For pure polynomial models, the simpler and more universal bounds obtained from $2$-to-$2$ amplitudes can be used to rule out almost all possible options. This means that we must restrict ourselves to quadratic functions of the form
\begin{align}\label{eq:quad_alive}
    F(\Phi)=F_0+F_1\Phi+F_2 \Phi^2,\qquad F_2>-{\cal O}\left(\frac{1}{M_*^2}\right),
\end{align}
with (almost) positive second derivative. The reasoning is simple -- higher powers of $\Phi$ would imply the presence of extra extremes where, due to continuity, $F''(\Phi)$ must change sign. This will automatically imply the violation of positivity bounds in at least one of the extremes, unless the scales are larger that $M^{-1}_*$, as we have discussed.  In other words,  it is not possible for a continuous function to have positive second derivative in all its extremes.  See e.g. the case of the cubic model above. Thus, we are forced to conclude that the only surviving SGBG model that (naively) satisfies positivity bounds is the quadratic model. A possible quartic term $F_4 \Phi^4$ such that $F_2 F_4>0$ is in principle allowed, since it bans the presence of additional extremes, but it is severely constrained once the bounds for $n$-to-$n$ amplitudes are taken into account.

\section{Conclusions and discussion}\label{sec:conclusions}

In this work we have studied the consistency of SGBG by demanding its compatibility with a UV completion satisfying reasonable assumptions of Lorentz invariance, causality, locality and unitarity. Since the action of SGBG contains interactions with coupling constants of positive length dimension, it must be interpreted as an EFT with a cut-off $\Lambda$ which, for consistency, must be significantly smaller than the Planck mass or any other scale at which quantum gravitational effects start dominating the dynamics of the theory. Here we have contemplated the possibility of this energy scale being lower than $M_P$ -- but still significantly large, and larger than $\Lambda$ -- through an additional mass scale $M_*$.

Even though SGBG has found its place on describing effective models for black-holes, which are highly curved space-times, the EFT logic implies that given its Lagrangian, any process happening at $E\ll \Lambda$ must be effectively described by the same action. This includes the case of particles described as perturbations around flat space, and allows us to borrow a large collection of recent results on constraining effective field theories from UV consistency. In particular, through this work we have focused on \emph{positivity bounds}, which constrain the sign of combinations of couplings in the EFT Lagrangian to be positive, and are derived by looking at scattering amplitudes.

We have shown that by demanding the reasonable existence of the UV completion, standard positivity bounds on the $2$-to-$2$ scattering amplitude can constrain the value of the second derivative of the function $F(\Phi)$ appearing in the Lagrangian. However, the information that we can easily extract from them ends here. In order to obtain more efficient bounds, we have turned our attention to the case of $n$-to-$n$ scattering processes of scalar fields around flat space. Under certain reasonable assumptions about the form of the Uv completion, we have shown that thanks to the large symmetry and simple form of the Lagrangian, these seemingly complicated processes can be computed, at leading order in scale suppression in the positivity bounds, in a compact and universal way, which after regularization and renormalization provides constraints for all even derivatives of the function $F(\Phi)$, starting with the fourth derivative.

Our main result is summarized in expressions \eqref{eq:bounds_fbar}, from which several conclusions can be extracted. A first straightforward one is that although scalarization of black holes is allowed by our bounds, spin-induced scalarization is highly constrained and possibly completely forbidden. In more general grounds, we can understand the consequences of our bounds in two steps. First, let us assume that all energy scales involved in a process are extremely small with respect to any gravitational scale. Thus, we can formally take the limit $(M_*,M_P)\rightarrow \infty$ and our positivity bounds reduce to
\begin{align}
    \left. \frac{d^2F}{d\Phi^2}\right|_{\Phi=\Phi_0}>0,\qquad \left. \frac{d^{2n}F}{d\Phi^{2n}}\right|_{\Phi=\Phi_0}<0,\quad n\geq 2.
\end{align}

Focusing on polynomial models, and in general on any function with more than one extreme $\Phi_0$, these conditions automatically rule out all models of these kind provided that all derivatives are always continuous, since they eventually have to flip sign between extremes. The only way out is to acknowledge that, in the presence of gravity, and thus of finite $M_*$ and $M_P$, positivity can be violated by terms controlled by UV gravitational effects, thus arriving to the original bounds \eqref{eq:bounds_fbar}. However, these imply that at least some of the length scales in the problem, given by the squared root of the derivatives, are small enough so that quantum gravitational effects cannot be ignored anymore. This puts into question the validity of SGBG models for scalarization, where the typical scale of the problem is macroscopic and comparable to the radius of the horizon of the black hole. Only the simple model \eqref{eq:quad_alive} seems to survive the constraints presented here, albeit with the value of its second derivative restricted. However, and although it exhibits scalarization through a tachyonic instability for a large class of black holes, this simple function cannot provide a mechanism to stabilize hairy black holes after quenching the instability, and thus its phenomenological interest is limited. There lingers the open challenge to find a non-linear function beyond \eqref{eq:quad_alive} that satisfies the bounds \eqref{eq:bounds_fbar} while providing both the possibility of scalarization and stable hairy black holes after the process.

\section*{Acknowledgements}
I am grateful to R. Santos-García and A. Tokareva for their collaboration at early stages of this project. I also want to thank E. Barausse, A. Dima and N. Franchini for valuable discussions and comments on the manuscript. My work has been supported by  the European Union’s H2020 ERC Consolidator Grant “GRavity from Astrophysical to Microscopic Scales” grant agreement no. GRAMS-815673. Part of the computations presented here have been performed by using the Mathematica packages xAct \cite{Brizuela:2008ra,Nutma:2013zea} and Package-X \cite{Patel:2015tea}.

%%%%%%%%%%%%%%%%%%%%%%%%%%%%%%%%%%%%%%%%%%%%%%%%%%%%%%%%%%%%%%%%%%%%%%
%%%%%%%%%%%%%%%%%%%%%%%%%%%%%%%%%%%%%%%%%%%%%%%%%%%%%%%%%%%%%%%%%%%%%%
%%%%%%%%%%%%%%%%%%%%%%%%%%%%%%%%%%%%%%%%%%%%%%%%%%%%%%%%%%%%%%%%%%%%%%
%%%%%%%%%%%%%%%%%%%%%%%%%%%%%%%%%%%%%%%%%%%%%%%%%%%%%%%%%%%%%%%%%%%%%%
%%%%%%%%%%%%%%%%%%%%%%%%%%%%%%%%%%%%%%%%%%%%%%%%%%%%%%%%%%%%%%%%%%%%%%
%%%%%%%%%%%%%%%%%%%%%%%%%%%%%%%%%%%%%%%%%%%%%%%%%%%%%%%%%%%%%%%%%%%%%%
\appendix

\section{Perturbative Expansion}\label{app:perturbation}
We summarize here the derivation of all the pieces required to evaluate scattering amplitudes within the action \eqref{eq:starting_action}. We thus start by expanding the fields as
\begin{align}
    g_{\m\n}=\eta_{\m\n}+h_{\m\n},\quad \phi=\phi_0 + \sigma,
\end{align}
where we stress that a flat metric $\eta_{\m\n}$ is a vacuum solution of the theory as long as $\phi_0\equiv {\rm constant}$ satisfies $f'(\phi_0/\Lambda)=0$. We will retain only terms with up to two gravitational perturbations $h_{\m\n}$, which will be enough for our purposes here. Thus, we neglect gravitational self-interactions. 

In order to be able to obtain the gravitational propagator, we need to fix the gauge invariance of the theory, corresponding to linearised diffeomorphisms
\begin{align}
    h_{\m\n}\rightarrow \partial_\m \xi_\n + \partial_\n\xi_\m,
\end{align}
where $\xi^\m$ is the generator of infinitesimal transformations. We choose the standard De Donder gauge
\begin{align}
    &S_{\rm gf}=\frac{M_P^2 \theta }{2}\int d^4x\  \left(\partial^\n h_{\m\n}-\frac{1}{2}\partial_\m h\right)\left(\partial^\b h_{\b}^\m-\frac{1}{2} \partial^\m h\right),
\end{align}
where $\theta$ is a constant that can be chosen at convenience. 

In principle, this gauge fixing should come accompanied by a ghost action. However, here we focus on amplitudes with only scalar fields in the external legs. Since we also restrain ourselves to one-loop computations, this implies that no ghost will propagate inside the Feynman diagrams that we consider. Therefore we refrain from including them.

The graviton propagator then becomes
\begin{align}
    \nonumber &\langle h_{\m\n}(q) h_{\rho\sigma}(-q)\rangle\\
    &=\frac{2}{M_P^2}\frac{i}{q^2-i\epsilon}\left[\eta_{\m\rho}\eta_{\n\sigma}+\eta_{\m\sigma}\eta_{\n\rho} -\eta_{\m\n}\eta_{\rho\sigma}-\frac{1+2\theta}{2q^2}\left(\eta_{\m\rho}q_\n q_\sigma + \eta_{\m\sigma}q_\n q_\rho + \eta_{\n\rho} q_\m q_\sigma + \eta_{\n\sigma} q_\m q_\rho\right)\right],
\end{align}
where we have used the usual $i\epsilon$ prescription. From now on we will choose $\theta=-1/2$ for simplicity. The scalar propagator is
\begin{align}
    \langle \phi(q)\phi(-q)\rangle=\frac{i}{q^2-i\epsilon}.
\end{align}

Regarding interactions, note that we are expanding around flat space. Thus, the GB term vanishes at zeroth and linear orders in the graviton expansion and there are no self-interactions of the scalar field. On the other hand, we have interactions between gravitons and the scalar field, which come in two flavours. From the kinetic term we obtain terms with one and two gravitons -- and in general more, but we are restraining us here to second order in $h_{\m\n}$ -- and two scalar fields. They read
\begin{align}
    \langle \phi (q_1) \phi(q_2) h^{\m\n}(q_3)\rangle &=\frac{i}{2}\left(q_1^\m q_2^\n + q_1^\n q_2^\m - \eta^{\m\n}  q_1\cdot q_2\right),\\
    \langle \phi (q_1) \phi(q_2) h^{\a\b}(q_3) h^{\m\n}(q_4)\rangle &= \frac{1}{4} g^{\m\n} q_{1}^{\b} q_{2}^{\a} -  \frac{1}{4} g^{\b\n} q_{1}^{\m} q_{2}^{\a} -  \frac{1}{4} g^{\b\m} q_{1}^{\n} q_{2}^{\a} + \frac{1}{4} g^{\m\n} q_{1}^{\a} q_{2}^{\b} -  \frac{1}{4} g^{\a\n} q_{1}^{\m} q_{2}^{\b}\nonumber \\
&-  \frac{1}{4} g^{\a\m} q_{1}^{\n} q_{2}^{\b} -  \frac{1}{4} g^{\b\n} q_{1}^{\a} q_{2}^{\m} -  \frac{1}{4} g^{\a\n} q_{1}^{\b} q_{2}^{\m} + \frac{1}{4} g^{\a\b} q_{1}^{\n} q_{2}^{\m} -  \frac{1}{4} g^{\b\m} q_{1}^{\a} q_{2}^{\n}  \nonumber \\
&-  \frac{1}{4} g^{\a\m} q_{1}^{\b} q_{2}^{\n} + \frac{1}{4} g^{\a\b} q_{1}^{\m} q_{2}^{\n}+ \frac{1}{4} g^{\a\n} g^{\b\m} q_{1}^{\delta}q_{2\delta} + \frac{1}{4} g^{\a\m} g^{\b\n} q_{1}^{\delta} q_{2\delta} \\
&-  \frac{1}{4} g^{\a\b} g^{\m\n} q_{1}^{\delta} q_{2\delta}.
\end{align}

On the other hand, from the GB term we obtain a vertex with two gravitons and any number of scalar legs, starting from two. We can pack all of them it in the expression
\begin{align}
    \langle \phi_1(q_1) \phi_2(q_2)\dots \phi_n(q_n) h^{\a\b}(q_m)h^{\m\n}(q_l)\rangle= i f_n {\cal T}^{\a\b\m\n},
\end{align}
where the tensor ${\cal T}^{\a\b\m\n}$ is given by
\begin{align}
{\cal T}^{\a\b\m\n}&=- \frac{1}{4} q_m^2 q_l^2 g^{\a\n} g^{\m\b} -  \frac{1}{4} q_m^2 q_l^2 g^{\a\m} g^{\b\n} + \frac{1}{2} q_m^2 q_l^2 g^{\a\b} g^{\m\n} -  \frac{1}{2} q_l^2 g^{\m\n} q_m^{\a} q_m^{\b} + \frac{1}{4} q_l^2 g^{\b\n} q_m^{\a} q_m^{\mu} \nonumber \\ 
& + \frac{1}{4} q_l^2 g^{\a\n} q_m^{\b} q_m^{\mu} + \frac{1}{4} q_l^2 g^{\m\b} q_m^{\a} q_m^{\nu} + \frac{1}{4} q_l^2 g^{\a\m} q_m^{\b} q_m^{\nu} -  \frac{1}{2} q_l^2 g^{\a\b} q_m^{\mu} q_m^{\nu} -  \frac{1}{2} q_m^2 g^{\m\n} q_l^{\a} q_l^{\b} \nonumber \\ 
& + \frac{1}{2} q_m^{\mu} q_m^{\nu} q_l^{\a} q_l^{\b} + \frac{1}{4} q_m^2 g^{\b\n} q_l^{\a} q_l^{\mu} -  \frac{1}{4} q_m^{\b} q_m^{\nu} q_l^{\a} q_l^{\mu} + \frac{1}{4} q_m^2 g^{\a\n} q_l^{\b} q_l^{\mu} -  \frac{1}{4} q_m^{\a} q_m^{\nu} q_l^{\b} q_l^{\mu} \nonumber \\ 
& + \frac{1}{4} q_m^2 g^{\m\b} q_l^{\a} q_l^{\nu} -  \frac{1}{4} q_m^{\b} q_m^{\mu} q_l^{\a} q_l^{\nu} + \frac{1}{4} q_m^2 g^{\a\m} q_l^{\b} q_l^{\nu} -  \frac{1}{4} q_m^{\a} q_m^{\mu} q_l^{\b} q_l^{\nu} -  \frac{1}{2} q_m^2 g^{\a\b}q_l^{\mu} q_l^{\nu} \nonumber \\ 
& + \frac{1}{2} q_m^{\a} q_m^{\b} q_l^{\mu} q_l^{\nu} + \frac{1}{2} g^{\m\n} q_m^{\b} q_m^{\delta} q_l^{\a} q_{l\delta} -  \frac{1}{4} g^{\b\n} q_m^{\mu} q_m^{\delta} q_l^{\a} q_{l\delta} -  \frac{1}{4} g^{\m\b} q_m^{\nu} q_m^{\delta} q_l^{\a} q_{l\delta} \nonumber \\ 
& + \frac{1}{2} g^{\m\n} q_m^{\a} q_m^{\delta} q_l^{\b} q_{l\delta} -  \frac{1}{4} g^{\a\n} q_m^{\mu} q_m^{\delta} q_l^{\b} q_{l\delta} -  \frac{1}{4} g^{\a\m} q_m^{\nu} q_m^{\delta} q_l^{\b} q_{l\delta} -  \frac{1}{4} g^{\b\n} q_m^{\a} q_m^{\delta} q_l^{\mu} q_{l\delta} \nonumber \\ 
& -  \frac{1}{4} g^{\a\n} q_m^{\b} q_m^{\delta} q_l^{\mu} q_{l\delta} + \frac{1}{2} g^{\a\b} q_m^{\nu} q_m^{\delta} q_l^{\mu} q_{l\delta} -  \frac{1}{4} g^{\m\b} q_m^{\a}q_m^{\delta} q_l^{\nu} q_{l\delta} -  \frac{1}{4} g^{\a\m} q_m^{\b} q_m^{\delta} q_l^{\nu} q_{l\delta} \nonumber \\ 
& + \frac{1}{2} g^{\a\b} q_m^{\mu} q_m^{\delta} q_l^{\nu} q_{l\delta} + \frac{1}{4} g^{\a\n} g^{\m\b} q_m^{\delta} q_m^{\rho} q_{l\delta} q_{l\rho} + \frac{1}{4} g^{\a\m} g^{\b\n} q_m^{\delta} q_m^{\rho} q_{l\delta} q_{l\rho} \nonumber \\ 
&-  \frac{1}{2} g^{\a\b} g^{\m\n} q_m^{\delta} q_m^{\rho} q_{l\delta} q_{l\rho}
\end{align}

Vertices with a larger number of graviton legs can also be obtained from the GB term. However, they will not enter into our computation.

Finally, we will also need the three graviton vertex from the Einstein-Hilbert Lagrangian. We however refrain to show it here due to its length.

%%%%%%%%%%%%%%%%%%%%%%%%%%%%%%%%%%
%%%%%%%%%%%%%%%%%%%%%%%%%%%%%%%%%%
%%%%%%%%%%%%%%%%%%%%%%%%%%%%%%%%%
\section{Kinematics of a 3-to-3 scattering process in the forward limit}\label{app:example}

We work out here a specific example of the kinematics in the forward limit described in section \ref{sec:nton}. We will particularize it to a 3-to-3 scattering amplitude and we will use the forward limit  \eqref{eq:forward_limit}. For $n=3$ we have 6 independent momenta, out of which we can construct 15 different $s_{ij}$ invariants. They can be arranged in a triangular matrix
\begin{align}
    S=\begin{pmatrix}
0 & s_{12} & s_{13} & s_{14} & s_{15} & s_{16}\\
0 & 0 & s_{23} & s_{24} & s_{25} & s_{26}\\
0 & 0 & 0 & s_{34} & s_{35} & s_{36}\\
0 &0 & 0 &0 & s_{45} & s_{56}\\
0 & 0 & 0 &0 & 0 & s_{56}
\end{pmatrix},
\end{align}
where the diagonal elements $s_{ii}=4 p_i^2$ vanish for massless states.

In the forward limit \eqref{eq:forward_limit} the momenta are identified in pairs
\begin{align}
    p_4=-p_1,\quad p_5=-p_2,\quad p_6=-p_3,
\end{align}
so that the invariants $s_{14}$, $s_{25}$, and $s_{36}$ vanish, and we find the relations
\begin{align}
    \nonumber &s_{45}=s_{12}, \quad s_{46}=s_{13}, \quad s_{56}=s_{23},\\
    \nonumber &s_{15}=-s_{12} \quad s_{16}=-s_{13}, \quad s_{24}=-s_{12},\\
    &s_{26}=-s_{23}, \quad s_{34}=-s_{13}, \quad s_{35}=-s_{23}.
\end{align}

Finally, we choose the kinematic configuration $s_{ij}=s/3$ for $i,j\leq 3$ such that energy is equally distributed through the $3$ incoming momenta. The matrix S thus reads
\begin{align}
 S=\frac{s}{3}\cdot \begin{pmatrix}
0 & 1 & 1 & 0 & -1 & -1\\
0 & 0 & 1 &-1 & 0 & -1\\
0 & 0 & 0 & -1 & -1 & 0\\
0 &0 & 0 &0 & 1 & 1\\
0 & 0 & 0 &0 & 0 & 1
\end{pmatrix}.
\end{align}

As discussed in the main text, we find that $n=3$ invariants vanish, while we get $n(n-1)=6$ of them to be of the form $s_{ij}=s/3$, and another $n(n-1)=6$ of them to be $s_{ij}=-s/3$. This configuration can now be used to perform an analytic continuation of the amplitude in the variable $s$ and derive positivity bounds.

%%%%%%%%%%%%%%%%%%%%%%%%%%%%%%%%%%
%%%%%%%%%%%%%%%%%%%%%%%%%%%%%%%%%%
%%%%%%%%%%%%%%%%%%%%%%%%%%%%%%%%%

\section{Scattering Amplitudes}\label{app:amplitudes}
\subsection{2-to-2 Amplitude}
We summarize here the computation of the $
\sigma h\rightarrow \sigma h$ amplitude. The leading contribution is given by tree-level exchange of a graviton\footnote{There is also a possible contribution coming from exchange of a scalar, however it vanishes once the polarizations are contracted in the external legs.}, plus a contact interaction. Both contain $t$ and $u$ channels and read
\begin{align}
\nonumber \\
%%%%%%%%%%%%%%%%%%%%%%%%%%%%%%%%%%%%%
\begin{fmffile}{tree-exchange} 
\parbox{15mm}{
\begin{fmfgraph*}(50,50) 
\fmfleft{l1,l2} 
\fmfright{r1,r2} 
\fmf{photon}{l1,c1}  
\fmf{plain}{l2,c2}  
\fmf{photon}{c1,c2}
\fmf{photon}{c1,r1}
\fmf{plain}{c2,r2}
\end{fmfgraph*}
}\end{fmffile}\quad \quad+ \quad \quad \begin{fmffile}{tree-exchangeu} 
\parbox{15mm}{
\begin{fmfgraph*}(50,50) 
\fmfleft{i1,i2}
\fmfright{o1,o2}
\fmf{photon}{i1,v1}
\fmf{phantom}{v1,o1}
\fmf{plain}{i2,v2}
\fmf{phantom}{v2,o2}
\fmf{photon}{v1,v2}
\fmf{photon,tension=0}{v1,o2}
\fmf{plain,tension=0}{v2,o1}
\end{fmfgraph*}
}\end{fmffile}\quad \quad =
\frac{is(s^2 + 3 st +3 t^2)}{M_P^2 t (s+t)}(\epsilon^*_{\pm})_{\m\n}(\epsilon_{\pm})^{\m\n},
\end{align}

\begin{align}
\nonumber \\
%%%%%%%%%%%%%%%%%%%%%%%%%%%%%%%%%%%%%
\begin{fmffile}{tree-contact} 
\parbox{15mm}{
\begin{fmfgraph*}(50,50) 
\fmfleft{l1,l2} 
\fmfright{r1,r2} 
\fmf{photon}{l1,c1}  
\fmf{plain}{l2,c1}  
\fmf{photon}{c1,r1}
\fmf{plain}{c1,r2}
\end{fmfgraph*}
}\end{fmffile}\quad \quad  +\quad \quad  \begin{fmffile}{tree-contactu} 
\parbox{15mm}{
\begin{fmfgraph*}(50,50) 
\fmfleft{l1,l2} 
\fmfright{r1,r2} 
\fmf{photon}{l1,c1}  
\fmf{plain}{l2,c1}  
\fmf{plain}{c1,r1}
\fmf{photon}{c1,r2}
\end{fmfgraph*}
}\end{fmffile}\quad \quad =  \frac{i \left(f_2 s^2 + 2 f_2 t^2 + 2 s (f_2 t -1)\right)}{M_P^2}(\epsilon^*_{\pm})_{\m\n}(\epsilon_{\pm})^{\m\n},
\end{align}
where $s,t$ and $u$ are here the standard Maldestam variables for a $2$-to-$2$ process and  $\epsilon_{\pm}$ is the polarization tensor of transverse traceless gravitons, satisfying $\epsilon^*_{\pm}=\epsilon_{\mp}$. The product $(\epsilon^*_{\pm})_{\m\n}(\epsilon_{\pm})^{\m\n}$ vanishes for gravitons of different helicity in the final states. Otherwise it sums to unity. We thus hereinafter consider the scattering of gravitons with the same helicity and omit the polarizations in the amplitude.

Finally, the scattering amplitude is defined through the sum of all diagrams in the standard way
\begin{align}
    i{\cal A}=\sum ({\rm diagrams}).
\end{align}

\subsection{$n$-to-$n$ amplitudes}
We compute now the one-loop diagrams displayed in figure \ref{fig:diagrams} and proceed with their renormalization. We start by computing the 12-channel individually in every diagram.

\begin{align}
    \begin{fmffile}{fishF} \parbox{15mm}{
\begin{fmfgraph*}(80,80) 
\fmfleft{l1,l2} 
\fmfright{r1,r2} 
\fmf{plain}{c3,l1}
\fmf{plain}{c3,l2}
\fmf{phantom}{r1,c1}
\fmf{phantom}{r2,c1}
\fmfblob{.15w}{c1}
\fmf{photon,left,tension=.3}{c1,c3,c1}
\end{fmfgraph*}}
\end{fmffile}\quad \quad \quad \quad = \frac{i f_n s_{12}^2}{12 \pi^2 M_P^4}\left[2(\mathfrak{m}^2 -s_{12})\left( \frac{1}{\varepsilon}-\gamma_E +\log(4\pi)\right)+\frac{1}{3}\left(20\mathfrak{m}^2 - 23 s_{12} -12 (\mathfrak{m}^2-s_{12})\log s_{12}\right)\right],
\end{align}
\begin{align}
    \begin{fmffile}{triF} 
    \parbox{15mm}{
\begin{fmfgraph*}(80,80) 
\fmfleft{l1,l2} 
\fmfright{r1,r2} 
\fmf{phantom}{r1,c3}
\fmf{phantom}{r2,c3}
\fmf{plain}{c1,l1}
\fmf{plain}{c2,l2}
\fmf{plain, tension=.2}{c1,c2}
\fmf{photon,tension=.4}{c1,c3}
\fmf{photon,tension=.4}{c2,c3}
\fmfblob{.15w}{c3}
\end{fmfgraph*}}
\end{fmffile}\quad \quad \quad \quad &=\frac{i f_n \mathfrak{m}^2 s_{12} }{64\pi^2 M_P^4}\left[(4 \mathfrak{m}^2 -31 s_{12})\left( \frac{1}{\varepsilon}-\gamma_E +\log(4\pi)\right)- 65 s_{12} -12 \mathfrak{m}^2 -22s_{12} \log s_{12} \right.\nonumber \\
&+(31s_{12}-4\mathfrak{m}^2)\log \mathfrak{m}^2 -4 s_{12} (\mathfrak{m}^2+3s_{12}) C_0(\mathfrak{m}^2,\mathfrak{m}^2,s_{12},0,\mathfrak{m},0)\bigg].
\end{align}
where we are using dimensional regularization in $d=4-\varepsilon$ dimensions, and $\gamma_E$ is the Euler-Mascheroni constant. The function $C_0(s_1,s_2,s_3,m_1,m_2,m_3)$ is defined by the integral
\begin{align}
\nonumber &C_0(s_1,s_2,s_3,m_1,m_2,m_3)=\\
&\lim_{\tau\rightarrow 0^+}\int_0^1 dy \int_0^{1-y}dx \left[s_1 y^2 +s_2 z^2 +(s_1+s_2+s_3)yz+(-s1+m_2^2-m_3^2)y +(-s_2+m_1^2-m_3^2)z + m_3^2 +i\tau\right]^{-1}.
\end{align}

Note however that the contribution from the triangle vanishes polynomially in the limit $\mathfrak{m}\rightarrow 0$. Therefore, it will not contribute to the amplitude in the case of pure SGBG and we neglect it from now on.

In the limit $\mathfrak{m}\rightarrow 0$ we thus get
\begin{align}
    {\cal A}_{12}(s_{12})=-\frac{f_n s_{12}^3}{6\pi^2 M_P^4 }\left( \frac{1}{\varepsilon}-\gamma_E +\log(4\pi)\right)-\frac{f_n s_{12}^3 \left(23-12\log s_{12}\right)}{72\pi^2 M_P^4}.
\end{align}

The total amplitude is given simply by performing crossing-symmetry on this channel, as explained in the main text
\begin{align}
    {\cal A}(s_{ij})=\sum_{s_{ij}}{\cal A}_{12}(s_{ij}).
\end{align}

The divergence in $\varepsilon^{-1}$ can be absorbed by appending the action with a counter-term which generates a higher order derivative interaction for the scalar field of the form
\begin{align}\label{eq:counter-term}
S_{\rm counter-term}=\int d^4x \sqrt{|g|}\ C f(\phi/\Lambda) \partial_\m \square \phi \partial^\m \square\phi,
\end{align}
as can be confirmed from dimensional analysis. By choosing $C$ appropriately in the $\overline{MS}$ scheme
\begin{align}
C\propto -\frac{1}{6\pi^2 M_P^4 }\left( \frac{1}{\varepsilon}-\gamma_E +\log(4\pi)+\log\mu_R^2\right),
\end{align}
we can cancel the divergences for all $n$, being left with a finite renormalized amplitude
\begin{align}
   {\cal A}^n(s_{ij})=-\frac{f_n}{72\pi^2 M_P^4}\sum_{s_{ij}} s_{ij}^3 \left(23-12\log \left(\frac{s_{ij}}{\mu_R^2}\right)\right),
\end{align}
where $\mu_R$ is the renormalization scale, which we leave arbitrary.

Note that the required counter-term precisely takes the form of a higher derivative operator suppressed by a large scale. This signals not only that the theory is non-renormalizable, as we knew already, but that it indeed behaves as an EFT, as we have advertised through this text. Indeed, the operator in \eqref{eq:counter-term} can also be thought as driving the induced effect on the scalar field after integrating out the gravitational interaction, which becomes only important at scales larger than $\Lambda$.

%%%%%%%%%%%%%%%%%%%%%%%%%%%%%%%%%%%%%%%%%%%
%%%%%%%%%%%%%%%%%%%%%%%%%%%%%%%%%%%%%%%%%%%
%%%%%%%%%%%%%%%%%%%%%%%%%%%%%%%%%%%%%%%%%%%
\bibliography{biblio}{}

\begin{thebibliography}{79}%
\makeatletter
\providecommand \@ifxundefined [1]{%
 \@ifx{#1\undefined}
}%
\providecommand \@ifnum [1]{%
 \ifnum #1\expandafter \@firstoftwo
 \else \expandafter \@secondoftwo
 \fi
}%
\providecommand \@ifx [1]{%
 \ifx #1\expandafter \@firstoftwo
 \else \expandafter \@secondoftwo
 \fi
}%
\providecommand \natexlab [1]{#1}%
\providecommand \enquote  [1]{``#1''}%
\providecommand \bibnamefont  [1]{#1}%
\providecommand \bibfnamefont [1]{#1}%
\providecommand \citenamefont [1]{#1}%
\providecommand \href@noop [0]{\@secondoftwo}%
\providecommand \href [0]{\begingroup \@sanitize@url \@href}%
\providecommand \@href[1]{\@@startlink{#1}\@@href}%
\providecommand \@@href[1]{\endgroup#1\@@endlink}%
\providecommand \@sanitize@url [0]{\catcode `\\12\catcode `\$12\catcode
  `\&12\catcode `\#12\catcode `\^12\catcode `\_12\catcode `\%12\relax}%
\providecommand \@@startlink[1]{}%
\providecommand \@@endlink[0]{}%
\providecommand \url  [0]{\begingroup\@sanitize@url \@url }%
\providecommand \@url [1]{\endgroup\@href {#1}{\urlprefix }}%
\providecommand \urlprefix  [0]{URL }%
\providecommand \Eprint [0]{\href }%
\providecommand \doibase [0]{https://doi.org/}%
\providecommand \selectlanguage [0]{\@gobble}%
\providecommand \bibinfo  [0]{\@secondoftwo}%
\providecommand \bibfield  [0]{\@secondoftwo}%
\providecommand \translation [1]{[#1]}%
\providecommand \BibitemOpen [0]{}%
\providecommand \bibitemStop [0]{}%
\providecommand \bibitemNoStop [0]{.\EOS\space}%
\providecommand \EOS [0]{\spacefactor3000\relax}%
\providecommand \BibitemShut  [1]{\csname bibitem#1\endcsname}%
\let\auto@bib@innerbib\@empty
%</preamble>
\bibitem [{\citenamefont {Cohen}(2019)}]{Cohen:2019wxr}%
  \BibitemOpen
  \bibfield  {author} {\bibinfo {author} {\bibfnamefont {T.}~\bibnamefont
  {Cohen}},\ }\bibfield  {title} {\bibinfo {title} {{As Scales Become
  Separated: Lectures on Effective Field Theory}},\ }\href@noop {} {\bibfield
  {journal} {\bibinfo  {journal} {PoS}\ }\textbf {\bibinfo {volume}
  {TASI2018}},\ \bibinfo {pages} {011} (\bibinfo {year} {2019})},\ \Eprint
  {https://arxiv.org/abs/1903.03622} {arXiv:1903.03622 [hep-ph]} \BibitemShut
  {NoStop}%
\bibitem [{\citenamefont {Abbott}\ \emph
  {et~al.}(2016{\natexlab{a}})\citenamefont {Abbott} \emph
  {et~al.}}]{Abbott:2016blz}%
  \BibitemOpen
  \bibfield  {author} {\bibinfo {author} {\bibfnamefont {B.~P.}\ \bibnamefont
  {Abbott}} \emph {et~al.} (\bibinfo {collaboration} {LIGO Scientific,
  Virgo}),\ }\bibfield  {title} {\bibinfo {title} {{Observation of
  Gravitational Waves from a Binary Black Hole Merger}},\ }\href
  {https://doi.org/10.1103/PhysRevLett.116.061102} {\bibfield  {journal}
  {\bibinfo  {journal} {Phys. Rev. Lett.}\ }\textbf {\bibinfo {volume} {116}},\
  \bibinfo {pages} {061102} (\bibinfo {year} {2016}{\natexlab{a}})},\ \Eprint
  {https://arxiv.org/abs/1602.03837} {arXiv:1602.03837 [gr-qc]} \BibitemShut
  {NoStop}%
\bibitem [{\citenamefont {Donoghue}(1994)}]{Donoghue:1994dn}%
  \BibitemOpen
  \bibfield  {author} {\bibinfo {author} {\bibfnamefont {J.~F.}\ \bibnamefont
  {Donoghue}},\ }\bibfield  {title} {\bibinfo {title} {{General relativity as
  an effective field theory: The leading quantum corrections}},\ }\href
  {https://doi.org/10.1103/PhysRevD.50.3874} {\bibfield  {journal} {\bibinfo
  {journal} {Phys. Rev. D}\ }\textbf {\bibinfo {volume} {50}},\ \bibinfo
  {pages} {3874} (\bibinfo {year} {1994})},\ \Eprint
  {https://arxiv.org/abs/gr-qc/9405057} {arXiv:gr-qc/9405057} \BibitemShut
  {NoStop}%
\bibitem [{\citenamefont {Barack}\ \emph {et~al.}(2019)\citenamefont {Barack}
  \emph {et~al.}}]{Barack:2018yly}%
  \BibitemOpen
  \bibfield  {author} {\bibinfo {author} {\bibfnamefont {L.}~\bibnamefont
  {Barack}} \emph {et~al.},\ }\bibfield  {title} {\bibinfo {title} {{Black
  holes, gravitational waves and fundamental physics: a roadmap}},\ }\href
  {https://doi.org/10.1088/1361-6382/ab0587} {\bibfield  {journal} {\bibinfo
  {journal} {Class. Quant. Grav.}\ }\textbf {\bibinfo {volume} {36}},\ \bibinfo
  {pages} {143001} (\bibinfo {year} {2019})},\ \Eprint
  {https://arxiv.org/abs/1806.05195} {arXiv:1806.05195 [gr-qc]} \BibitemShut
  {NoStop}%
\bibitem [{\citenamefont {Abbott}\ \emph
  {et~al.}(2016{\natexlab{b}})\citenamefont {Abbott} \emph
  {et~al.}}]{TheLIGOScientific:2016src}%
  \BibitemOpen
  \bibfield  {author} {\bibinfo {author} {\bibfnamefont {B.~P.}\ \bibnamefont
  {Abbott}} \emph {et~al.} (\bibinfo {collaboration} {LIGO Scientific,
  Virgo}),\ }\bibfield  {title} {\bibinfo {title} {{Tests of general relativity
  with GW150914}},\ }\href {https://doi.org/10.1103/PhysRevLett.116.221101}
  {\bibfield  {journal} {\bibinfo  {journal} {Phys. Rev. Lett.}\ }\textbf
  {\bibinfo {volume} {116}},\ \bibinfo {pages} {221101} (\bibinfo {year}
  {2016}{\natexlab{b}})},\ \bibinfo {note} {[Erratum: Phys.Rev.Lett. 121,
  129902 (2018)]},\ \Eprint {https://arxiv.org/abs/1602.03841}
  {arXiv:1602.03841 [gr-qc]} \BibitemShut {NoStop}%
\bibitem [{\citenamefont {Jordan}(1959)}]{Jordan:1959eg}%
  \BibitemOpen
  \bibfield  {author} {\bibinfo {author} {\bibfnamefont {P.}~\bibnamefont
  {Jordan}},\ }\bibfield  {title} {\bibinfo {title} {{The present state of
  Dirac's cosmological hypothesis}},\ }\href
  {https://doi.org/10.1007/BF01375155} {\bibfield  {journal} {\bibinfo
  {journal} {Z. Phys.}\ }\textbf {\bibinfo {volume} {157}},\ \bibinfo {pages}
  {112} (\bibinfo {year} {1959})}\BibitemShut {NoStop}%
\bibitem [{\citenamefont {Fierz}(1956)}]{Fierz:1956zz}%
  \BibitemOpen
  \bibfield  {author} {\bibinfo {author} {\bibfnamefont {M.}~\bibnamefont
  {Fierz}},\ }\bibfield  {title} {\bibinfo {title} {{On the physical
  interpretation of P.Jordan's extended theory of gravitation}},\ }\href@noop
  {} {\bibfield  {journal} {\bibinfo  {journal} {Helv. Phys. Acta}\ }\textbf
  {\bibinfo {volume} {29}},\ \bibinfo {pages} {128} (\bibinfo {year}
  {1956})}\BibitemShut {NoStop}%
\bibitem [{\citenamefont {Brans}\ and\ \citenamefont
  {Dicke}(1961)}]{Brans:1961sx}%
  \BibitemOpen
  \bibfield  {author} {\bibinfo {author} {\bibfnamefont {C.}~\bibnamefont
  {Brans}}\ and\ \bibinfo {author} {\bibfnamefont {R.~H.}\ \bibnamefont
  {Dicke}},\ }\bibfield  {title} {\bibinfo {title} {{Mach's principle and a
  relativistic theory of gravitation}},\ }\href
  {https://doi.org/10.1103/PhysRev.124.925} {\bibfield  {journal} {\bibinfo
  {journal} {Phys. Rev.}\ }\textbf {\bibinfo {volume} {124}},\ \bibinfo {pages}
  {925} (\bibinfo {year} {1961})}\BibitemShut {NoStop}%
\bibitem [{\citenamefont {Langlois}\ and\ \citenamefont
  {Noui}(2016)}]{Langlois:2015cwa}%
  \BibitemOpen
  \bibfield  {author} {\bibinfo {author} {\bibfnamefont {D.}~\bibnamefont
  {Langlois}}\ and\ \bibinfo {author} {\bibfnamefont {K.}~\bibnamefont
  {Noui}},\ }\bibfield  {title} {\bibinfo {title} {{Degenerate higher
  derivative theories beyond Horndeski: evading the Ostrogradski
  instability}},\ }\href {https://doi.org/10.1088/1475-7516/2016/02/034}
  {\bibfield  {journal} {\bibinfo  {journal} {JCAP}\ }\textbf {\bibinfo
  {volume} {02}},\ \bibinfo {pages} {034}},\ \Eprint
  {https://arxiv.org/abs/1510.06930} {arXiv:1510.06930 [gr-qc]} \BibitemShut
  {NoStop}%
\bibitem [{\citenamefont {Langlois}(2017)}]{Langlois:2017mdk}%
  \BibitemOpen
  \bibfield  {author} {\bibinfo {author} {\bibfnamefont {D.}~\bibnamefont
  {Langlois}},\ }\bibfield  {title} {\bibinfo {title} {{Degenerate Higher-Order
  Scalar-Tensor (DHOST) theories}},\ }in\ \href@noop {} {\emph {\bibinfo
  {booktitle} {{52nd Rencontres de Moriond on Gravitation}}}}\ (\bibinfo {year}
  {2017})\ \Eprint {https://arxiv.org/abs/1707.03625} {arXiv:1707.03625
  [gr-qc]} \BibitemShut {NoStop}%
\bibitem [{\citenamefont {Horndeski}(1974)}]{Horndeski:1974wa}%
  \BibitemOpen
  \bibfield  {author} {\bibinfo {author} {\bibfnamefont {G.~W.}\ \bibnamefont
  {Horndeski}},\ }\bibfield  {title} {\bibinfo {title} {{Second-order
  scalar-tensor field equations in a four-dimensional space}},\ }\href
  {https://doi.org/10.1007/BF01807638} {\bibfield  {journal} {\bibinfo
  {journal} {Int. J. Theor. Phys.}\ }\textbf {\bibinfo {volume} {10}},\
  \bibinfo {pages} {363} (\bibinfo {year} {1974})}\BibitemShut {NoStop}%
\bibitem [{\citenamefont {Gleyzes}\ \emph {et~al.}(2015)\citenamefont
  {Gleyzes}, \citenamefont {Langlois}, \citenamefont {Piazza},\ and\
  \citenamefont {Vernizzi}}]{Gleyzes:2014dya}%
  \BibitemOpen
  \bibfield  {author} {\bibinfo {author} {\bibfnamefont {J.}~\bibnamefont
  {Gleyzes}}, \bibinfo {author} {\bibfnamefont {D.}~\bibnamefont {Langlois}},
  \bibinfo {author} {\bibfnamefont {F.}~\bibnamefont {Piazza}},\ and\ \bibinfo
  {author} {\bibfnamefont {F.}~\bibnamefont {Vernizzi}},\ }\bibfield  {title}
  {\bibinfo {title} {{Healthy theories beyond Horndeski}},\ }\href
  {https://doi.org/10.1103/PhysRevLett.114.211101} {\bibfield  {journal}
  {\bibinfo  {journal} {Phys. Rev. Lett.}\ }\textbf {\bibinfo {volume} {114}},\
  \bibinfo {pages} {211101} (\bibinfo {year} {2015})},\ \Eprint
  {https://arxiv.org/abs/1404.6495} {arXiv:1404.6495 [hep-th]} \BibitemShut
  {NoStop}%
\bibitem [{\citenamefont {Clifton}\ \emph {et~al.}(2012)\citenamefont
  {Clifton}, \citenamefont {Ferreira}, \citenamefont {Padilla},\ and\
  \citenamefont {Skordis}}]{Clifton:2011jh}%
  \BibitemOpen
  \bibfield  {author} {\bibinfo {author} {\bibfnamefont {T.}~\bibnamefont
  {Clifton}}, \bibinfo {author} {\bibfnamefont {P.~G.}\ \bibnamefont
  {Ferreira}}, \bibinfo {author} {\bibfnamefont {A.}~\bibnamefont {Padilla}},\
  and\ \bibinfo {author} {\bibfnamefont {C.}~\bibnamefont {Skordis}},\
  }\bibfield  {title} {\bibinfo {title} {{Modified Gravity and Cosmology}},\
  }\href {https://doi.org/10.1016/j.physrep.2012.01.001} {\bibfield  {journal}
  {\bibinfo  {journal} {Phys. Rept.}\ }\textbf {\bibinfo {volume} {513}},\
  \bibinfo {pages} {1} (\bibinfo {year} {2012})},\ \Eprint
  {https://arxiv.org/abs/1106.2476} {arXiv:1106.2476 [astro-ph.CO]}
  \BibitemShut {NoStop}%
\bibitem [{\citenamefont {Ezquiaga}\ and\ \citenamefont
  {Zumalac\'arregui}(2017)}]{Ezquiaga:2017ekz}%
  \BibitemOpen
  \bibfield  {author} {\bibinfo {author} {\bibfnamefont {J.~M.}\ \bibnamefont
  {Ezquiaga}}\ and\ \bibinfo {author} {\bibfnamefont {M.}~\bibnamefont
  {Zumalac\'arregui}},\ }\bibfield  {title} {\bibinfo {title} {{Dark Energy
  After GW170817: Dead Ends and the Road Ahead}},\ }\href
  {https://doi.org/10.1103/PhysRevLett.119.251304} {\bibfield  {journal}
  {\bibinfo  {journal} {Phys. Rev. Lett.}\ }\textbf {\bibinfo {volume} {119}},\
  \bibinfo {pages} {251304} (\bibinfo {year} {2017})},\ \Eprint
  {https://arxiv.org/abs/1710.05901} {arXiv:1710.05901 [astro-ph.CO]}
  \BibitemShut {NoStop}%
\bibitem [{\citenamefont {Baker}\ \emph {et~al.}(2017)\citenamefont {Baker},
  \citenamefont {Bellini}, \citenamefont {Ferreira}, \citenamefont {Lagos},
  \citenamefont {Noller},\ and\ \citenamefont {Sawicki}}]{Baker:2017hug}%
  \BibitemOpen
  \bibfield  {author} {\bibinfo {author} {\bibfnamefont {T.}~\bibnamefont
  {Baker}}, \bibinfo {author} {\bibfnamefont {E.}~\bibnamefont {Bellini}},
  \bibinfo {author} {\bibfnamefont {P.~G.}\ \bibnamefont {Ferreira}}, \bibinfo
  {author} {\bibfnamefont {M.}~\bibnamefont {Lagos}}, \bibinfo {author}
  {\bibfnamefont {J.}~\bibnamefont {Noller}},\ and\ \bibinfo {author}
  {\bibfnamefont {I.}~\bibnamefont {Sawicki}},\ }\bibfield  {title} {\bibinfo
  {title} {{Strong constraints on cosmological gravity from GW170817 and GRB
  170817A}},\ }\href {https://doi.org/10.1103/PhysRevLett.119.251301}
  {\bibfield  {journal} {\bibinfo  {journal} {Phys. Rev. Lett.}\ }\textbf
  {\bibinfo {volume} {119}},\ \bibinfo {pages} {251301} (\bibinfo {year}
  {2017})},\ \Eprint {https://arxiv.org/abs/1710.06394} {arXiv:1710.06394
  [astro-ph.CO]} \BibitemShut {NoStop}%
\bibitem [{\citenamefont {Creminelli}\ and\ \citenamefont
  {Vernizzi}(2017)}]{Creminelli:2017sry}%
  \BibitemOpen
  \bibfield  {author} {\bibinfo {author} {\bibfnamefont {P.}~\bibnamefont
  {Creminelli}}\ and\ \bibinfo {author} {\bibfnamefont {F.}~\bibnamefont
  {Vernizzi}},\ }\bibfield  {title} {\bibinfo {title} {{Dark Energy after
  GW170817 and GRB170817A}},\ }\href
  {https://doi.org/10.1103/PhysRevLett.119.251302} {\bibfield  {journal}
  {\bibinfo  {journal} {Phys. Rev. Lett.}\ }\textbf {\bibinfo {volume} {119}},\
  \bibinfo {pages} {251302} (\bibinfo {year} {2017})},\ \Eprint
  {https://arxiv.org/abs/1710.05877} {arXiv:1710.05877 [astro-ph.CO]}
  \BibitemShut {NoStop}%
\bibitem [{\citenamefont {Abbott}\ \emph
  {et~al.}(2017{\natexlab{a}})\citenamefont {Abbott} \emph
  {et~al.}}]{Monitor:2017mdv}%
  \BibitemOpen
  \bibfield  {author} {\bibinfo {author} {\bibfnamefont {B.~P.}\ \bibnamefont
  {Abbott}} \emph {et~al.} (\bibinfo {collaboration} {LIGO Scientific, Virgo,
  Fermi-GBM, INTEGRAL}),\ }\bibfield  {title} {\bibinfo {title} {{Gravitational
  Waves and Gamma-rays from a Binary Neutron Star Merger: GW170817 and GRB
  170817A}},\ }\href {https://doi.org/10.3847/2041-8213/aa920c} {\bibfield
  {journal} {\bibinfo  {journal} {Astrophys. J. Lett.}\ }\textbf {\bibinfo
  {volume} {848}},\ \bibinfo {pages} {L13} (\bibinfo {year}
  {2017}{\natexlab{a}})},\ \Eprint {https://arxiv.org/abs/1710.05834}
  {arXiv:1710.05834 [astro-ph.HE]} \BibitemShut {NoStop}%
\bibitem [{\citenamefont {Abbott}\ \emph
  {et~al.}(2017{\natexlab{b}})\citenamefont {Abbott} \emph
  {et~al.}}]{TheLIGOScientific:2017qsa}%
  \BibitemOpen
  \bibfield  {author} {\bibinfo {author} {\bibfnamefont {B.~P.}\ \bibnamefont
  {Abbott}} \emph {et~al.} (\bibinfo {collaboration} {LIGO Scientific,
  Virgo}),\ }\bibfield  {title} {\bibinfo {title} {{GW170817: Observation of
  Gravitational Waves from a Binary Neutron Star Inspiral}},\ }\href
  {https://doi.org/10.1103/PhysRevLett.119.161101} {\bibfield  {journal}
  {\bibinfo  {journal} {Phys. Rev. Lett.}\ }\textbf {\bibinfo {volume} {119}},\
  \bibinfo {pages} {161101} (\bibinfo {year} {2017}{\natexlab{b}})},\ \Eprint
  {https://arxiv.org/abs/1710.05832} {arXiv:1710.05832 [gr-qc]} \BibitemShut
  {NoStop}%
\bibitem [{\citenamefont {Creminelli}\ \emph {et~al.}(2018)\citenamefont
  {Creminelli}, \citenamefont {Lewandowski}, \citenamefont {Tambalo},\ and\
  \citenamefont {Vernizzi}}]{Creminelli:2018xsv}%
  \BibitemOpen
  \bibfield  {author} {\bibinfo {author} {\bibfnamefont {P.}~\bibnamefont
  {Creminelli}}, \bibinfo {author} {\bibfnamefont {M.}~\bibnamefont
  {Lewandowski}}, \bibinfo {author} {\bibfnamefont {G.}~\bibnamefont
  {Tambalo}},\ and\ \bibinfo {author} {\bibfnamefont {F.}~\bibnamefont
  {Vernizzi}},\ }\bibfield  {title} {\bibinfo {title} {{Gravitational Wave
  Decay into Dark Energy}},\ }\href
  {https://doi.org/10.1088/1475-7516/2018/12/025} {\bibfield  {journal}
  {\bibinfo  {journal} {JCAP}\ }\textbf {\bibinfo {volume} {12}},\ \bibinfo
  {pages} {025}},\ \Eprint {https://arxiv.org/abs/1809.03484} {arXiv:1809.03484
  [astro-ph.CO]} \BibitemShut {NoStop}%
\bibitem [{\citenamefont {Creminelli}\ \emph {et~al.}(2020)\citenamefont
  {Creminelli}, \citenamefont {Tambalo}, \citenamefont {Vernizzi},\ and\
  \citenamefont {Yingcharoenrat}}]{Creminelli:2019kjy}%
  \BibitemOpen
  \bibfield  {author} {\bibinfo {author} {\bibfnamefont {P.}~\bibnamefont
  {Creminelli}}, \bibinfo {author} {\bibfnamefont {G.}~\bibnamefont {Tambalo}},
  \bibinfo {author} {\bibfnamefont {F.}~\bibnamefont {Vernizzi}},\ and\
  \bibinfo {author} {\bibfnamefont {V.}~\bibnamefont {Yingcharoenrat}},\
  }\bibfield  {title} {\bibinfo {title} {{Dark-Energy Instabilities induced by
  Gravitational Waves}},\ }\href
  {https://doi.org/10.1088/1475-7516/2020/05/002} {\bibfield  {journal}
  {\bibinfo  {journal} {JCAP}\ }\textbf {\bibinfo {volume} {05}},\ \bibinfo
  {pages} {002}},\ \Eprint {https://arxiv.org/abs/1910.14035} {arXiv:1910.14035
  [gr-qc]} \BibitemShut {NoStop}%
\bibitem [{\citenamefont {Bezares}\ \emph {et~al.}(2021)\citenamefont
  {Bezares}, \citenamefont {Crisostomi}, \citenamefont {Palenzuela},\ and\
  \citenamefont {Barausse}}]{Bezares:2020wkn}%
  \BibitemOpen
  \bibfield  {author} {\bibinfo {author} {\bibfnamefont {M.}~\bibnamefont
  {Bezares}}, \bibinfo {author} {\bibfnamefont {M.}~\bibnamefont {Crisostomi}},
  \bibinfo {author} {\bibfnamefont {C.}~\bibnamefont {Palenzuela}},\ and\
  \bibinfo {author} {\bibfnamefont {E.}~\bibnamefont {Barausse}},\ }\bibfield
  {title} {\bibinfo {title} {{K-dynamics: well-posed 1+1 evolutions in
  K-essence}},\ }\href {https://doi.org/10.1088/1475-7516/2021/03/072}
  {\bibfield  {journal} {\bibinfo  {journal} {JCAP}\ }\textbf {\bibinfo
  {volume} {03}},\ \bibinfo {pages} {072}},\ \Eprint
  {https://arxiv.org/abs/2008.07546} {arXiv:2008.07546 [gr-qc]} \BibitemShut
  {NoStop}%
\bibitem [{\citenamefont {Barausse}(2019)}]{Barausse:2019pri}%
  \BibitemOpen
  \bibfield  {author} {\bibinfo {author} {\bibfnamefont {E.}~\bibnamefont
  {Barausse}},\ }\bibfield  {title} {\bibinfo {title} {{Black holes in General
  Relativity and beyond}},\ }\href
  {https://doi.org/10.3390/proceedings2019017001} {\bibfield  {journal}
  {\bibinfo  {journal} {MDPI Proc.}\ }\textbf {\bibinfo {volume} {17}},\
  \bibinfo {pages} {1} (\bibinfo {year} {2019})},\ \Eprint
  {https://arxiv.org/abs/1902.09199} {arXiv:1902.09199 [gr-qc]} \BibitemShut
  {NoStop}%
\bibitem [{\citenamefont {Sotiriou}(2015)}]{Sotiriou:2015pka}%
  \BibitemOpen
  \bibfield  {author} {\bibinfo {author} {\bibfnamefont {T.~P.}\ \bibnamefont
  {Sotiriou}},\ }\bibfield  {title} {\bibinfo {title} {{Black Holes and Scalar
  Fields}},\ }\href {https://doi.org/10.1088/0264-9381/32/21/214002} {\bibfield
   {journal} {\bibinfo  {journal} {Class. Quant. Grav.}\ }\textbf {\bibinfo
  {volume} {32}},\ \bibinfo {pages} {214002} (\bibinfo {year} {2015})},\
  \Eprint {https://arxiv.org/abs/1505.00248} {arXiv:1505.00248 [gr-qc]}
  \BibitemShut {NoStop}%
\bibitem [{\citenamefont {Damour}\ and\ \citenamefont
  {Taylor}(1992)}]{Damour:1991rd}%
  \BibitemOpen
  \bibfield  {author} {\bibinfo {author} {\bibfnamefont {T.}~\bibnamefont
  {Damour}}\ and\ \bibinfo {author} {\bibfnamefont {J.~H.}\ \bibnamefont
  {Taylor}},\ }\bibfield  {title} {\bibinfo {title} {{Strong field tests of
  relativistic gravity and binary pulsars}},\ }\href
  {https://doi.org/10.1103/PhysRevD.45.1840} {\bibfield  {journal} {\bibinfo
  {journal} {Phys. Rev. D}\ }\textbf {\bibinfo {volume} {45}},\ \bibinfo
  {pages} {1840} (\bibinfo {year} {1992})}\BibitemShut {NoStop}%
\bibitem [{\citenamefont {Will}(2014)}]{Will:2014kxa}%
  \BibitemOpen
  \bibfield  {author} {\bibinfo {author} {\bibfnamefont {C.~M.}\ \bibnamefont
  {Will}},\ }\bibfield  {title} {\bibinfo {title} {{The Confrontation between
  General Relativity and Experiment}},\ }\href
  {https://doi.org/10.12942/lrr-2014-4} {\bibfield  {journal} {\bibinfo
  {journal} {Living Rev. Rel.}\ }\textbf {\bibinfo {volume} {17}},\ \bibinfo
  {pages} {4} (\bibinfo {year} {2014})},\ \Eprint
  {https://arxiv.org/abs/1403.7377} {arXiv:1403.7377 [gr-qc]} \BibitemShut
  {NoStop}%
\bibitem [{\citenamefont {Yunes}\ and\ \citenamefont
  {Siemens}(2013)}]{Yunes:2013dva}%
  \BibitemOpen
  \bibfield  {author} {\bibinfo {author} {\bibfnamefont {N.}~\bibnamefont
  {Yunes}}\ and\ \bibinfo {author} {\bibfnamefont {X.}~\bibnamefont
  {Siemens}},\ }\bibfield  {title} {\bibinfo {title} {{Gravitational-Wave Tests
  of General Relativity with Ground-Based Detectors and Pulsar
  Timing-Arrays}},\ }\href {https://doi.org/10.12942/lrr-2013-9} {\bibfield
  {journal} {\bibinfo  {journal} {Living Rev. Rel.}\ }\textbf {\bibinfo
  {volume} {16}},\ \bibinfo {pages} {9} (\bibinfo {year} {2013})},\ \Eprint
  {https://arxiv.org/abs/1304.3473} {arXiv:1304.3473 [gr-qc]} \BibitemShut
  {NoStop}%
\bibitem [{\citenamefont {Barausse}\ \emph {et~al.}(2016)\citenamefont
  {Barausse}, \citenamefont {Yunes},\ and\ \citenamefont
  {Chamberlain}}]{Barausse:2016eii}%
  \BibitemOpen
  \bibfield  {author} {\bibinfo {author} {\bibfnamefont {E.}~\bibnamefont
  {Barausse}}, \bibinfo {author} {\bibfnamefont {N.}~\bibnamefont {Yunes}},\
  and\ \bibinfo {author} {\bibfnamefont {K.}~\bibnamefont {Chamberlain}},\
  }\bibfield  {title} {\bibinfo {title} {{Theory-Agnostic Constraints on
  Black-Hole Dipole Radiation with Multiband Gravitational-Wave
  Astrophysics}},\ }\href {https://doi.org/10.1103/PhysRevLett.116.241104}
  {\bibfield  {journal} {\bibinfo  {journal} {Phys. Rev. Lett.}\ }\textbf
  {\bibinfo {volume} {116}},\ \bibinfo {pages} {241104} (\bibinfo {year}
  {2016})},\ \Eprint {https://arxiv.org/abs/1603.04075} {arXiv:1603.04075
  [gr-qc]} \BibitemShut {NoStop}%
\bibitem [{\citenamefont {Toubiana}\ \emph {et~al.}(2020)\citenamefont
  {Toubiana}, \citenamefont {Marsat}, \citenamefont {Barausse}, \citenamefont
  {Babak},\ and\ \citenamefont {Baker}}]{Toubiana:2020vtf}%
  \BibitemOpen
  \bibfield  {author} {\bibinfo {author} {\bibfnamefont {A.}~\bibnamefont
  {Toubiana}}, \bibinfo {author} {\bibfnamefont {S.}~\bibnamefont {Marsat}},
  \bibinfo {author} {\bibfnamefont {E.}~\bibnamefont {Barausse}}, \bibinfo
  {author} {\bibfnamefont {S.}~\bibnamefont {Babak}},\ and\ \bibinfo {author}
  {\bibfnamefont {J.}~\bibnamefont {Baker}},\ }\bibfield  {title} {\bibinfo
  {title} {{Tests of general relativity with stellar-mass black hole binaries
  observed by LISA}},\ }\href {https://doi.org/10.1103/PhysRevD.101.104038}
  {\bibfield  {journal} {\bibinfo  {journal} {Phys. Rev. D}\ }\textbf {\bibinfo
  {volume} {101}},\ \bibinfo {pages} {104038} (\bibinfo {year} {2020})},\
  \Eprint {https://arxiv.org/abs/2004.03626} {arXiv:2004.03626 [gr-qc]}
  \BibitemShut {NoStop}%
\bibitem [{\citenamefont {Berti}\ \emph {et~al.}(2016)\citenamefont {Berti},
  \citenamefont {Sesana}, \citenamefont {Barausse}, \citenamefont {Cardoso},\
  and\ \citenamefont {Belczynski}}]{Berti:2016lat}%
  \BibitemOpen
  \bibfield  {author} {\bibinfo {author} {\bibfnamefont {E.}~\bibnamefont
  {Berti}}, \bibinfo {author} {\bibfnamefont {A.}~\bibnamefont {Sesana}},
  \bibinfo {author} {\bibfnamefont {E.}~\bibnamefont {Barausse}}, \bibinfo
  {author} {\bibfnamefont {V.}~\bibnamefont {Cardoso}},\ and\ \bibinfo {author}
  {\bibfnamefont {K.}~\bibnamefont {Belczynski}},\ }\bibfield  {title}
  {\bibinfo {title} {{Spectroscopy of Kerr black holes with Earth- and
  space-based interferometers}},\ }\href
  {https://doi.org/10.1103/PhysRevLett.117.101102} {\bibfield  {journal}
  {\bibinfo  {journal} {Phys. Rev. Lett.}\ }\textbf {\bibinfo {volume} {117}},\
  \bibinfo {pages} {101102} (\bibinfo {year} {2016})},\ \Eprint
  {https://arxiv.org/abs/1605.09286} {arXiv:1605.09286 [gr-qc]} \BibitemShut
  {NoStop}%
\bibitem [{\citenamefont {Akiyama}\ \emph {et~al.}(2019)\citenamefont {Akiyama}
  \emph {et~al.}}]{Akiyama:2019cqa}%
  \BibitemOpen
  \bibfield  {author} {\bibinfo {author} {\bibfnamefont {K.}~\bibnamefont
  {Akiyama}} \emph {et~al.} (\bibinfo {collaboration} {Event Horizon
  Telescope}),\ }\bibfield  {title} {\bibinfo {title} {{First M87 Event Horizon
  Telescope Results. I. The Shadow of the Supermassive Black Hole}},\ }\href
  {https://doi.org/10.3847/2041-8213/ab0ec7} {\bibfield  {journal} {\bibinfo
  {journal} {Astrophys. J. Lett.}\ }\textbf {\bibinfo {volume} {875}},\
  \bibinfo {pages} {L1} (\bibinfo {year} {2019})},\ \Eprint
  {https://arxiv.org/abs/1906.11238} {arXiv:1906.11238 [astro-ph.GA]}
  \BibitemShut {NoStop}%
\bibitem [{\citenamefont {Psaltis}\ \emph {et~al.}(2020)\citenamefont {Psaltis}
  \emph {et~al.}}]{Psaltis:2020lvx}%
  \BibitemOpen
  \bibfield  {author} {\bibinfo {author} {\bibfnamefont {D.}~\bibnamefont
  {Psaltis}} \emph {et~al.} (\bibinfo {collaboration} {Event Horizon
  Telescope}),\ }\bibfield  {title} {\bibinfo {title} {{Gravitational Test
  Beyond the First Post-Newtonian Order with the Shadow of the M87 Black
  Hole}},\ }\href {https://doi.org/10.1103/PhysRevLett.125.141104} {\bibfield
  {journal} {\bibinfo  {journal} {Phys. Rev. Lett.}\ }\textbf {\bibinfo
  {volume} {125}},\ \bibinfo {pages} {141104} (\bibinfo {year} {2020})},\
  \Eprint {https://arxiv.org/abs/2010.01055} {arXiv:2010.01055 [gr-qc]}
  \BibitemShut {NoStop}%
\bibitem [{\citenamefont {V\"olkel}\ \emph {et~al.}(2020)\citenamefont
  {V\"olkel}, \citenamefont {Barausse}, \citenamefont {Franchini},\ and\
  \citenamefont {Broderick}}]{Volkel:2020xlc}%
  \BibitemOpen
  \bibfield  {author} {\bibinfo {author} {\bibfnamefont {S.~H.}\ \bibnamefont
  {V\"olkel}}, \bibinfo {author} {\bibfnamefont {E.}~\bibnamefont {Barausse}},
  \bibinfo {author} {\bibfnamefont {N.}~\bibnamefont {Franchini}},\ and\
  \bibinfo {author} {\bibfnamefont {A.~E.}\ \bibnamefont {Broderick}},\
  }\bibfield  {title} {\bibinfo {title} {{EHT tests of the strong-field regime
  of General Relativity}},\ }\href@noop {} {\  (\bibinfo {year} {2020})},\
  \Eprint {https://arxiv.org/abs/2011.06812} {arXiv:2011.06812 [gr-qc]}
  \BibitemShut {NoStop}%
\bibitem [{\citenamefont {Kanti}\ \emph {et~al.}(1996)\citenamefont {Kanti},
  \citenamefont {Mavromatos}, \citenamefont {Rizos}, \citenamefont {Tamvakis},\
  and\ \citenamefont {Winstanley}}]{Kanti:1995vq}%
  \BibitemOpen
  \bibfield  {author} {\bibinfo {author} {\bibfnamefont {P.}~\bibnamefont
  {Kanti}}, \bibinfo {author} {\bibfnamefont {N.~E.}\ \bibnamefont
  {Mavromatos}}, \bibinfo {author} {\bibfnamefont {J.}~\bibnamefont {Rizos}},
  \bibinfo {author} {\bibfnamefont {K.}~\bibnamefont {Tamvakis}},\ and\
  \bibinfo {author} {\bibfnamefont {E.}~\bibnamefont {Winstanley}},\ }\bibfield
   {title} {\bibinfo {title} {{Dilatonic black holes in higher curvature string
  gravity}},\ }\href {https://doi.org/10.1103/PhysRevD.54.5049} {\bibfield
  {journal} {\bibinfo  {journal} {Phys. Rev. D}\ }\textbf {\bibinfo {volume}
  {54}},\ \bibinfo {pages} {5049} (\bibinfo {year} {1996})},\ \Eprint
  {https://arxiv.org/abs/hep-th/9511071} {arXiv:hep-th/9511071} \BibitemShut
  {NoStop}%
\bibitem [{\citenamefont {Sotiriou}\ and\ \citenamefont
  {Zhou}(2014{\natexlab{a}})}]{Sotiriou:2013qea}%
  \BibitemOpen
  \bibfield  {author} {\bibinfo {author} {\bibfnamefont {T.~P.}\ \bibnamefont
  {Sotiriou}}\ and\ \bibinfo {author} {\bibfnamefont {S.-Y.}\ \bibnamefont
  {Zhou}},\ }\bibfield  {title} {\bibinfo {title} {{Black hole hair in
  generalized scalar-tensor gravity}},\ }\href
  {https://doi.org/10.1103/PhysRevLett.112.251102} {\bibfield  {journal}
  {\bibinfo  {journal} {Phys. Rev. Lett.}\ }\textbf {\bibinfo {volume} {112}},\
  \bibinfo {pages} {251102} (\bibinfo {year} {2014}{\natexlab{a}})},\ \Eprint
  {https://arxiv.org/abs/1312.3622} {arXiv:1312.3622 [gr-qc]} \BibitemShut
  {NoStop}%
\bibitem [{\citenamefont {Sotiriou}\ and\ \citenamefont
  {Zhou}(2014{\natexlab{b}})}]{Sotiriou:2014pfa}%
  \BibitemOpen
  \bibfield  {author} {\bibinfo {author} {\bibfnamefont {T.~P.}\ \bibnamefont
  {Sotiriou}}\ and\ \bibinfo {author} {\bibfnamefont {S.-Y.}\ \bibnamefont
  {Zhou}},\ }\bibfield  {title} {\bibinfo {title} {{Black hole hair in
  generalized scalar-tensor gravity: An explicit example}},\ }\href
  {https://doi.org/10.1103/PhysRevD.90.124063} {\bibfield  {journal} {\bibinfo
  {journal} {Phys. Rev. D}\ }\textbf {\bibinfo {volume} {90}},\ \bibinfo
  {pages} {124063} (\bibinfo {year} {2014}{\natexlab{b}})},\ \Eprint
  {https://arxiv.org/abs/1408.1698} {arXiv:1408.1698 [gr-qc]} \BibitemShut
  {NoStop}%
\bibitem [{\citenamefont {Juli\'e}\ and\ \citenamefont
  {Berti}(2019)}]{Julie:2019sab}%
  \BibitemOpen
  \bibfield  {author} {\bibinfo {author} {\bibfnamefont {F.-L.}\ \bibnamefont
  {Juli\'e}}\ and\ \bibinfo {author} {\bibfnamefont {E.}~\bibnamefont
  {Berti}},\ }\bibfield  {title} {\bibinfo {title} {{Post-Newtonian dynamics
  and black hole thermodynamics in Einstein-scalar-Gauss-Bonnet gravity}},\
  }\href {https://doi.org/10.1103/PhysRevD.100.104061} {\bibfield  {journal}
  {\bibinfo  {journal} {Phys. Rev. D}\ }\textbf {\bibinfo {volume} {100}},\
  \bibinfo {pages} {104061} (\bibinfo {year} {2019})},\ \Eprint
  {https://arxiv.org/abs/1909.05258} {arXiv:1909.05258 [gr-qc]} \BibitemShut
  {NoStop}%
\bibitem [{\citenamefont {East}\ and\ \citenamefont
  {Ripley}(2021)}]{East:2021bqk}%
  \BibitemOpen
  \bibfield  {author} {\bibinfo {author} {\bibfnamefont {W.~E.}\ \bibnamefont
  {East}}\ and\ \bibinfo {author} {\bibfnamefont {J.~L.}\ \bibnamefont
  {Ripley}},\ }\bibfield  {title} {\bibinfo {title} {{Dynamics of spontaneous
  black hole scalarization and mergers in Einstein-scalar-Gauss-Bonnet
  gravity}},\ }\href@noop {} {\  (\bibinfo {year} {2021})},\ \Eprint
  {https://arxiv.org/abs/2105.08571} {arXiv:2105.08571 [gr-qc]} \BibitemShut
  {NoStop}%
\bibitem [{\citenamefont {Antoniou}\ \emph
  {et~al.}(2018{\natexlab{a}})\citenamefont {Antoniou}, \citenamefont
  {Bakopoulos},\ and\ \citenamefont {Kanti}}]{Antoniou:2017acq}%
  \BibitemOpen
  \bibfield  {author} {\bibinfo {author} {\bibfnamefont {G.}~\bibnamefont
  {Antoniou}}, \bibinfo {author} {\bibfnamefont {A.}~\bibnamefont
  {Bakopoulos}},\ and\ \bibinfo {author} {\bibfnamefont {P.}~\bibnamefont
  {Kanti}},\ }\bibfield  {title} {\bibinfo {title} {{Evasion of No-Hair
  Theorems and Novel Black-Hole Solutions in Gauss-Bonnet Theories}},\ }\href
  {https://doi.org/10.1103/PhysRevLett.120.131102} {\bibfield  {journal}
  {\bibinfo  {journal} {Phys. Rev. Lett.}\ }\textbf {\bibinfo {volume} {120}},\
  \bibinfo {pages} {131102} (\bibinfo {year} {2018}{\natexlab{a}})},\ \Eprint
  {https://arxiv.org/abs/1711.03390} {arXiv:1711.03390 [hep-th]} \BibitemShut
  {NoStop}%
\bibitem [{\citenamefont {Silva}\ \emph {et~al.}(2018)\citenamefont {Silva},
  \citenamefont {Sakstein}, \citenamefont {Gualtieri}, \citenamefont
  {Sotiriou},\ and\ \citenamefont {Berti}}]{Silva:2017uqg}%
  \BibitemOpen
  \bibfield  {author} {\bibinfo {author} {\bibfnamefont {H.~O.}\ \bibnamefont
  {Silva}}, \bibinfo {author} {\bibfnamefont {J.}~\bibnamefont {Sakstein}},
  \bibinfo {author} {\bibfnamefont {L.}~\bibnamefont {Gualtieri}}, \bibinfo
  {author} {\bibfnamefont {T.~P.}\ \bibnamefont {Sotiriou}},\ and\ \bibinfo
  {author} {\bibfnamefont {E.}~\bibnamefont {Berti}},\ }\bibfield  {title}
  {\bibinfo {title} {{Spontaneous scalarization of black holes and compact
  stars from a Gauss-Bonnet coupling}},\ }\href
  {https://doi.org/10.1103/PhysRevLett.120.131104} {\bibfield  {journal}
  {\bibinfo  {journal} {Phys. Rev. Lett.}\ }\textbf {\bibinfo {volume} {120}},\
  \bibinfo {pages} {131104} (\bibinfo {year} {2018})},\ \Eprint
  {https://arxiv.org/abs/1711.02080} {arXiv:1711.02080 [gr-qc]} \BibitemShut
  {NoStop}%
\bibitem [{\citenamefont {Doneva}\ and\ \citenamefont
  {Yazadjiev}(2018)}]{Doneva:2017bvd}%
  \BibitemOpen
  \bibfield  {author} {\bibinfo {author} {\bibfnamefont {D.~D.}\ \bibnamefont
  {Doneva}}\ and\ \bibinfo {author} {\bibfnamefont {S.~S.}\ \bibnamefont
  {Yazadjiev}},\ }\bibfield  {title} {\bibinfo {title} {{New Gauss-Bonnet Black
  Holes with Curvature-Induced Scalarization in Extended Scalar-Tensor
  Theories}},\ }\href {https://doi.org/10.1103/PhysRevLett.120.131103}
  {\bibfield  {journal} {\bibinfo  {journal} {Phys. Rev. Lett.}\ }\textbf
  {\bibinfo {volume} {120}},\ \bibinfo {pages} {131103} (\bibinfo {year}
  {2018})},\ \Eprint {https://arxiv.org/abs/1711.01187} {arXiv:1711.01187
  [gr-qc]} \BibitemShut {NoStop}%
\bibitem [{\citenamefont {Herdeiro}\ \emph {et~al.}(2021)\citenamefont
  {Herdeiro}, \citenamefont {Radu}, \citenamefont {Silva}, \citenamefont
  {Sotiriou},\ and\ \citenamefont {Yunes}}]{Herdeiro:2020wei}%
  \BibitemOpen
  \bibfield  {author} {\bibinfo {author} {\bibfnamefont {C.~A.~R.}\
  \bibnamefont {Herdeiro}}, \bibinfo {author} {\bibfnamefont {E.}~\bibnamefont
  {Radu}}, \bibinfo {author} {\bibfnamefont {H.~O.}\ \bibnamefont {Silva}},
  \bibinfo {author} {\bibfnamefont {T.~P.}\ \bibnamefont {Sotiriou}},\ and\
  \bibinfo {author} {\bibfnamefont {N.}~\bibnamefont {Yunes}},\ }\bibfield
  {title} {\bibinfo {title} {{Spin-induced scalarized black holes}},\ }\href
  {https://doi.org/10.1103/PhysRevLett.126.011103} {\bibfield  {journal}
  {\bibinfo  {journal} {Phys. Rev. Lett.}\ }\textbf {\bibinfo {volume} {126}},\
  \bibinfo {pages} {011103} (\bibinfo {year} {2021})},\ \Eprint
  {https://arxiv.org/abs/2009.03904} {arXiv:2009.03904 [gr-qc]} \BibitemShut
  {NoStop}%
\bibitem [{\citenamefont {Berti}\ \emph {et~al.}(2021)\citenamefont {Berti},
  \citenamefont {Collodel}, \citenamefont {Kleihaus},\ and\ \citenamefont
  {Kunz}}]{Berti:2020kgk}%
  \BibitemOpen
  \bibfield  {author} {\bibinfo {author} {\bibfnamefont {E.}~\bibnamefont
  {Berti}}, \bibinfo {author} {\bibfnamefont {L.~G.}\ \bibnamefont {Collodel}},
  \bibinfo {author} {\bibfnamefont {B.}~\bibnamefont {Kleihaus}},\ and\
  \bibinfo {author} {\bibfnamefont {J.}~\bibnamefont {Kunz}},\ }\bibfield
  {title} {\bibinfo {title} {{Spin-induced black-hole scalarization in
  Einstein-scalar-Gauss-Bonnet theory}},\ }\href
  {https://doi.org/10.1103/PhysRevLett.126.011104} {\bibfield  {journal}
  {\bibinfo  {journal} {Phys. Rev. Lett.}\ }\textbf {\bibinfo {volume} {126}},\
  \bibinfo {pages} {011104} (\bibinfo {year} {2021})},\ \Eprint
  {https://arxiv.org/abs/2009.03905} {arXiv:2009.03905 [gr-qc]} \BibitemShut
  {NoStop}%
\bibitem [{\citenamefont {Barausse}\ \emph {et~al.}(2013)\citenamefont
  {Barausse}, \citenamefont {Palenzuela}, \citenamefont {Ponce},\ and\
  \citenamefont {Lehner}}]{Barausse:2012da}%
  \BibitemOpen
  \bibfield  {author} {\bibinfo {author} {\bibfnamefont {E.}~\bibnamefont
  {Barausse}}, \bibinfo {author} {\bibfnamefont {C.}~\bibnamefont
  {Palenzuela}}, \bibinfo {author} {\bibfnamefont {M.}~\bibnamefont {Ponce}},\
  and\ \bibinfo {author} {\bibfnamefont {L.}~\bibnamefont {Lehner}},\
  }\bibfield  {title} {\bibinfo {title} {{Neutron-star mergers in scalar-tensor
  theories of gravity}},\ }\href {https://doi.org/10.1103/PhysRevD.87.081506}
  {\bibfield  {journal} {\bibinfo  {journal} {Phys. Rev. D}\ }\textbf {\bibinfo
  {volume} {87}},\ \bibinfo {pages} {081506} (\bibinfo {year} {2013})},\
  \Eprint {https://arxiv.org/abs/1212.5053} {arXiv:1212.5053 [gr-qc]}
  \BibitemShut {NoStop}%
\bibitem [{\citenamefont {Palenzuela}\ \emph {et~al.}(2014)\citenamefont
  {Palenzuela}, \citenamefont {Barausse}, \citenamefont {Ponce},\ and\
  \citenamefont {Lehner}}]{Palenzuela:2013hsa}%
  \BibitemOpen
  \bibfield  {author} {\bibinfo {author} {\bibfnamefont {C.}~\bibnamefont
  {Palenzuela}}, \bibinfo {author} {\bibfnamefont {E.}~\bibnamefont
  {Barausse}}, \bibinfo {author} {\bibfnamefont {M.}~\bibnamefont {Ponce}},\
  and\ \bibinfo {author} {\bibfnamefont {L.}~\bibnamefont {Lehner}},\
  }\bibfield  {title} {\bibinfo {title} {{Dynamical scalarization of neutron
  stars in scalar-tensor gravity theories}},\ }\href
  {https://doi.org/10.1103/PhysRevD.89.044024} {\bibfield  {journal} {\bibinfo
  {journal} {Phys. Rev. D}\ }\textbf {\bibinfo {volume} {89}},\ \bibinfo
  {pages} {044024} (\bibinfo {year} {2014})},\ \Eprint
  {https://arxiv.org/abs/1310.4481} {arXiv:1310.4481 [gr-qc]} \BibitemShut
  {NoStop}%
\bibitem [{\citenamefont {Sennett}\ and\ \citenamefont
  {Buonanno}(2016)}]{Sennett:2016rwa}%
  \BibitemOpen
  \bibfield  {author} {\bibinfo {author} {\bibfnamefont {N.}~\bibnamefont
  {Sennett}}\ and\ \bibinfo {author} {\bibfnamefont {A.}~\bibnamefont
  {Buonanno}},\ }\bibfield  {title} {\bibinfo {title} {{Modeling dynamical
  scalarization with a resummed post-Newtonian expansion}},\ }\href
  {https://doi.org/10.1103/PhysRevD.93.124004} {\bibfield  {journal} {\bibinfo
  {journal} {Phys. Rev. D}\ }\textbf {\bibinfo {volume} {93}},\ \bibinfo
  {pages} {124004} (\bibinfo {year} {2016})},\ \Eprint
  {https://arxiv.org/abs/1603.03300} {arXiv:1603.03300 [gr-qc]} \BibitemShut
  {NoStop}%
\bibitem [{\citenamefont {Shibata}\ \emph {et~al.}(2014)\citenamefont
  {Shibata}, \citenamefont {Taniguchi}, \citenamefont {Okawa},\ and\
  \citenamefont {Buonanno}}]{Shibata:2013pra}%
  \BibitemOpen
  \bibfield  {author} {\bibinfo {author} {\bibfnamefont {M.}~\bibnamefont
  {Shibata}}, \bibinfo {author} {\bibfnamefont {K.}~\bibnamefont {Taniguchi}},
  \bibinfo {author} {\bibfnamefont {H.}~\bibnamefont {Okawa}},\ and\ \bibinfo
  {author} {\bibfnamefont {A.}~\bibnamefont {Buonanno}},\ }\bibfield  {title}
  {\bibinfo {title} {{Coalescence of binary neutron stars in a scalar-tensor
  theory of gravity}},\ }\href {https://doi.org/10.1103/PhysRevD.89.084005}
  {\bibfield  {journal} {\bibinfo  {journal} {Phys. Rev. D}\ }\textbf {\bibinfo
  {volume} {89}},\ \bibinfo {pages} {084005} (\bibinfo {year} {2014})},\
  \Eprint {https://arxiv.org/abs/1310.0627} {arXiv:1310.0627 [gr-qc]}
  \BibitemShut {NoStop}%
\bibitem [{\citenamefont {Silva}\ \emph {et~al.}(2019)\citenamefont {Silva},
  \citenamefont {Macedo}, \citenamefont {Sotiriou}, \citenamefont {Gualtieri},
  \citenamefont {Sakstein},\ and\ \citenamefont {Berti}}]{Silva:2018qhn}%
  \BibitemOpen
  \bibfield  {author} {\bibinfo {author} {\bibfnamefont {H.~O.}\ \bibnamefont
  {Silva}}, \bibinfo {author} {\bibfnamefont {C.~F.~B.}\ \bibnamefont
  {Macedo}}, \bibinfo {author} {\bibfnamefont {T.~P.}\ \bibnamefont
  {Sotiriou}}, \bibinfo {author} {\bibfnamefont {L.}~\bibnamefont {Gualtieri}},
  \bibinfo {author} {\bibfnamefont {J.}~\bibnamefont {Sakstein}},\ and\
  \bibinfo {author} {\bibfnamefont {E.}~\bibnamefont {Berti}},\ }\bibfield
  {title} {\bibinfo {title} {{Stability of scalarized black hole solutions in
  scalar-Gauss-Bonnet gravity}},\ }\href
  {https://doi.org/10.1103/PhysRevD.99.064011} {\bibfield  {journal} {\bibinfo
  {journal} {Phys. Rev. D}\ }\textbf {\bibinfo {volume} {99}},\ \bibinfo
  {pages} {064011} (\bibinfo {year} {2019})},\ \Eprint
  {https://arxiv.org/abs/1812.05590} {arXiv:1812.05590 [gr-qc]} \BibitemShut
  {NoStop}%
\bibitem [{\citenamefont {Macedo}\ \emph {et~al.}(2019)\citenamefont {Macedo},
  \citenamefont {Sakstein}, \citenamefont {Berti}, \citenamefont {Gualtieri},
  \citenamefont {Silva},\ and\ \citenamefont {Sotiriou}}]{Macedo:2019sem}%
  \BibitemOpen
  \bibfield  {author} {\bibinfo {author} {\bibfnamefont {C.~F.~B.}\
  \bibnamefont {Macedo}}, \bibinfo {author} {\bibfnamefont {J.}~\bibnamefont
  {Sakstein}}, \bibinfo {author} {\bibfnamefont {E.}~\bibnamefont {Berti}},
  \bibinfo {author} {\bibfnamefont {L.}~\bibnamefont {Gualtieri}}, \bibinfo
  {author} {\bibfnamefont {H.~O.}\ \bibnamefont {Silva}},\ and\ \bibinfo
  {author} {\bibfnamefont {T.~P.}\ \bibnamefont {Sotiriou}},\ }\bibfield
  {title} {\bibinfo {title} {{Self-interactions and Spontaneous Black Hole
  Scalarization}},\ }\href {https://doi.org/10.1103/PhysRevD.99.104041}
  {\bibfield  {journal} {\bibinfo  {journal} {Phys. Rev. D}\ }\textbf {\bibinfo
  {volume} {99}},\ \bibinfo {pages} {104041} (\bibinfo {year} {2019})},\
  \Eprint {https://arxiv.org/abs/1903.06784} {arXiv:1903.06784 [gr-qc]}
  \BibitemShut {NoStop}%
\bibitem [{\citenamefont {Bl\'azquez-Salcedo}\ \emph
  {et~al.}(2018)\citenamefont {Bl\'azquez-Salcedo}, \citenamefont {Doneva},
  \citenamefont {Kunz},\ and\ \citenamefont
  {Yazadjiev}}]{Blazquez-Salcedo:2018jnn}%
  \BibitemOpen
  \bibfield  {author} {\bibinfo {author} {\bibfnamefont {J.~L.}\ \bibnamefont
  {Bl\'azquez-Salcedo}}, \bibinfo {author} {\bibfnamefont {D.~D.}\ \bibnamefont
  {Doneva}}, \bibinfo {author} {\bibfnamefont {J.}~\bibnamefont {Kunz}},\ and\
  \bibinfo {author} {\bibfnamefont {S.~S.}\ \bibnamefont {Yazadjiev}},\
  }\bibfield  {title} {\bibinfo {title} {{Radial perturbations of the
  scalarized Einstein-Gauss-Bonnet black holes}},\ }\href
  {https://doi.org/10.1103/PhysRevD.98.084011} {\bibfield  {journal} {\bibinfo
  {journal} {Phys. Rev. D}\ }\textbf {\bibinfo {volume} {98}},\ \bibinfo
  {pages} {084011} (\bibinfo {year} {2018})},\ \Eprint
  {https://arxiv.org/abs/1805.05755} {arXiv:1805.05755 [gr-qc]} \BibitemShut
  {NoStop}%
\bibitem [{\citenamefont {Nicolis}\ \emph {et~al.}(2010)\citenamefont
  {Nicolis}, \citenamefont {Rattazzi},\ and\ \citenamefont
  {Trincherini}}]{Nicolis:2009qm}%
  \BibitemOpen
  \bibfield  {author} {\bibinfo {author} {\bibfnamefont {A.}~\bibnamefont
  {Nicolis}}, \bibinfo {author} {\bibfnamefont {R.}~\bibnamefont {Rattazzi}},\
  and\ \bibinfo {author} {\bibfnamefont {E.}~\bibnamefont {Trincherini}},\
  }\bibfield  {title} {\bibinfo {title} {{Energy's and amplitudes'
  positivity}},\ }\href {https://doi.org/10.1007/JHEP05(2010)095} {\bibfield
  {journal} {\bibinfo  {journal} {JHEP}\ }\textbf {\bibinfo {volume} {05}},\
  \bibinfo {pages} {095}},\ \bibinfo {note} {[Erratum: JHEP 11, 128 (2011)]},\
  \Eprint {https://arxiv.org/abs/0912.4258} {arXiv:0912.4258 [hep-th]}
  \BibitemShut {NoStop}%
\bibitem [{\citenamefont {Adams}\ \emph {et~al.}(2006)\citenamefont {Adams},
  \citenamefont {Arkani-Hamed}, \citenamefont {Dubovsky}, \citenamefont
  {Nicolis},\ and\ \citenamefont {Rattazzi}}]{Adams:2006sv}%
  \BibitemOpen
  \bibfield  {author} {\bibinfo {author} {\bibfnamefont {A.}~\bibnamefont
  {Adams}}, \bibinfo {author} {\bibfnamefont {N.}~\bibnamefont {Arkani-Hamed}},
  \bibinfo {author} {\bibfnamefont {S.}~\bibnamefont {Dubovsky}}, \bibinfo
  {author} {\bibfnamefont {A.}~\bibnamefont {Nicolis}},\ and\ \bibinfo {author}
  {\bibfnamefont {R.}~\bibnamefont {Rattazzi}},\ }\bibfield  {title} {\bibinfo
  {title} {{Causality, analyticity and an IR obstruction to UV completion}},\
  }\href {https://doi.org/10.1088/1126-6708/2006/10/014} {\bibfield  {journal}
  {\bibinfo  {journal} {JHEP}\ }\textbf {\bibinfo {volume} {10}},\ \bibinfo
  {pages} {014}},\ \Eprint {https://arxiv.org/abs/hep-th/0602178}
  {arXiv:hep-th/0602178} \BibitemShut {NoStop}%
\bibitem [{\citenamefont {de~Rham}\ \emph
  {et~al.}(2017{\natexlab{a}})\citenamefont {de~Rham}, \citenamefont
  {Melville}, \citenamefont {Tolley},\ and\ \citenamefont
  {Zhou}}]{deRham:2017avq}%
  \BibitemOpen
  \bibfield  {author} {\bibinfo {author} {\bibfnamefont {C.}~\bibnamefont
  {de~Rham}}, \bibinfo {author} {\bibfnamefont {S.}~\bibnamefont {Melville}},
  \bibinfo {author} {\bibfnamefont {A.~J.}\ \bibnamefont {Tolley}},\ and\
  \bibinfo {author} {\bibfnamefont {S.-Y.}\ \bibnamefont {Zhou}},\ }\bibfield
  {title} {\bibinfo {title} {{Positivity bounds for scalar field theories}},\
  }\href {https://doi.org/10.1103/PhysRevD.96.081702} {\bibfield  {journal}
  {\bibinfo  {journal} {Phys. Rev. D}\ }\textbf {\bibinfo {volume} {96}},\
  \bibinfo {pages} {081702} (\bibinfo {year} {2017}{\natexlab{a}})},\ \Eprint
  {https://arxiv.org/abs/1702.06134} {arXiv:1702.06134 [hep-th]} \BibitemShut
  {NoStop}%
\bibitem [{\citenamefont {Bellazzini}\ \emph {et~al.}(2020)\citenamefont
  {Bellazzini}, \citenamefont {Elias~Mir\'o}, \citenamefont {Rattazzi},
  \citenamefont {Riembau},\ and\ \citenamefont {Riva}}]{Bellazzini:2020cot}%
  \BibitemOpen
  \bibfield  {author} {\bibinfo {author} {\bibfnamefont {B.}~\bibnamefont
  {Bellazzini}}, \bibinfo {author} {\bibfnamefont {J.}~\bibnamefont
  {Elias~Mir\'o}}, \bibinfo {author} {\bibfnamefont {R.}~\bibnamefont
  {Rattazzi}}, \bibinfo {author} {\bibfnamefont {M.}~\bibnamefont {Riembau}},\
  and\ \bibinfo {author} {\bibfnamefont {F.}~\bibnamefont {Riva}},\ }\bibfield
  {title} {\bibinfo {title} {{Positive Moments for Scattering Amplitudes}},\
  }\href@noop {} {\  (\bibinfo {year} {2020})},\ \Eprint
  {https://arxiv.org/abs/2011.00037} {arXiv:2011.00037 [hep-th]} \BibitemShut
  {NoStop}%
\bibitem [{\citenamefont {Elvang}\ \emph {et~al.}(2012)\citenamefont {Elvang},
  \citenamefont {Freedman}, \citenamefont {Hung}, \citenamefont {Kiermaier},
  \citenamefont {Myers},\ and\ \citenamefont {Theisen}}]{Elvang:2012st}%
  \BibitemOpen
  \bibfield  {author} {\bibinfo {author} {\bibfnamefont {H.}~\bibnamefont
  {Elvang}}, \bibinfo {author} {\bibfnamefont {D.~Z.}\ \bibnamefont
  {Freedman}}, \bibinfo {author} {\bibfnamefont {L.-Y.}\ \bibnamefont {Hung}},
  \bibinfo {author} {\bibfnamefont {M.}~\bibnamefont {Kiermaier}}, \bibinfo
  {author} {\bibfnamefont {R.~C.}\ \bibnamefont {Myers}},\ and\ \bibinfo
  {author} {\bibfnamefont {S.}~\bibnamefont {Theisen}},\ }\bibfield  {title}
  {\bibinfo {title} {{On renormalization group flows and the a-theorem in
  6d}},\ }\href {https://doi.org/10.1007/JHEP10(2012)011} {\bibfield  {journal}
  {\bibinfo  {journal} {JHEP}\ }\textbf {\bibinfo {volume} {10}},\ \bibinfo
  {pages} {011}},\ \Eprint {https://arxiv.org/abs/1205.3994} {arXiv:1205.3994
  [hep-th]} \BibitemShut {NoStop}%
\bibitem [{\citenamefont {Chandrasekaran}\ \emph {et~al.}(2018)\citenamefont
  {Chandrasekaran}, \citenamefont {Remmen},\ and\ \citenamefont
  {Shahbazi-Moghaddam}}]{Chandrasekaran:2018qmx}%
  \BibitemOpen
  \bibfield  {author} {\bibinfo {author} {\bibfnamefont {V.}~\bibnamefont
  {Chandrasekaran}}, \bibinfo {author} {\bibfnamefont {G.~N.}\ \bibnamefont
  {Remmen}},\ and\ \bibinfo {author} {\bibfnamefont {A.}~\bibnamefont
  {Shahbazi-Moghaddam}},\ }\bibfield  {title} {\bibinfo {title} {{Higher-Point
  Positivity}},\ }\href {https://doi.org/10.1007/JHEP11(2018)015} {\bibfield
  {journal} {\bibinfo  {journal} {JHEP}\ }\textbf {\bibinfo {volume} {11}},\
  \bibinfo {pages} {015}},\ \Eprint {https://arxiv.org/abs/1804.03153}
  {arXiv:1804.03153 [hep-th]} \BibitemShut {NoStop}%
\bibitem [{\citenamefont {Bellazzini}\ \emph {et~al.}(2018)\citenamefont
  {Bellazzini}, \citenamefont {Riva}, \citenamefont {Serra},\ and\
  \citenamefont {Sgarlata}}]{Bellazzini:2017fep}%
  \BibitemOpen
  \bibfield  {author} {\bibinfo {author} {\bibfnamefont {B.}~\bibnamefont
  {Bellazzini}}, \bibinfo {author} {\bibfnamefont {F.}~\bibnamefont {Riva}},
  \bibinfo {author} {\bibfnamefont {J.}~\bibnamefont {Serra}},\ and\ \bibinfo
  {author} {\bibfnamefont {F.}~\bibnamefont {Sgarlata}},\ }\bibfield  {title}
  {\bibinfo {title} {{Beyond Positivity Bounds and the Fate of Massive
  Gravity}},\ }\href {https://doi.org/10.1103/PhysRevLett.120.161101}
  {\bibfield  {journal} {\bibinfo  {journal} {Phys. Rev. Lett.}\ }\textbf
  {\bibinfo {volume} {120}},\ \bibinfo {pages} {161101} (\bibinfo {year}
  {2018})},\ \Eprint {https://arxiv.org/abs/1710.02539} {arXiv:1710.02539
  [hep-th]} \BibitemShut {NoStop}%
\bibitem [{\citenamefont {Hamada}\ \emph {et~al.}(2019)\citenamefont {Hamada},
  \citenamefont {Noumi},\ and\ \citenamefont {Shiu}}]{Hamada:2018dde}%
  \BibitemOpen
  \bibfield  {author} {\bibinfo {author} {\bibfnamefont {Y.}~\bibnamefont
  {Hamada}}, \bibinfo {author} {\bibfnamefont {T.}~\bibnamefont {Noumi}},\ and\
  \bibinfo {author} {\bibfnamefont {G.}~\bibnamefont {Shiu}},\ }\bibfield
  {title} {\bibinfo {title} {{Weak Gravity Conjecture from Unitarity and
  Causality}},\ }\href {https://doi.org/10.1103/PhysRevLett.123.051601}
  {\bibfield  {journal} {\bibinfo  {journal} {Phys. Rev. Lett.}\ }\textbf
  {\bibinfo {volume} {123}},\ \bibinfo {pages} {051601} (\bibinfo {year}
  {2019})},\ \Eprint {https://arxiv.org/abs/1810.03637} {arXiv:1810.03637
  [hep-th]} \BibitemShut {NoStop}%
\bibitem [{\citenamefont {Tokuda}\ \emph {et~al.}(2020)\citenamefont {Tokuda},
  \citenamefont {Aoki},\ and\ \citenamefont {Hirano}}]{Tokuda:2020mlf}%
  \BibitemOpen
  \bibfield  {author} {\bibinfo {author} {\bibfnamefont {J.}~\bibnamefont
  {Tokuda}}, \bibinfo {author} {\bibfnamefont {K.}~\bibnamefont {Aoki}},\ and\
  \bibinfo {author} {\bibfnamefont {S.}~\bibnamefont {Hirano}},\ }\bibfield
  {title} {\bibinfo {title} {{Gravitational positivity bounds}},\ }\href
  {https://doi.org/10.1007/JHEP11(2020)054} {\bibfield  {journal} {\bibinfo
  {journal} {JHEP}\ }\textbf {\bibinfo {volume} {11}},\ \bibinfo {pages}
  {054}},\ \Eprint {https://arxiv.org/abs/2007.15009} {arXiv:2007.15009
  [hep-th]} \BibitemShut {NoStop}%
\bibitem [{\citenamefont {Herrero-Valea}\ \emph {et~al.}(2020)\citenamefont
  {Herrero-Valea}, \citenamefont {Santos-Garcia},\ and\ \citenamefont
  {Tokareva}}]{Herrero-Valea:2020wxz}%
  \BibitemOpen
  \bibfield  {author} {\bibinfo {author} {\bibfnamefont {M.}~\bibnamefont
  {Herrero-Valea}}, \bibinfo {author} {\bibfnamefont {R.}~\bibnamefont
  {Santos-Garcia}},\ and\ \bibinfo {author} {\bibfnamefont {A.}~\bibnamefont
  {Tokareva}},\ }\bibfield  {title} {\bibinfo {title} {{Massless Positivity in
  Graviton Exchange}},\ }\href@noop {} {\  (\bibinfo {year} {2020})},\ \Eprint
  {https://arxiv.org/abs/2011.11652} {arXiv:2011.11652 [hep-th]} \BibitemShut
  {NoStop}%
\bibitem [{\citenamefont {Alberte}\ \emph
  {et~al.}(2020{\natexlab{a}})\citenamefont {Alberte}, \citenamefont {de~Rham},
  \citenamefont {Jaitly},\ and\ \citenamefont {Tolley}}]{Alberte:2020jsk}%
  \BibitemOpen
  \bibfield  {author} {\bibinfo {author} {\bibfnamefont {L.}~\bibnamefont
  {Alberte}}, \bibinfo {author} {\bibfnamefont {C.}~\bibnamefont {de~Rham}},
  \bibinfo {author} {\bibfnamefont {S.}~\bibnamefont {Jaitly}},\ and\ \bibinfo
  {author} {\bibfnamefont {A.~J.}\ \bibnamefont {Tolley}},\ }\bibfield  {title}
  {\bibinfo {title} {{Positivity Bounds and the Massless Spin-2 Pole}},\ }\href
  {https://doi.org/10.1103/PhysRevD.102.125023} {\bibfield  {journal} {\bibinfo
   {journal} {Phys. Rev. D}\ }\textbf {\bibinfo {volume} {102}},\ \bibinfo
  {pages} {125023} (\bibinfo {year} {2020}{\natexlab{a}})},\ \Eprint
  {https://arxiv.org/abs/2007.12667} {arXiv:2007.12667 [hep-th]} \BibitemShut
  {NoStop}%
\bibitem [{\citenamefont {Bellazzini}\ \emph {et~al.}(2019)\citenamefont
  {Bellazzini}, \citenamefont {Lewandowski},\ and\ \citenamefont
  {Serra}}]{Bellazzini:2019xts}%
  \BibitemOpen
  \bibfield  {author} {\bibinfo {author} {\bibfnamefont {B.}~\bibnamefont
  {Bellazzini}}, \bibinfo {author} {\bibfnamefont {M.}~\bibnamefont
  {Lewandowski}},\ and\ \bibinfo {author} {\bibfnamefont {J.}~\bibnamefont
  {Serra}},\ }\bibfield  {title} {\bibinfo {title} {{Positivity of Amplitudes,
  Weak Gravity Conjecture, and Modified Gravity}},\ }\href
  {https://doi.org/10.1103/PhysRevLett.123.251103} {\bibfield  {journal}
  {\bibinfo  {journal} {Phys. Rev. Lett.}\ }\textbf {\bibinfo {volume} {123}},\
  \bibinfo {pages} {251103} (\bibinfo {year} {2019})},\ \Eprint
  {https://arxiv.org/abs/1902.03250} {arXiv:1902.03250 [hep-th]} \BibitemShut
  {NoStop}%
\bibitem [{\citenamefont {Alberte}\ \emph
  {et~al.}(2020{\natexlab{b}})\citenamefont {Alberte}, \citenamefont {de~Rham},
  \citenamefont {Jaitly},\ and\ \citenamefont {Tolley}}]{Alberte:2020bdz}%
  \BibitemOpen
  \bibfield  {author} {\bibinfo {author} {\bibfnamefont {L.}~\bibnamefont
  {Alberte}}, \bibinfo {author} {\bibfnamefont {C.}~\bibnamefont {de~Rham}},
  \bibinfo {author} {\bibfnamefont {S.}~\bibnamefont {Jaitly}},\ and\ \bibinfo
  {author} {\bibfnamefont {A.~J.}\ \bibnamefont {Tolley}},\ }\bibfield  {title}
  {\bibinfo {title} {{QED positivity bounds}},\ }\href@noop {} {\  (\bibinfo
  {year} {2020}{\natexlab{b}})},\ \Eprint {https://arxiv.org/abs/2012.05798}
  {arXiv:2012.05798 [hep-th]} \BibitemShut {NoStop}%
\bibitem [{\citenamefont {Aoki}\ \emph {et~al.}(2021)\citenamefont {Aoki},
  \citenamefont {Loc}, \citenamefont {Noumi},\ and\ \citenamefont
  {Tokuda}}]{Aoki:2021ckh}%
  \BibitemOpen
  \bibfield  {author} {\bibinfo {author} {\bibfnamefont {K.}~\bibnamefont
  {Aoki}}, \bibinfo {author} {\bibfnamefont {T.~Q.}\ \bibnamefont {Loc}},
  \bibinfo {author} {\bibfnamefont {T.}~\bibnamefont {Noumi}},\ and\ \bibinfo
  {author} {\bibfnamefont {J.}~\bibnamefont {Tokuda}},\ }\bibfield  {title}
  {\bibinfo {title} {{Is the Standard Model in the Swampland?}},\ }\href@noop
  {} {\  (\bibinfo {year} {2021})},\ \Eprint {https://arxiv.org/abs/2104.09682}
  {arXiv:2104.09682 [hep-th]} \BibitemShut {NoStop}%
\bibitem [{\citenamefont {Dvali}\ \emph {et~al.}(2012)\citenamefont {Dvali},
  \citenamefont {Franca},\ and\ \citenamefont {Gomez}}]{Dvali:2012zc}%
  \BibitemOpen
  \bibfield  {author} {\bibinfo {author} {\bibfnamefont {G.}~\bibnamefont
  {Dvali}}, \bibinfo {author} {\bibfnamefont {A.}~\bibnamefont {Franca}},\ and\
  \bibinfo {author} {\bibfnamefont {C.}~\bibnamefont {Gomez}},\ }\bibfield
  {title} {\bibinfo {title} {{Road Signs for UV-Completion}},\ }\href@noop {}
  {\  (\bibinfo {year} {2012})},\ \Eprint {https://arxiv.org/abs/1204.6388}
  {arXiv:1204.6388 [hep-th]} \BibitemShut {NoStop}%
\bibitem [{\citenamefont {de~Rham}\ \emph
  {et~al.}(2017{\natexlab{b}})\citenamefont {de~Rham}, \citenamefont
  {Melville}, \citenamefont {Tolley},\ and\ \citenamefont
  {Zhou}}]{deRham:2017imi}%
  \BibitemOpen
  \bibfield  {author} {\bibinfo {author} {\bibfnamefont {C.}~\bibnamefont
  {de~Rham}}, \bibinfo {author} {\bibfnamefont {S.}~\bibnamefont {Melville}},
  \bibinfo {author} {\bibfnamefont {A.~J.}\ \bibnamefont {Tolley}},\ and\
  \bibinfo {author} {\bibfnamefont {S.-Y.}\ \bibnamefont {Zhou}},\ }\bibfield
  {title} {\bibinfo {title} {{Massive Galileon Positivity Bounds}},\ }\href
  {https://doi.org/10.1007/JHEP09(2017)072} {\bibfield  {journal} {\bibinfo
  {journal} {JHEP}\ }\textbf {\bibinfo {volume} {09}},\ \bibinfo {pages}
  {072}},\ \Eprint {https://arxiv.org/abs/1702.08577} {arXiv:1702.08577
  [hep-th]} \BibitemShut {NoStop}%
\bibitem [{\citenamefont {Melville}\ and\ \citenamefont
  {Noller}(2020)}]{Melville:2019wyy}%
  \BibitemOpen
  \bibfield  {author} {\bibinfo {author} {\bibfnamefont {S.}~\bibnamefont
  {Melville}}\ and\ \bibinfo {author} {\bibfnamefont {J.}~\bibnamefont
  {Noller}},\ }\bibfield  {title} {\bibinfo {title} {{Positivity in the Sky:
  Constraining dark energy and modified gravity from the UV}},\ }\href
  {https://doi.org/10.1103/PhysRevD.101.021502} {\bibfield  {journal} {\bibinfo
   {journal} {Phys. Rev. D}\ }\textbf {\bibinfo {volume} {101}},\ \bibinfo
  {pages} {021502} (\bibinfo {year} {2020})},\ \bibinfo {note} {[Erratum:
  Phys.Rev.D 102, 049902 (2020)]},\ \Eprint {https://arxiv.org/abs/1904.05874}
  {arXiv:1904.05874 [astro-ph.CO]} \BibitemShut {NoStop}%
\bibitem [{\citenamefont {de~Rham}\ \emph {et~al.}(2021)\citenamefont
  {de~Rham}, \citenamefont {Melville},\ and\ \citenamefont
  {Noller}}]{deRham:2021fpu}%
  \BibitemOpen
  \bibfield  {author} {\bibinfo {author} {\bibfnamefont {C.}~\bibnamefont
  {de~Rham}}, \bibinfo {author} {\bibfnamefont {S.}~\bibnamefont {Melville}},\
  and\ \bibinfo {author} {\bibfnamefont {J.}~\bibnamefont {Noller}},\
  }\bibfield  {title} {\bibinfo {title} {{Positivity bounds on dark energy:
  when matter matters}},\ }\href
  {https://doi.org/10.1088/1475-7516/2021/08/018} {\bibfield  {journal}
  {\bibinfo  {journal} {JCAP}\ }\textbf {\bibinfo {volume} {08}},\ \bibinfo
  {pages} {018}},\ \Eprint {https://arxiv.org/abs/2103.06855} {arXiv:2103.06855
  [astro-ph.CO]} \BibitemShut {NoStop}%
\bibitem [{\citenamefont {Davis}\ and\ \citenamefont
  {Melville}(2021)}]{Davis:2021oce}%
  \BibitemOpen
  \bibfield  {author} {\bibinfo {author} {\bibfnamefont {A.-C.}\ \bibnamefont
  {Davis}}\ and\ \bibinfo {author} {\bibfnamefont {S.}~\bibnamefont
  {Melville}},\ }\bibfield  {title} {\bibinfo {title} {{Scalar fields near
  compact objects: resummation versus UV completion}},\ }\href
  {https://doi.org/10.1088/1475-7516/2021/11/012} {\bibfield  {journal}
  {\bibinfo  {journal} {JCAP}\ }\textbf {\bibinfo {volume} {11}},\ \bibinfo
  {pages} {012}},\ \Eprint {https://arxiv.org/abs/2107.00010} {arXiv:2107.00010
  [gr-qc]} \BibitemShut {NoStop}%
\bibitem [{\citenamefont {Cheung}\ and\ \citenamefont
  {Remmen}(2016)}]{Cheung:2016yqr}%
  \BibitemOpen
  \bibfield  {author} {\bibinfo {author} {\bibfnamefont {C.}~\bibnamefont
  {Cheung}}\ and\ \bibinfo {author} {\bibfnamefont {G.~N.}\ \bibnamefont
  {Remmen}},\ }\bibfield  {title} {\bibinfo {title} {{Positive Signs in Massive
  Gravity}},\ }\href {https://doi.org/10.1007/JHEP04(2016)002} {\bibfield
  {journal} {\bibinfo  {journal} {JHEP}\ }\textbf {\bibinfo {volume} {04}},\
  \bibinfo {pages} {002}},\ \Eprint {https://arxiv.org/abs/1601.04068}
  {arXiv:1601.04068 [hep-th]} \BibitemShut {NoStop}%
\bibitem [{\citenamefont {de~Rham}\ \emph {et~al.}(2018)\citenamefont
  {de~Rham}, \citenamefont {Melville},\ and\ \citenamefont
  {Tolley}}]{deRham:2017xox}%
  \BibitemOpen
  \bibfield  {author} {\bibinfo {author} {\bibfnamefont {C.}~\bibnamefont
  {de~Rham}}, \bibinfo {author} {\bibfnamefont {S.}~\bibnamefont {Melville}},\
  and\ \bibinfo {author} {\bibfnamefont {A.~J.}\ \bibnamefont {Tolley}},\
  }\bibfield  {title} {\bibinfo {title} {{Improved Positivity Bounds and
  Massive Gravity}},\ }\href {https://doi.org/10.1007/JHEP04(2018)083}
  {\bibfield  {journal} {\bibinfo  {journal} {JHEP}\ }\textbf {\bibinfo
  {volume} {04}},\ \bibinfo {pages} {083}},\ \Eprint
  {https://arxiv.org/abs/1710.09611} {arXiv:1710.09611 [hep-th]} \BibitemShut
  {NoStop}%
\bibitem [{\citenamefont {Deser}(1957)}]{Deser:1957zz}%
  \BibitemOpen
  \bibfield  {author} {\bibinfo {author} {\bibfnamefont {S.}~\bibnamefont
  {Deser}},\ }\bibfield  {title} {\bibinfo {title} {{General Relativity and the
  Divergence Problem in Quantum Field Theory}},\ }\href
  {https://doi.org/10.1103/RevModPhys.29.417} {\bibfield  {journal} {\bibinfo
  {journal} {Rev. Mod. Phys.}\ }\textbf {\bibinfo {volume} {29}},\ \bibinfo
  {pages} {417} (\bibinfo {year} {1957})}\BibitemShut {NoStop}%
\bibitem [{\citenamefont {Dima}\ \emph {et~al.}(2020)\citenamefont {Dima},
  \citenamefont {Barausse}, \citenamefont {Franchini},\ and\ \citenamefont
  {Sotiriou}}]{Dima:2020yac}%
  \BibitemOpen
  \bibfield  {author} {\bibinfo {author} {\bibfnamefont {A.}~\bibnamefont
  {Dima}}, \bibinfo {author} {\bibfnamefont {E.}~\bibnamefont {Barausse}},
  \bibinfo {author} {\bibfnamefont {N.}~\bibnamefont {Franchini}},\ and\
  \bibinfo {author} {\bibfnamefont {T.~P.}\ \bibnamefont {Sotiriou}},\
  }\bibfield  {title} {\bibinfo {title} {{Spin-induced black hole spontaneous
  scalarization}},\ }\href {https://doi.org/10.1103/PhysRevLett.125.231101}
  {\bibfield  {journal} {\bibinfo  {journal} {Phys. Rev. Lett.}\ }\textbf
  {\bibinfo {volume} {125}},\ \bibinfo {pages} {231101} (\bibinfo {year}
  {2020})},\ \Eprint {https://arxiv.org/abs/2006.03095} {arXiv:2006.03095
  [gr-qc]} \BibitemShut {NoStop}%
\bibitem [{\citenamefont {Chaichian}\ and\ \citenamefont
  {Fischer}(1988)}]{Chaichian:1987zt}%
  \BibitemOpen
  \bibfield  {author} {\bibinfo {author} {\bibfnamefont {M.}~\bibnamefont
  {Chaichian}}\ and\ \bibinfo {author} {\bibfnamefont {J.}~\bibnamefont
  {Fischer}},\ }\bibfield  {title} {\bibinfo {title} {{Higher Dimensional
  Space-time and Unitarity Bound on the Scattering Amplitude}},\ }\href
  {https://doi.org/10.1016/0550-3213(88)90394-X} {\bibfield  {journal}
  {\bibinfo  {journal} {Nucl. Phys. B}\ }\textbf {\bibinfo {volume} {303}},\
  \bibinfo {pages} {557} (\bibinfo {year} {1988})}\BibitemShut {NoStop}%
\bibitem [{\citenamefont {Froissart}(1961)}]{Froissart:1961ux}%
  \BibitemOpen
  \bibfield  {author} {\bibinfo {author} {\bibfnamefont {M.}~\bibnamefont
  {Froissart}},\ }\bibfield  {title} {\bibinfo {title} {{Asymptotic behavior
  and subtractions in the Mandelstam representation}},\ }\href
  {https://doi.org/10.1103/PhysRev.123.1053} {\bibfield  {journal} {\bibinfo
  {journal} {Phys. Rev.}\ }\textbf {\bibinfo {volume} {123}},\ \bibinfo {pages}
  {1053} (\bibinfo {year} {1961})}\BibitemShut {NoStop}%
\bibitem [{\citenamefont {Peskin}\ and\ \citenamefont
  {Schroeder}(1995)}]{Peskin:1995ev}%
  \BibitemOpen
  \bibfield  {author} {\bibinfo {author} {\bibfnamefont {M.~E.}\ \bibnamefont
  {Peskin}}\ and\ \bibinfo {author} {\bibfnamefont {D.~V.}\ \bibnamefont
  {Schroeder}},\ }\href@noop {} {\emph {\bibinfo {title} {{An Introduction to
  quantum field theory}}}}\ (\bibinfo  {publisher} {Addison-Wesley},\ \bibinfo
  {address} {Reading, USA},\ \bibinfo {year} {1995})\BibitemShut {NoStop}%
\bibitem [{\citenamefont {Antoniou}\ \emph
  {et~al.}(2018{\natexlab{b}})\citenamefont {Antoniou}, \citenamefont
  {Bakopoulos},\ and\ \citenamefont {Kanti}}]{Antoniou:2017hxj}%
  \BibitemOpen
  \bibfield  {author} {\bibinfo {author} {\bibfnamefont {G.}~\bibnamefont
  {Antoniou}}, \bibinfo {author} {\bibfnamefont {A.}~\bibnamefont
  {Bakopoulos}},\ and\ \bibinfo {author} {\bibfnamefont {P.}~\bibnamefont
  {Kanti}},\ }\bibfield  {title} {\bibinfo {title} {{Black-Hole Solutions with
  Scalar Hair in Einstein-Scalar-Gauss-Bonnet Theories}},\ }\href
  {https://doi.org/10.1103/PhysRevD.97.084037} {\bibfield  {journal} {\bibinfo
  {journal} {Phys. Rev. D}\ }\textbf {\bibinfo {volume} {97}},\ \bibinfo
  {pages} {084037} (\bibinfo {year} {2018}{\natexlab{b}})},\ \Eprint
  {https://arxiv.org/abs/1711.07431} {arXiv:1711.07431 [hep-th]} \BibitemShut
  {NoStop}%
\bibitem [{\citenamefont {Brizuela}\ \emph {et~al.}(2009)\citenamefont
  {Brizuela}, \citenamefont {Martin-Garcia},\ and\ \citenamefont
  {Mena~Marugan}}]{Brizuela:2008ra}%
  \BibitemOpen
  \bibfield  {author} {\bibinfo {author} {\bibfnamefont {D.}~\bibnamefont
  {Brizuela}}, \bibinfo {author} {\bibfnamefont {J.~M.}\ \bibnamefont
  {Martin-Garcia}},\ and\ \bibinfo {author} {\bibfnamefont {G.~A.}\
  \bibnamefont {Mena~Marugan}},\ }\bibfield  {title} {\bibinfo {title} {{xPert:
  Computer algebra for metric perturbation theory}},\ }\href
  {https://doi.org/10.1007/s10714-009-0773-2} {\bibfield  {journal} {\bibinfo
  {journal} {Gen. Rel. Grav.}\ }\textbf {\bibinfo {volume} {41}},\ \bibinfo
  {pages} {2415} (\bibinfo {year} {2009})},\ \Eprint
  {https://arxiv.org/abs/0807.0824} {arXiv:0807.0824 [gr-qc]} \BibitemShut
  {NoStop}%
\bibitem [{\citenamefont {Nutma}(2014)}]{Nutma:2013zea}%
  \BibitemOpen
  \bibfield  {author} {\bibinfo {author} {\bibfnamefont {T.}~\bibnamefont
  {Nutma}},\ }\bibfield  {title} {\bibinfo {title} {{xTras : A field-theory
  inspired xAct package for mathematica}},\ }\href
  {https://doi.org/10.1016/j.cpc.2014.02.006} {\bibfield  {journal} {\bibinfo
  {journal} {Comput. Phys. Commun.}\ }\textbf {\bibinfo {volume} {185}},\
  \bibinfo {pages} {1719} (\bibinfo {year} {2014})},\ \Eprint
  {https://arxiv.org/abs/1308.3493} {arXiv:1308.3493 [cs.SC]} \BibitemShut
  {NoStop}%
\bibitem [{\citenamefont {Patel}(2015)}]{Patel:2015tea}%
  \BibitemOpen
  \bibfield  {author} {\bibinfo {author} {\bibfnamefont {H.~H.}\ \bibnamefont
  {Patel}},\ }\bibfield  {title} {\bibinfo {title} {{Package-X: A Mathematica
  package for the analytic calculation of one-loop integrals}},\ }\href
  {https://doi.org/10.1016/j.cpc.2015.08.017} {\bibfield  {journal} {\bibinfo
  {journal} {Comput. Phys. Commun.}\ }\textbf {\bibinfo {volume} {197}},\
  \bibinfo {pages} {276} (\bibinfo {year} {2015})},\ \Eprint
  {https://arxiv.org/abs/1503.01469} {arXiv:1503.01469 [hep-ph]} \BibitemShut
  {NoStop}%
\end{thebibliography}%
\end{document}